\documentclass[12pt]{article}
\pdfoutput=1
\usepackage{amssymb,cite,graphicx}
\usepackage{slashed}
\usepackage{amsmath,bm,bbm}
\usepackage{amsfonts}
\usepackage[titletoc,title]{appendix}
\usepackage[small]{caption}
\usepackage[margin=1in]{geometry}
\usepackage[multiple]{footmisc}
\usepackage{mathtools}
\usepackage{slashed}
\usepackage[nottoc]{tocbibind}
\usepackage{xcolor}

\definecolor{mediumblue}{rgb}{0,0,0.8}
\usepackage{hyperref}
\hypersetup{
  linktocpage=true,
  colorlinks=true,
  citecolor=mediumblue,
  filecolor=mediumblue,
  linkcolor=mediumblue,
  urlcolor=mediumblue,
}

\graphicspath{{./}{./figures/}}
\numberwithin{equation}{section}
\allowdisplaybreaks%

\newcommand{\be}{\begin{equation}}
\newcommand{\ee}{\end{equation}}
\newcommand{\bea}{\begin{eqnarray}}
\newcommand{\eea}{\end{eqnarray}}

\begin{document}

\begin{titlepage}
  \begin{flushright}
    \texttt{CTPU-17-38}
  \end{flushright}
  \medskip

  \begin{center}
    {\Large\bf\boldmath
      $B$-meson anomalies and Higgs physics\\[0.2cm]
      in flavored $U(1)'$ model}
    \bigskip

    {\bf Ligong Bian$^{1,2,*}$, Hyun Min Lee$^{2,\dagger}$ and Chan Beom Park$^{3,4,\ddagger}$}

    {\it $^1$Department of Physics,
      Chongqing University, Chongqing 401331, China}\\[0.2cm]
    {\it $^2$Department of Physics, Chung-Ang University, Seoul
      06974, Korea}\\[0.2cm]
    {\it $^3$Center for Theoretical Physics of the Universe,\\
      Institute for Basic Science (IBS), Daejeon 34051, Korea}\\[0.2cm]
    {\it $^4$School of Physics, Korea Institute for Advanced Study,
      Seoul 02455, Korea}
  \end{center}

  \bigskip

  \begin{abstract}
    \noindent
    We consider a simple extension of the Standard Model with
    flavor-dependent $U(1)'$, that has been proposed to explain some of
    $B$-meson anomalies recently reported at LHCb. The
    $U(1)'$ charge is chosen as a linear combination of anomaly-free
    $B_3-L_3$ and $L_\mu-L_\tau$. In this model, the flavor structure in the SM is restricted due to flavor-dependent $U(1)'$ charges, in particular,  quark mixings are
    induced by a small vacuum expectation value of the extra Higgs
    doublet. As a result, it is natural to get sizable flavor-violating Yukawa couplings of
    heavy Higgs bosons involving the bottom quark.
    In this article, we focus on the
    phenomenology of the Higgs sector of the model including extra Higgs
    doublet and singlet scalars. We impose various bounds on
    the extended Higgs sector from Higgs and electroweak precision
    data, $B$-meson mixings and decays as well as unitarity and
    stability bounds, then discuss the productions and decays of heavy
    Higgs bosons at the LHC\@.
  \end{abstract}

  \vspace*{\fill}
  \begin{flushleft}
    $^*$Email: ligongbian@gmail.com  \\
    $^\dagger$Email: hminlee@cau.ac.kr \\
    $^\ddagger$Email: cbpark@ibs.re.kr
  \end{flushleft}
\end{titlepage}

\section{Introduction}

The observed fermion masses and mixing angles are well parametrized by
the Higgs Yukawa couplings in the Standard Model (SM). However, the
neutrino masses and mixing angles call for the addition of
right-handed (RH) neutrinos or physics beyond the SM and, moreover, the flavor
structures of quarks and leptons are not understood yet.
As there is no flavor changing neutral current at tree level in the SM
due to the GIM mechanism, the observation of flavor violation is an
important probe of new physics up to very high energy scales and it
can be complementary to direct searches at the LHC\@.
In particular, the violation of lepton flavor universality would be a
strong hint at new physics.

Recently, there have been interesting reports on the anomalies in rare
semileptonic $B$-meson decays at LHCb such as $R_K$~\cite{RK},
$R_{K^*}$~\cite{RKs}, $P'_5$~\cite{P5}.
The reported value of $R_K={\cal B}(B\rightarrow K\mu^+\mu^-)/{\cal
  B}(B\rightarrow Ke^+e^-)$ is
\begin{equation}
  R_K=0.745^{+0.097}_{-0.082}, \quad 1\,{\rm GeV}^2<q^2 <6\,{\rm GeV}^2,
\end{equation}
which deviates from the SM prediction by $2.6\sigma$.
On the other hand for vector $B$-mesons, $R_{K^*}={\cal
  B}(B\rightarrow K^*\mu^+\mu^-)/{\cal B}(B\rightarrow
K^*e^+e^-)$ is
\begin{align}
  R_{K^*}&= 0.66^{+0.11}_{-0.07}(\rm stat)\pm 0.03({\rm syst}), \quad 0.045\,{\rm GeV}^2<q^2 <1.1\,{\rm GeV}^2,\nonumber\\
  R_{K^*}&= 0.69^{+0.11}_{-0.07}({\rm stat})\pm 0.05({\rm syst}), \quad 1.1\,{\rm GeV}^2<q^2 <6.0\,{\rm GeV}^2,
\end{align}
which again differs from the SM prediction by 2.1--$2.3\sigma$ and
2.4--$2.5\sigma$, depending on the energy bins.
Explaining the $B$-meson anomalies would require new physics
violating the lepton flavor universality at a few 100~GeV up to a few
10~TeV, depending on the coupling strength of new particles to the
SM\@.  We also note that there have been interesting anomalies in
$B\rightarrow D^{(*)}\tau\nu$ decays, the so called $R_{D^{(*)}}={\cal
  B}(B\rightarrow D^{(*)}\tau\nu)/{\cal B}(B\rightarrow D^{(*)} \ell
\nu)$ with $\ell=e$, $\mu$, whose experimental values are deviated
from the SM values by more than $2\sigma$~\cite{RDexp,RDexp2,RDsexp,RDsexp2}.

Motivated by the $B$-anomalies $R_{K^{(*)}}$, some of the authors recently proposed
a simple extension of the SM with extra $U(1)'$ gauge symmetry with
flavor-dependent couplings~\cite{Bian:2017rpg}.
The $U(1)'$ symmetry is taken as a linear combination of
${U(1)}_{L_\mu-L_\tau}$ and ${U(1)}_{B_3-L_3}$, which might be a good
symmetry at low energy and originated from enhanced gauge symmetries
such as in the $U(1)$ clockwork framework~\cite{u1cw}.
In this model, the quark mixings and
neutrino masses/mixings require an extended Higgs sector, which has
one extra Higgs doublet and multiple singlet scalars beyond the SM\@.
As a result, nonzero off-diagonal components of quark mass
matrices are obtained from the vacuum expectation value (VEV) of the
extra Higgs doublet and correct electroweak symmetry breaking is
ensured by the VEV of one of the singlet scalars.

In this paper, we study the phenomenology of the heavy Higgs bosons
in the flavored $U(1)'$ model mentioned above.
We first show that the correct flavor structure of the SM is well reproduced
in the presence of the VEV of the extra Higgs doublet. In particular, in the
case with a small VEV of the extra Higgs doublet or small $\tan\beta$,
we find that the heavy Higgs bosons have sizable flavor-violating
couplings to the bottom quark and reduced flavor-conserving Yukawa
couplings to the top quark such that
LHC searches for heavy Higgs bosons can be affected by extra or
modified production and decay channels. We also briefly mention the implication of our extended Higgs sector for $R_{D^{(*)}}$ anomalies. We discuss various constraints on the extended Higgs sector
from Higgs and electroweak precision data,
flavor data such as the $B$-meson mixings and decays, as well as
unitarity and stability bounds. For certain benchmark points that
can evade such bounds, we study the productions and decays of the heavy
Higgs bosons at the LHC and show distinct features of the model with
flavor-violating interactions in the Higgs sector.

This paper is organized as follows.
First, we begin with a summary of the $U(1)'$ model with the extended
Higgs sector and new interactions.
The Higgs spectrum and Yukawa couplings for heavy Higgs bosons
are presented in Sec.~\ref{sec:higgs_spectrum}.
We then discuss various theoretical and phenomenological constraints
on the Higgs sector are studied in Sec.~\ref{sec:constraints}, and
collider signatures of the heavy Higgs bosons at the LHC are studied
in Sec.~\ref{sec:higgs_lhc}.
Finally, conclusions are drawn. There are four
appendices dealing with the extended Higgs sector, unitarity bounds,
quark Yukawa couplings, and the $U(1)'$ interactions.

\section{Flavored \boldmath{$U(1)'$} model}

We consider a simple extension of the SM with $U(1)'$, where a new gauge
boson $Z^\prime$ couples specifically to heavy flavors.
It is taken as a linear combination of ${U(1)}_{L_\mu-L_\tau}$ and
${U(1)}_{B_3-L_3}$ with
\begin{equation*}
  Q' \equiv y(L_\mu-L_\tau)+x(B_3-L_3)
\end{equation*}
for real parameters $x$ and $y$~\cite{Bian:2017rpg}.\footnote{We
note that we can take two independent parameters for the $Z'$
couplings to be either $(x g_{Z'}, \, y g_{Z'})$ or $(x/y, \,
g_{Z'})$ by absorbing $y$ into $g_{Z'}$. Our following discussion
does not depend on the choice of the $Z'$ couplings.}
Introducing two Higgs doublets $H_{1,2}$ is necessary to have right
quark masses and mixings.
We add one complex singlet scalar $S$ for a correct vacuum to break
electroweak symmetry and $U(1)'$.
Moreover, in order to cancel the anomalies,
the fermion sector is required to include at least two RH neutrinos
$\nu_{iR}$ ($i = 2$, 3).
One more RH neutrino $\nu_{1R}$ with zero $U(1)'$ charge as well as
extra singlet scalars, $\Phi_a$ $(a=1,\,2,\,3)$, with $U(1)'$ charges
of $-y$, $x+y$, $x$, respectively, are also necessary for neutrino
masses and mixings.
As $L_\mu-L_\tau$ is extended to RH neutrinos, $L_\mu-L_\tau$ and
$L_2-L_3$ can be used interchangeably in our model.
The $U(1)'$ charge assignments are given in Table~\ref{modelA}.
\begin{table}[hbt!]
  \begin{center}
    \begin{tabular}{c|ccccccccc}
      \hline\hline
      &&&&&&&&&\\[-2mm]
      & $q_{3L}$ & $u_{3R}$  &  $d_{3R}$ & $\ell_{2L}$  & $e_{2R}$
      & $\nu_{2R}$ & $\ell_{3L}$ & $e_{3R}$ & $\nu_{3R}$\\[2mm]
      \hline
      &&&&&&&&&\\[-2mm]
      $Q'$ & $\frac{1}{3}x$ & $\frac{1}{3}x$ & $\frac{1}{3}x$
            & $y$ & $y$ & $y$ & $-x-y$ & $-x-y$ & $-x-y$\\[2mm]
      \hline\hline
    \end{tabular}\\[2mm]
    \begin{tabular}{c|cccccc}
      \hline\hline
      &&&&&&\\[-2mm]
      & $S$  & $H_1$ & $H_2$ & $\Phi_1$ & $\Phi_2$ & $\Phi_3$\\[2mm]
      \hline
      &&&&&&\\[-2mm]
      $Q'$ &  $\frac{1}{3}x$ & $0$ & $-\frac{1}{3}x$  & $-y$ & $x+y$
                     & $x$\\[2mm]
      \hline\hline
    \end{tabular}
  \end{center}
    \caption{$U(1)'$ charges of fermions and scalars.\label{modelA}}
\end{table}

The Lagrangian of the model is given as
\begin{equation}
  {\cal L}=-\frac{1}{4}Z'_{\mu\nu} Z^{\prime\mu\nu}-\frac{1}{2}
  \sin\xi \, Z'_{\mu\nu} B^{\mu\nu}+ {\cal L}_S + {\cal L}_Y
\end{equation}
with
\begin{equation}
  {\cal L}_S= |D_\mu H_1|^2+|D_\mu H_2|^2+|D_\mu S|^2+\sum_{a=1}^3|D_\mu\Phi_a| - V(\phi_i),
\end{equation}
where $Z'_{\mu\nu}=\partial_\mu Z'_\nu -\partial_\nu Z'_\mu$ is the
field strength of the $U(1)'$ gauge boson, $\sin\xi$ is
the gauge kinetic mixing between $U(1)'$ and SM hypercharge,
and $D_\mu \phi_i=(\partial_\mu -ig_{Z'}Q'_{\phi_i} Z'_\mu)\phi_i$
are covariant derivatives. Here $Q'_{\phi_i}$ is the $U(1)'$ charge of
$\phi_i$, $g_{Z'}$ is the extra gauge coupling.
The scalar potential $V(\phi_i)$ is given by $V=V_1+V_2$ with
\begin{align}
  V_1=
  &~\mu^2_1 |H_1|^2 + \mu^2_2 |H_2|^2- \left( \mu S H^\dagger_1 H_2
    + \mathrm{h.c.}\right) \nonumber \\
  &+\lambda_1 |H_1|^4+\lambda_2 |H_2|^4 + 2\lambda_3
    |H_1|^2|H_2|^2+2\lambda_4 (H^\dagger_1 H_2)(H^\dagger_2 H_1)
    \nonumber \\
  &+ 2 |S|^2(\kappa_1 |H_1|^2 +\kappa_2
    |H_2|^2)+m^2_{S}|S|^2+\lambda_{S}|S|^4,\label{eq:scalar_potential_1}\\
  V_2=
  &~\sum_{a=1}^3\Big( \mu^2_{\Phi_a} |\Phi_i|^2
    +\lambda_{\Phi_a}|\Phi_a|^4 \Big)+ \left(\lambda_{S3} S^3
    \Phi^\dagger_3 +\mu_4 \Phi_1 \Phi_2 \Phi^\dagger_3
    + \mathrm{h.c.} \right) \nonumber \\
  &+ 2\sum_{a=1}^3 |\Phi_a|^2(\beta_{a1} |H_1|^2 +\beta_{a2}  |H_2|^2+
    \beta_{a3} |S|^2)+2 \sum_{a<b}\lambda_{ab} |\Phi_a|^2 |\Phi_b|^2.
\end{align}
The extended Higgs sector is presented in the next section and
studied in more detail in Appendix~\ref{app:higgs_sector}.
For a set of quartic couplings for $S$ and $H_{1,2}$ that are relevant
for electroweak symmetry and $U(1)'$ breaking, we have collected
unitarity bounds in Appendix~\ref{app:unitarity_bounds},
which are used to constrain the parameter space of the Higgs sector in
Sec.~\ref{sec:constraints}.

The Yukawa Lagrangian for quarks and leptons is given by
\begin{align}
  -{\cal L}_Y=
  &~{\bar q}_i ( y^u_{ij}{\tilde H}_1+ h^u_{ij}{\tilde H}_2 ) u_j
    +{\bar q}_i ( y^d_{ij} {H}_1+h^d_{ij} {H}_2 ) d_j  \nonumber \\
  &+y^\ell_{ij} {\bar \ell}_i {H}_1 e_j + y^\nu_{ij} {\bar \ell}_i
    {\tilde H}_1 \nu_{jR} + \overline {({\nu_{iR}})^c}
    ( M_{ij}+\Phi_a z^{(a)}_{ij} )\nu_{jR} + \mathrm{h.c.}
\end{align}
with ${\tilde H}_{1,2}\equiv i\sigma_2 H^*_{1,2}$.
After electroweak symmetry and $U(1)'$ are broken by the VEVs of
scalar fields, $\langle H_{1,2}\rangle=v_{1,2} / \sqrt{2}$ with $v^2_1+v^2_2=v^2=(246\,{\rm GeV})^2$ ,
$\langle S\rangle=v_s / \sqrt{2}$ and
$\langle\Phi_a\rangle=\omega_a / \sqrt{2}$, the quark and lepton mass
terms are given as
\begin{equation}
  {\cal L}_Y=- {\bar u} M_u u-{\bar d} M_d d - {\bar \ell} M_\ell \ell
  - {\bar \ell} M_D \nu_R -  \overline{({\nu_{R}})^c}M_R \nu_R
  + \mathrm{h.c.}
\end{equation}
with the following flavor structure:
\begin{align}
  M_u
  &= \begin{pmatrix}
    y^u_{11}\langle  {\tilde H}_1\rangle &
    y^u_{12}\langle {\tilde H}_1\rangle & 0 \\
    y^u_{21} \langle {\tilde H}_1\rangle &
    y^u_{22} \langle {\tilde H}_1 \rangle &  0 \\
    h^u_{31} \langle {\tilde H}_2 \rangle &
    h^u_{32}\langle {\tilde H}_2\rangle &
    y^u_{33} \langle {\tilde H}_1 \rangle
  \end{pmatrix},\label{qmass1} \\
  M_d
  &= \begin{pmatrix}
    y^d_{11}\langle { H}_1\rangle &
    y^d_{12}\langle { H}_1\rangle &
    h^d_{13} \langle {H}_2\rangle \\
    y^d_{21} \langle { H}_1 \rangle &
    y^d_{22} \langle {H}_1\rangle &
    h^d_{23}\langle {H}_2\rangle \\
    0 &  0 & y^d_{33}  \langle { H}_1 \rangle
  \end{pmatrix},\label{qmass2} \\
  M_\ell
  &= \begin{pmatrix}
    y^\ell_{11} \langle { H}_1 \rangle & 0 & 0 \\
    0 & y^\ell_{22}  \langle { H}_1 \rangle & 0 \\
    0 & 0 & y^\ell_{33} \langle { H}_1\rangle
    \end{pmatrix},\label{cleptonmass} \\
  M_D
  &= \begin{pmatrix}
    y^\nu_{11} \langle {\tilde H}_1 \rangle & 0 & 0 \\
    0 & y^\nu_{22}\langle {\tilde H}_1 \rangle & 0 \\
    0 & 0 &  y^\nu_{33} \langle {\tilde H}_1 \rangle
    \end{pmatrix},\\
  M_R
  &= \begin{pmatrix}
    M_{11} & z^{(1)}_{12} \langle \Phi_1\rangle &
    z^{(2)}_{13} \langle\Phi_2\rangle \\
    z^{(1)}_{21}\langle\Phi_1\rangle & 0 &
    z^{(3)}_{23}\langle\Phi_3\rangle \\
    z^{(2)}_{31}\langle \Phi_2\rangle &
    z^{(3)}_{32} \langle\Phi_3\rangle & 0
  \end{pmatrix}.\label{RHmass}
\end{align}
Since the mass matrix for charged leptons is already diagonal, the
lepton mixings come from the mass matrix of RH neutrinos.
There are four other categories of neutrino mixing
matrices~\cite{neutrinomix}, that are compatible with neutrino
data. In all the cases, we need at least three complex scalar
fields with different $U(1)'$ charges, similarly to the case given
in~(\ref{RHmass}).
The quark Yukawa couplings to Higgs bosons are summarized in
Appendix~\ref{app:quark_yukawa}.

We find the $Z$-like ($Z_1$) and $Z'$-like ($Z_2$) masses as
\begin{equation}
  m^2_{Z_{1,2}}= \frac{1}{2} \Big(m^2_Z+m^2_{22}\mp
  \sqrt{(m^2_Z-m^2_{22})^2+4 m^4_{12}} \Big),
\end{equation}
where $m^2_Z\equiv (g^2+g^2_Y) v^2/4$, and
\begin{align}
  m^2_{22}
  &\equiv m^2_Z s^2_W t^2_\xi + m^2_{Z'}/c^2_\xi - c^{-1}_W e g_{Z'}
    Q'_{H_2} v^2_w t_\xi/c_\xi, \nonumber\\
  m^2_{12}
  &\equiv m^2_Z s_W t_\xi - \frac{1}{2} c^{-1}_W s^{-1}_W e g_{Z'}
    Q'_{H_2} v^2_2/c_\xi
\end{align}
with
\begin{equation}
  m^2_{Z'}=g^2_{Z'} \left(\frac{1}{9}x^2 v^2_s+ y^2 \omega^2_1+(x+y)^2
  \omega^2_2+ x^2 \omega^2_3 \right).
\end{equation}
Here $s_{\varphi} \equiv \sin\varphi$, $c_{\varphi} \equiv
\cos\varphi$, and $t_{\varphi} \equiv \tan\varphi$.
The modified $Z$ boson mass can receive constraints from electroweak
precision data, which is studied in Sec.~\ref{sec:constraints}.
We note that for a small mass mixing, the $Z'$-like mass is approximately
given by $m^2_{Z_2}\approx m^2_{Z'}$ and we can treat $m_{Z'}$ and
$g_{Z'}$ to be independent parameters due to the presence of nonzero
$\omega_i$'s. The $U(1)'$ interactions are collected in
Appendix~\ref{app:u1_ints}.

\section{Higgs spectrum and Yukawa couplings\label{sec:higgs_spectrum}}

We here specify the Higgs spectrum of our model and identify the quark
and lepton Yukawa couplings of neutral and charged Higgs bosons for
studies in next sections. The expressions are based on results in
Appendices~\ref{app:higgs_sector} and~\ref{app:quark_yukawa}.

\subsection{The Higgs spectrum}

The Higgs sector of our model has two Higgs doublets, which are
expressed in components as
\begin{equation}
  H_j = \begin{pmatrix}
    \phi^+_j \\
    (v_j+\rho_j+i\eta_j)/\sqrt{2}
  \end{pmatrix} \quad (j = 1, \, 2),
\end{equation}
and the complex singlet scalar decomposed into $S=\left(v_s+S_R+i
  S_I\right)/\sqrt{2}$.

In the limit of negligible mixing with the $CP$-even singlet scalar,
the mass eigenstates of $CP$-even neutral Higgs scalars, $h$ and $H$, are
given by
\begin{align}
  h &= - \sin\alpha \, \rho_1 + \cos\alpha \, \rho_2, \nonumber\\
  H &= \cos\alpha \, \rho_1 + \sin\alpha \, \rho_2. \label{base-h}
\end{align}
The general case where the $CP$-even part of the singlet scalar
$S$ mixes with the Higgs counterpart is considered in
Appendix~\ref{app:higgs_sector}. The mass eigenvalues of
$CP$-even neutral Higgs scalars are denoted as $m_{h_{1,2,3}}$ with
$m_{h_1}<m_{h_2}<m_{h_3}$, alternatively, $m_h\equiv m_{h_1}$, $m_H\equiv m_{h_2}$ and $m_s\equiv m_{h_3}$,  and there are three mixing angles,
$\alpha_{1,2,3}$: $\alpha_1=\alpha$ in the limit of a decoupled
$CP$-even singlet scalar, while $\alpha_2$ and $\alpha_3$ are mixing
angles between $\rho_{1,2}$ and $S_R$, respectively.
For $ 2\kappa_1 v_1 v_s\approx \mu v_2 / \sqrt{2}$ and $2\kappa_2 v_2
v_s\approx \mu v_1 / \sqrt{2}$, the mixing between $\rho_{1,2}$ and
$S_R$ can be neglected. For a later discussion, we focus mainly on
this case.

The $CP$-odd parts of the singlet scalars, $S$ and $\Phi_a$, can mix
with the Higgs counterpart due to a nonzero $U(1)'$ charge of the
second Higgs $H_2$, but for a small $x$ and small VEV of $H_2$,
the mixing effect is negligible.
In this case, the neutral Goldstone boson $G^0$ and the
$CP$-odd Higgs scalar $A^0$ are turned out to be
\begin{align}
    G^0 &= \cos\beta \, \eta_1 + \sin\beta \, \eta_2 , \nonumber\\
    A^0 &= \sin\beta \, \eta_1 - \cos\beta \, \eta_2 \label{base-A}
\end{align}
with $\tan\beta\equiv v_2/v_1$.
The massless combination of $\eta_1$ and $\eta_2$ is
eaten by the $Z$ boson, while a linear combination of $S_I$ and other
pseudoscalars of $\Phi_a$ is eaten by the $Z'$ boson if the $Z'$ mass
is determined dominantly by the VEV of $S$.
The other combination of the $CP$-odd scalars from two Higgs doublets
has the mass of
\begin{equation}
  m_A^2 = \frac{\mu \sin\beta \cos\beta}{\sqrt{2} v_s} \left( v^2 +
    \frac{v_s^2}{\sin^2\beta \cos^2\beta} \right). \label{A0mass}
\end{equation}

On the other hand, the charged Goldstone bosons $G^+$ and charged
Higgs scalar $H^+$ identified as
\begin{align}
  G^+ &= \cos\beta \, \phi_1^+ + \sin\beta \, \phi_2^+ ,\nonumber\\
  H^+ &= \sin\beta \, \phi_1^+ - \cos\beta \, \phi_2^+ \label{base-chargedH}
\end{align}
with nonzero mass eigenvalue given by
\begin{equation}
  m_{H^+}^2 = m_A^2 - \left ( \frac{\mu\sin\beta \cos\beta}{\sqrt{2}
      v_s} +\lambda_4 \right ) v^2.  \label{H+mass}
\end{equation}

We remark that in the limit of $\mu v_s\gg v^2$, the heavy scalars in
the Higgs doublets become almost degenerate as $m^2_A\approx
m^2_H\approx m^2_{H^+}\approx \mu v_s/(\sqrt{2}\sin\beta\cos\beta)$
and $m^2_s\approx  2\lambda_S v^2_s$ from Eqs.~(\ref{A0mass}),
(\ref{H+mass}) and (\ref{h0s}). In this limit, the mixing angles
between the SM-like Higgs and extra scalars can be negligibly small
and the resulting Higgs spectrum is consistent with Higgs data
and electroweak precision tests (EWPT) as will be discussed in
Subsec~\ref{sec:higgs_ewpd}. But, as $\mu v_s$ is constrained by
perturbativity and unitarity bounds on the quartic couplings with
Eq.~(\ref{lams0}) or (\ref{lams}), as will be discussed in
Sec.~\ref{sec:constraints}, the extra scalars in our model remain
non-decoupled. Since it is sufficient to take almost degenerate masses
for two of $m_A$, $m_H$, and $m_{H^+}$ for EWPT, we henceforth
consider more general scalar masses but with small mixings between the
SM-like Higgs and the extra neutral scalars.

\subsection{Quark mass matrices}

We now consider the quark mass matrices and their diagonalization.
After two Higgs doublets develop VEVs, we obtain the quark mass matrices
from Eqs.~(\ref{qmass1}) and~(\ref{qmass2}) as
\begin{align}
  {(M_u)}_{ij}
  &= \frac{1}{\sqrt{2}} v\cos\beta
    \begin{pmatrix}
      y^u_{11} & y^u_{12} & 0\\ y^u_{21} & y^u_{22}  & 0 \\ 0 & 0 & y^u_{33}
    \end{pmatrix} + \frac{1}{\sqrt{2}} v\sin\beta \begin{pmatrix}
      0 & 0 & 0\\ 0 & 0  & 0 \\ h^u_{31} & h^u_{32} & 0
    \end{pmatrix},  \nonumber\\
  {(M_d)}_{ij}
  &= \frac{1}{\sqrt{2}} v\cos\beta
    \begin{pmatrix}
      y^d_{11} & y^d_{12} & 0\\ y^d_{21} & y^d_{22}  & 0 \\ 0 & 0 & y^d_{33}
    \end{pmatrix} + \frac{1}{\sqrt{2}} v\sin\beta \begin{pmatrix}
      0 & 0 & h^d_{13}\\ 0 & 0  & h^d_{23} \\ 0 & 0 & 0
    \end{pmatrix}.
\end{align}
The quark mass matrices can be diagonalized by
\begin{equation}
  U^\dagger_L M_u U_R
  = M^D_u= \begin{pmatrix}
    m_u  & 0 & 0\\ 0 & m_c & 0 \\ 0 & 0 & m_t
  \end{pmatrix}, \quad
  D^\dagger_L M_d D_R
  = M^D_d= \begin{pmatrix}
    m_d  & 0 & 0\\ 0 & m_s & 0 \\ 0 & 0 & m_b
  \end{pmatrix},
\end{equation}
thus the CKM matrix is given as $V_\text{CKM}= U^\dagger_L D_L$.
We note that the Yukawa couplings of the second Higgs doublet are
sources of flavor violation, which could be important in meson
decays/mixings and collider searches for flavor-violating top decays
and/or heavy Higgs bosons~\cite{flavor,FChiggs,crivellin2017}. The
detailed derivation of flavor-violating Higgs couplings is
presented in the next section.

Since $h^u_{31}$ and  $h^u_{32}$ correspond to rotations of right-handed
up-type quarks, we can take $U_L=1$, so $V_\text{CKM}=D_L$. In this
case, we have an approximate relation for the down-type quark mass
matrix, $M_d\approx V_\text{CKM} M^D_d $, up to $m_{d,s}/m_b$
corrections. Then the Yukawa couplings between the third and first
two generations are given as follows.
\begin{equation}
  h^d_{13} =\frac{\sqrt{2} m_b}{v\sin\beta}\, V_{ub},\quad
  h^d_{23} =\frac{\sqrt{2} m_b}{v\sin\beta}\, V_{cb}.\label{hd}
\end{equation}
For $V_{ub}\simeq 0.004\ll V_{cb}\simeq 0.04$, we have $h^d_{13}\ll
h^d_{23}$. The down-type Yukawa couplings are determined as
\begin{align}
y^d_{11}&= \frac{\sqrt{2} m_d}{v\cos\beta}\, V_{ud}, \quad
y^d_{12}= \frac{\sqrt{2} m_s}{v\cos\beta}\, V_{us}, \nonumber\\
y^d_{21}&= \frac{\sqrt{2} m_d}{v\cos\beta}\, V_{cd}, \quad
y^d_{22}= \frac{\sqrt{2} m_s}{v\cos\beta}\, V_{cs}, \quad
y^d_{33}= \frac{\sqrt{2} m_b}{v\cos\beta}\, V_{tb}.
\end{align}

On the other hand, taking $U_L=1$ as above, we find another approximate
relation for the up-type quark mass matrix: $M_u=M^D_u U^\dagger_R$.
Then the rotation mass matrix for right-handed down-type quarks
becomes $U^\dagger_R={(M^D_u)}^{-1} M_u$, which is given as
\begin{equation}
  U^\dagger_R= \frac{1}{\sqrt{2}}
  \begin{pmatrix}
    \frac{v}{m_u}\cos\beta\, y^u_{11} &  \frac{v}{m_u}\cos\beta \, y^u_{12} & 0\\  \frac{v}{m_c}\cos\beta \, y^u_{21} &  \frac{v}{m_c}\cos\beta \, y^u_{22}  & 0 \\  \frac{v}{m_t}\sin\beta\,  h^u_{31} &  \frac{v}{m_t}\sin\beta\, h^u_{32} &  \frac{v}{m_t}\cos\beta\,  y^u_{33}
  \end{pmatrix}.
\end{equation}
From the unitarity condition of $U_R$ we further find the following
constraints on the up-type quark Yukawa couplings:
\begin{align}
  |y^u_{11}|^2 + |y^u_{12}|^2 &= \frac{2m^2_u}{v^2\cos^2\beta}, \label{UR1} \\
  |y^u_{21}|^2+ |y^u_{22}|^2 &=  \frac{2m^2_c}{v^2 \cos^2\beta},  
  \\
  |y^u_{33}|^2+ \tan^2\beta (|h^u_{31}|^2+|h^u_{32}|^2 )&= \frac{2m^2_t}{v^2\cos^2\beta},   \label{UR2}  \\
  y^u_{11}(y^u_{21})^* + y^u_{12} (y^u_{22})^* &= 0,  \label{UR3} \\
  y^u_{21} (h^u_{31})^*+ y^u_{22} (h^u_{32})^* &= 0, \label{UR4}  \\
  y^u_{11} (h^u_{31})^*+ y^u_{12} (h^u_{32})^* &= 0. \label{UR5}
\end{align}

\subsection{Quark Yukawa couplings}

Using the results in Appendix~\ref{app:quark_yukawa}, we get the
Yukawa interactions for the SM-like Higgs boson $h$ and heavy neutral
Higgs bosons $H$, $A$ as
\begin{align}
  -\mathcal{L}_{Y}^{h/H/A} =
  &~\frac{\cos(\alpha - \beta)}{\sqrt{2} \cos\beta} \bar b_R \left(
    \tilde h_{13}^{d \ast} d_L +  \tilde h_{23}^{d \ast} s_L  \right)
    h + \frac{\lambda_b^h}{\sqrt{2}} \bar b_R b_L h+ \frac{\lambda_t^h}{\sqrt{2}} \bar t_R t_L h
    \nonumber\\
  & + \frac{\sin(\alpha - \beta)}{\sqrt{2} \cos\beta} \bar b_R \left(
    \tilde h_{13}^{d \ast} d_L +  \tilde h_{23}^{d \ast} s_L  \right)
    H + \frac{\lambda_b^H}{\sqrt{2}} \bar b_R b_L H+ \frac{\lambda_t^H}{\sqrt{2}} \bar t_R t_L H
    \nonumber\\
  & - \frac{i}{\sqrt{2} \cos\beta} \bar b_R \left( \tilde h_{13}^{d \ast} d_L +  \tilde h_{23}^{d \ast} s_L \right) A +
    \frac{i \lambda_b^A}{\sqrt{2}} \bar b_R b_L A - \frac{i \lambda_t^A}{\sqrt{2}} \bar t_R t_L A     + \mathrm{h.c.}
     \label{FCYukawas}
\end{align}
where
\begin{align}
 \lambda_b^h &= -\frac{\sqrt{2} m_b \sin\alpha}{v \cos\beta} +
  \frac{\tilde h_{33}^d \cos(\alpha - \beta)}{\cos\beta},  \\
    \lambda_t^h &= -\frac{\sqrt{2} m_t \sin\alpha}{v \cos\beta} + \frac{\tilde
    h_{33}^u \cos(\alpha - \beta)}{\cos\beta} ,   \\
  \lambda_b^H &= \frac{\sqrt{2} m_b \cos\alpha}{v \cos\beta} +
  \frac{\tilde h_{33}^d \sin(\alpha - \beta)}{\cos\beta},   \label{eq:lambda_b} \\
     \lambda_t^H &= \frac{\sqrt{2} m_t \cos\alpha}{v \cos\beta} + \frac{\tilde
    h_{33}^u \sin(\alpha - \beta)}{\cos\beta} , \label{eq:lambda_t}  \\
  \lambda_b^A &=\frac{\sqrt{2} m_b \tan\beta}{v}
  - \frac{\tilde h_{33}^d}{\cos\beta}, \\
  \lambda_t^A &= \frac{\sqrt{2} m_t \tan\beta}{v} - \frac{\tilde
    h_{33}^u}{\cos\beta}.
\end{align}
We note that ${\tilde h}^d\equiv D^\dagger_L h^d D_R$ and
${\tilde h}^u\equiv U^\dagger_L h^u U_R$. Thus, by taking $U_L=1$ we
get ${\tilde h}^u=h^u U_R$ and ${\tilde h}^d=V^\dagger_{\rm CKM} h^d$.
In this case, as compared to two-Higgs-doublet model type I, extra
Yukawa couplings are given by
\begin{align}
  {\tilde h}^u_{33}
  &=\frac{\sqrt{2} m_t}{v\sin\beta}\Big(1-\frac{v^2\cos^2\beta}{2m^2_t}\,|y^u_{33}|^2 \Big),  \label{FV3} \\
  {\tilde h}^d_{13}
  &= 1.80\times 10^{-2}\Big(\frac{m_b}{v\sin\beta}\Big),  \label{ht13} \\
  {\tilde h}^d_{23}
  &=5.77\times 10^{-2}\Big(\frac{m_b}{v\sin\beta}\Big) , \label{ht23} \\
  {\tilde h}^d_{33}
  &= 2.41\times 10^{-3}\Big(\frac{m_b}{v\sin\beta}\Big).  \label{ht33}
\end{align}
We find that the flavor-violating couplings for light up-type
quarks vanish, while the top quark Yukawa can have a sizable
modification due to nonzero ${\tilde h}^u_{33}$.
On the other hand, the flavor-violating couplings for down-type quarks
can be large if $\tan\beta$ is small, even though the couplings have
the suppression factors of CKM mixing and smallness of bottom quark
mass.
The couplings can be constrained by bounds from $B$-meson mixings and
decays as is discussed in the next section.
We note that the flavor-violating interactions of the SM-like Higgs
boson are turned off in the alignment limit where $\alpha=\beta-\pi/2$.

The Yukawa terms of the charged Higgs boson are given as
\begin{equation}
  -\mathcal{L}_{Y}^{H^-} =
  {\bar b}(\lambda_{t_L}^{H^-} P_L + \lambda_{t_R}^{H^-} P_R) t H^-
  + {\bar b}(\lambda_{c_L}^{H^-} P_L + \lambda_{c_R}^{H^-} P_R ) c H^-+
    \lambda_{u_L}^{H^-} {\bar b}P_L u H^- + \mathrm{h.c.},
\end{equation}
where
\begin{align}
  \lambda_{t_L}^{H^-}
  &= \frac{\sqrt{2}m_b \tan\beta}{v}\, V^*_{tb}
    -\frac{(V_{\rm CKM} {\tilde h}^d)^*_{33}}{\cos\beta},
    \label{lamtL} \\
  \lambda_{t_R}^{H^-}
  &= -\left( \frac{\sqrt{2} m_t \tan\beta}{v}
    - \frac{\tilde h_{33}^u}{\cos\beta} \right)V^*_{tb},
    \label{lamtR} \\
  \lambda_{c_L}^{H^-}
  &= \frac{\sqrt{2}m_b \tan\beta}{v}\, V^*_{cb}
    -\frac{(V_{\rm CKM}{\tilde h}^d)^*_{23}}{\cos\beta},
    \label{lamcL} \\
  \lambda_{c_R}^{H^-}
  &= -\frac{\sqrt{2} m_c\tan\beta}{v}\, V^*_{cb},
    \label{lamcR} \\
  \lambda_{u_L}^{H^-}
  &= \frac{\sqrt{2}m_b \tan\beta}{v}\, V^*_{ub}
    -\frac{(V_{\rm CKM}{\tilde h}^d)^*_{13}}{\cos\beta}
\end{align}
with
\begin{equation}
  V_{\rm CKM}{\tilde h}^d=
  \begin{pmatrix}
    0 & 0 & V_{ud}{\tilde h}^d_{13}+ V_{us}{\tilde h}^d_{23}+V_{ub}{\tilde h}^d_{33} \\
    0 & 0 & V_{cd}{\tilde h}^d_{13}+ V_{cs}{\tilde h}^d_{23}+V_{cb}{\tilde h}^d_{33} \\
    0 & 0 & V_{td}{\tilde h}^d_{13}+ V_{ts}{\tilde h}^d_{23}+V_{tb}{\tilde h}^d_{33}
  \end{pmatrix}.
\end{equation}
If $y_{33}^u = y_t^\text{SM} = \sqrt{2} m_t / v$, the Higgs coupling
to top quark becomes
\begin{equation}
  \lambda_t^H = y_t^\text{SM} \cos (\alpha-\beta) ,
  \label{eq:lambda_t_SM}
\end{equation}
and $ \lambda_t^A =  \lambda_{t_R}^{H^-} = 0$.

\subsection{Lepton Yukawa couplings}

As seen in~(\ref{cleptonmass}), the mass matrix for charged leptons
$e_j$ is already diagonal due to the $U(1)'$ symmetry. Thus, the
lepton Yukawa couplings are in a flavor-diagonal form given by
\begin{align}
  -{\cal L}_{Y}^\ell =
  &-\frac{m_{e_j}\sin\alpha }{v\cos\beta}\, {\bar e}_j\, e_j \,h +
    \frac{m_{e_j}\cos\alpha }{v\cos\beta} \,{\bar e}_j\, e_j \,H +
    \frac{i m_{e_j}\tan\beta}{v}\, {\bar e}_j \gamma^5 e_j \,A^0
    \nonumber\\
  &+\frac{\sqrt{2}m_{e_j}\tan\beta}{v}\, \left({\bar \nu}_j\,P_R\,
    e_j \,H^+ + \mathrm{h.c.}\right)
\end{align}

\section{Constraints on the Higgs sector}
\label{sec:constraints}

In this section we consider various phenomenological constraints on
the model coming from $B$-meson mixings and decays as well as Higgs
and electroweak precision data on top of unitarity and stability
bounds on the Higgs sector.
We also show how to explain the deficits in $R_K$ and $R_{K^*}$ in
the $B$-meson decays at LHCb in our model, and discuss the predictions
for $R_D$ and $R_{D^*}$ through the charged Higgs exchange.

\subsection{Unitarity and stability bounds}
\label{sec:unitarity_bounds}

Before considering the phenomenological constraints, we consider
unitarity and stability bounds for the Higgs sector.
As derived in Appendix~\ref{app:unitarity_bounds}, the
conditions for perturbativity and unitarity are
\begin{align}
|\lambda_{1,2,3,S}|\leq 4\pi ,\qquad
|\kappa_{1,2}| \leq 4\pi ,\nonumber\\
|\lambda_3\pm\lambda_4| \leq 4\pi ,\quad
|\lambda_3+2\lambda_4| \leq 4\pi ,\quad
\sqrt{\lambda_3(\lambda_3+2\lambda_4)}\leq 4 \pi ,\nonumber\\
|\lambda_1+\lambda_2\pm\sqrt{(\lambda_1-\lambda_2)^2+4\lambda_4^2}| \leq 8 \pi \nonumber\\
a_{1,2,3} \leq 8 \pi,
\end{align}
where $a_{1,2,3}$ are the solutions to Eq.~(\ref{cubic}).
%
The vacuum stability conditions of the scalar potential can be obtained
by considering the potential to be bounded from below along
the directions of large Higgs doublet and singlet scalar fields.
Following
Refs.~\cite{ElKaffas:2006gdt,Grzadkowski:2009bt,Drozd:2014yla}, we
obtain the stability conditions as follows:
\begin{align}
&\lambda_{1,2,S}>0\nonumber\\
&\sqrt{\lambda_1 \lambda_2}+\lambda_3+\lambda_4>0,\nonumber\\
&\sqrt{\lambda_1 \lambda_2}+\lambda_3>0,\nonumber\\
&\sqrt{\lambda_1 \lambda_S}+\kappa_1>0,\nonumber\\
&\sqrt{\lambda_2 \lambda_S}+\kappa_2>0,\nonumber\\
&\sqrt{(\kappa_1^2-\lambda_1\lambda_S)(\kappa_2^2-\lambda_2\lambda_S)}+\lambda_3\lambda_S>\kappa_1\kappa_2,\nonumber\\
&\sqrt{(\kappa_1^2-\lambda_1\lambda_S)(\kappa_2^2-\lambda_2\lambda_S)}+(\lambda_3+\lambda_4)\lambda_S>\kappa_1\kappa_2.
\end{align}
The stability conditions along the other scalar fields $\Phi_a$ can be
obtained in the similar way, but they are not relevant for our study
because $\Phi_a$'s do not couple directly to Higgs doublets as long as
the extra quartic couplings for $\Phi_a$ are positive and large enough.

\begin{figure}[t!]
  \begin{center}
    \includegraphics[height=0.45\textwidth]{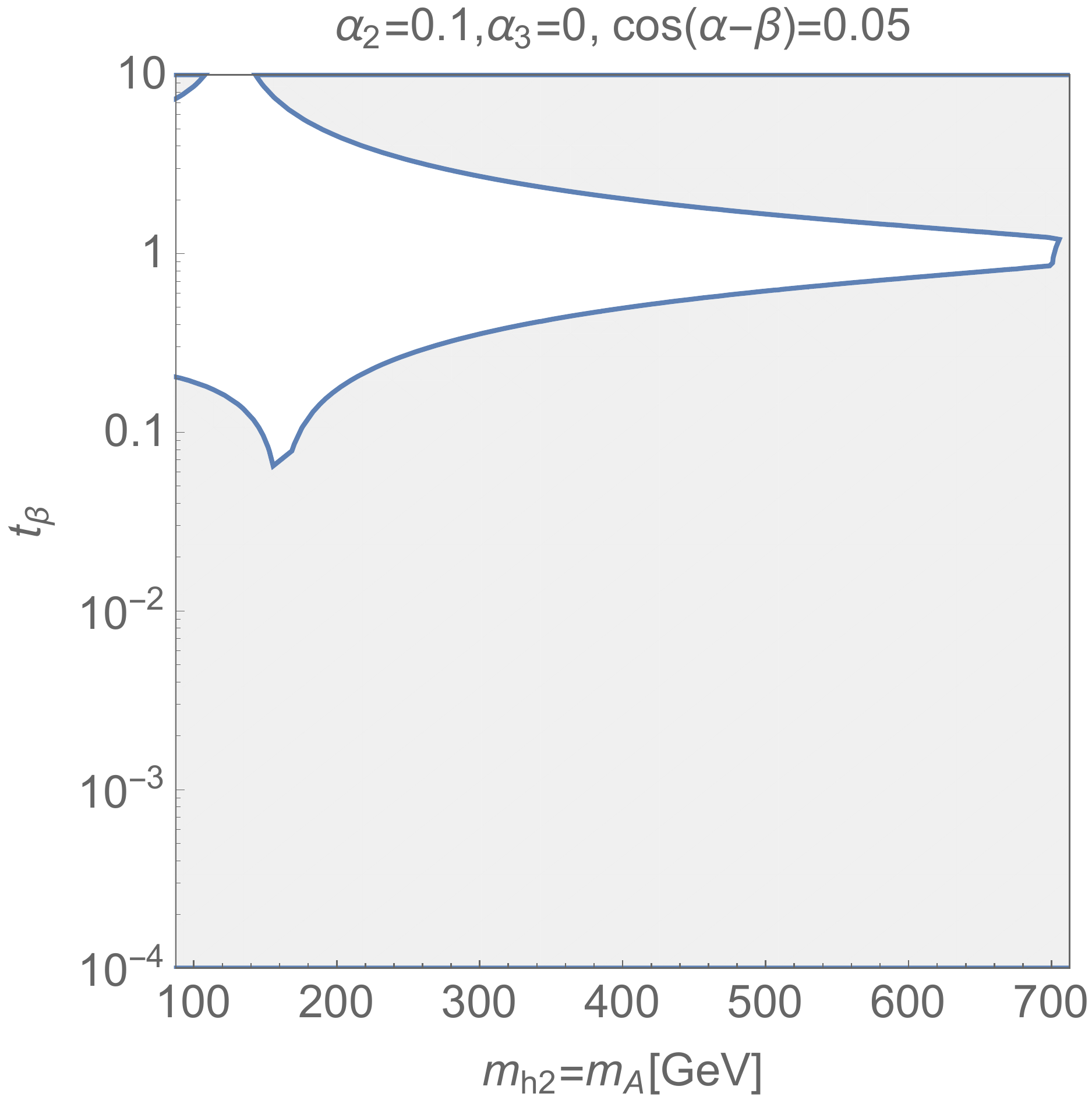}
    \includegraphics[height=0.45\textwidth]{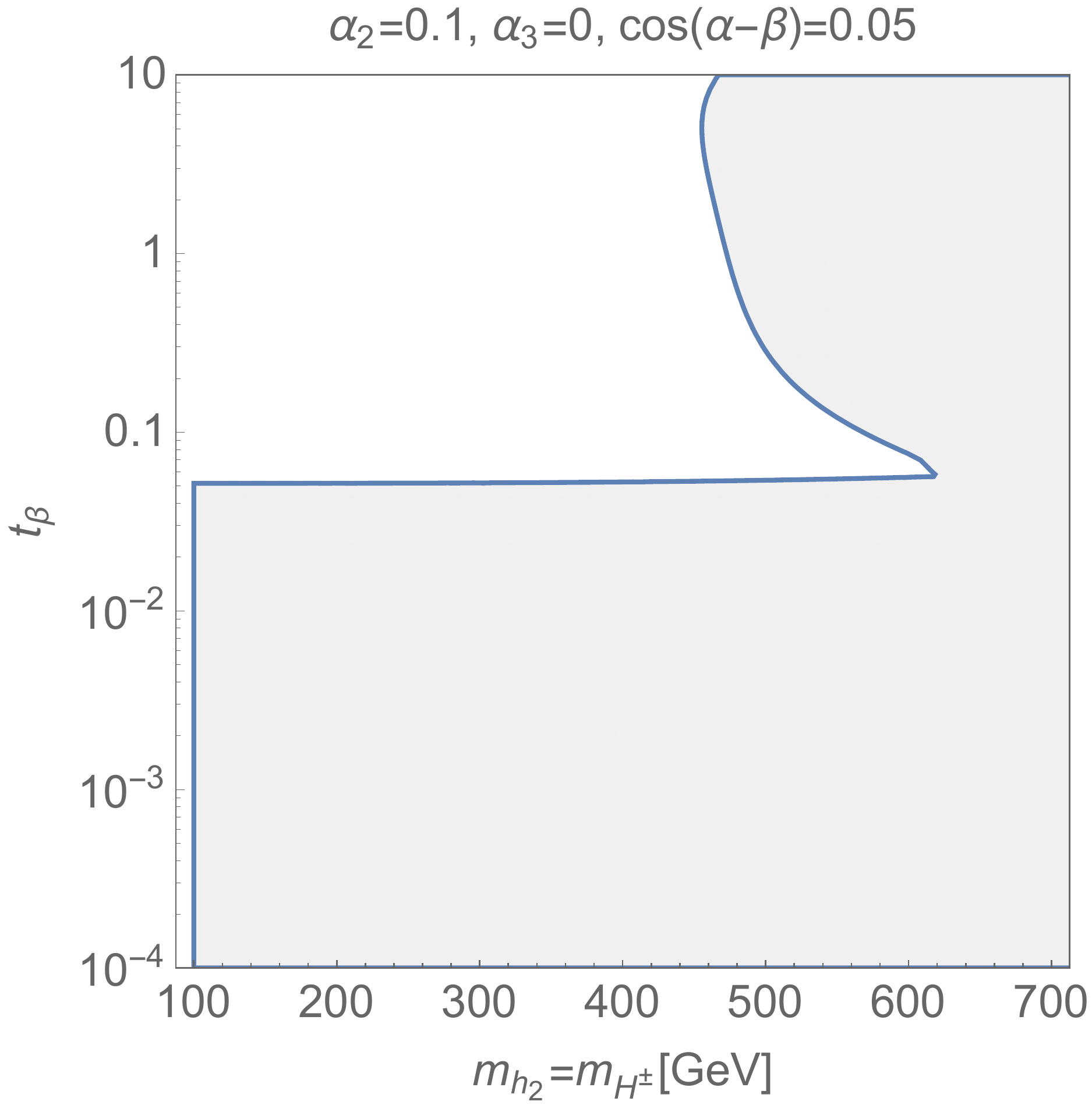}
  \end{center}
  \caption{Parameter space in terms of $m_{h_2}$ and $\tan\beta$.
    The gray regions are excluded by unitarity and stability bounds.
    $v_s=2m_{h_3}=1$ TeV and $\cos(\alpha-\beta)=0.05$ with
    $m_{h_2}=m_{A}$ and $m_{H^\pm}=500$~GeV in the left, and
    $m_{h_2}=m_{H^\pm}$ and $m_A=140$~GeV in the right
    panel. The mixing between heavy $CP$-even scalars is taken to be
    zero.\label{fig:unitarity1}}
\end{figure}
\begin{figure}[t!]
  \begin{center}
    \includegraphics[height=0.45\textwidth]{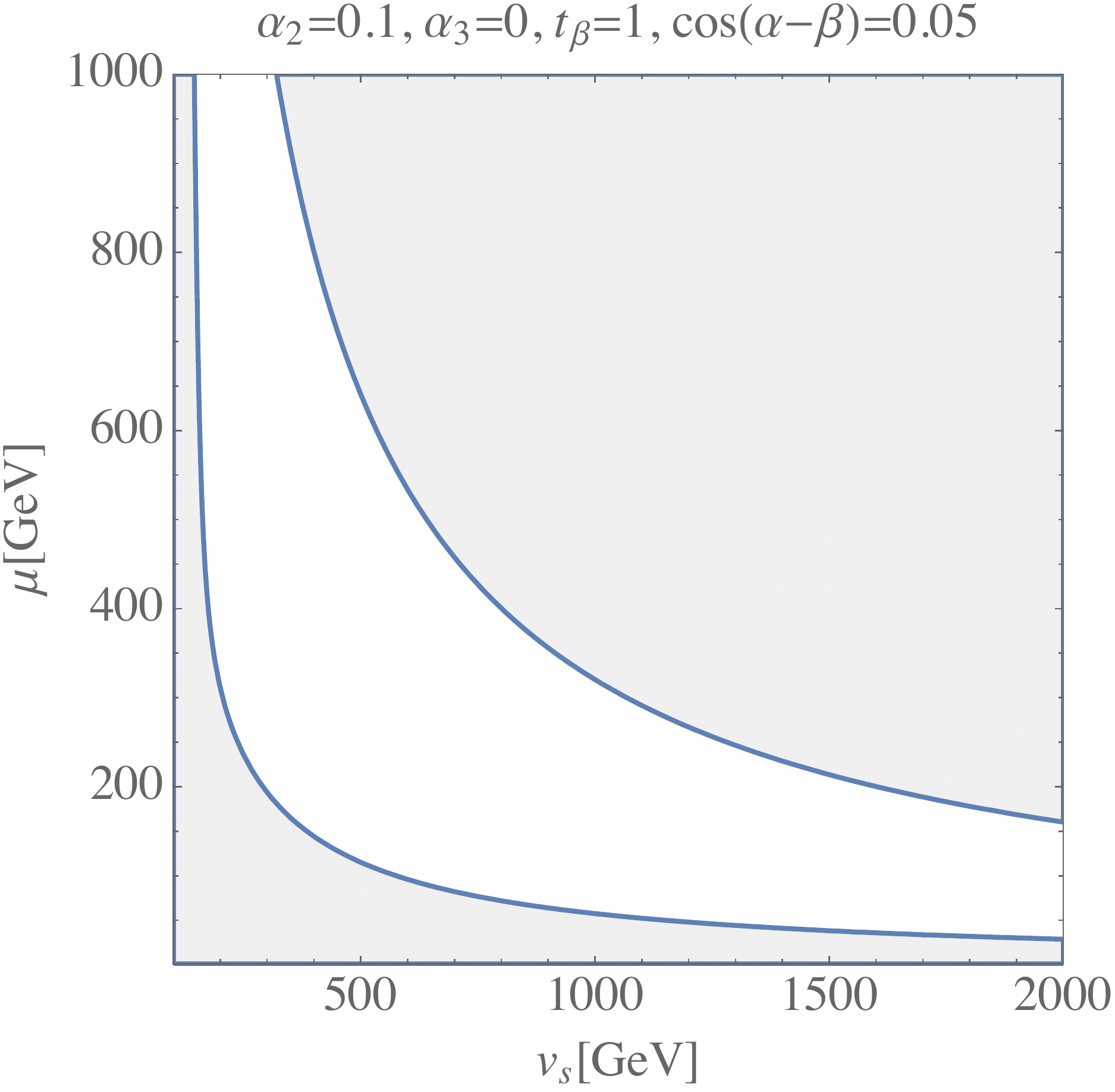}
    \includegraphics[height=0.45\textwidth]{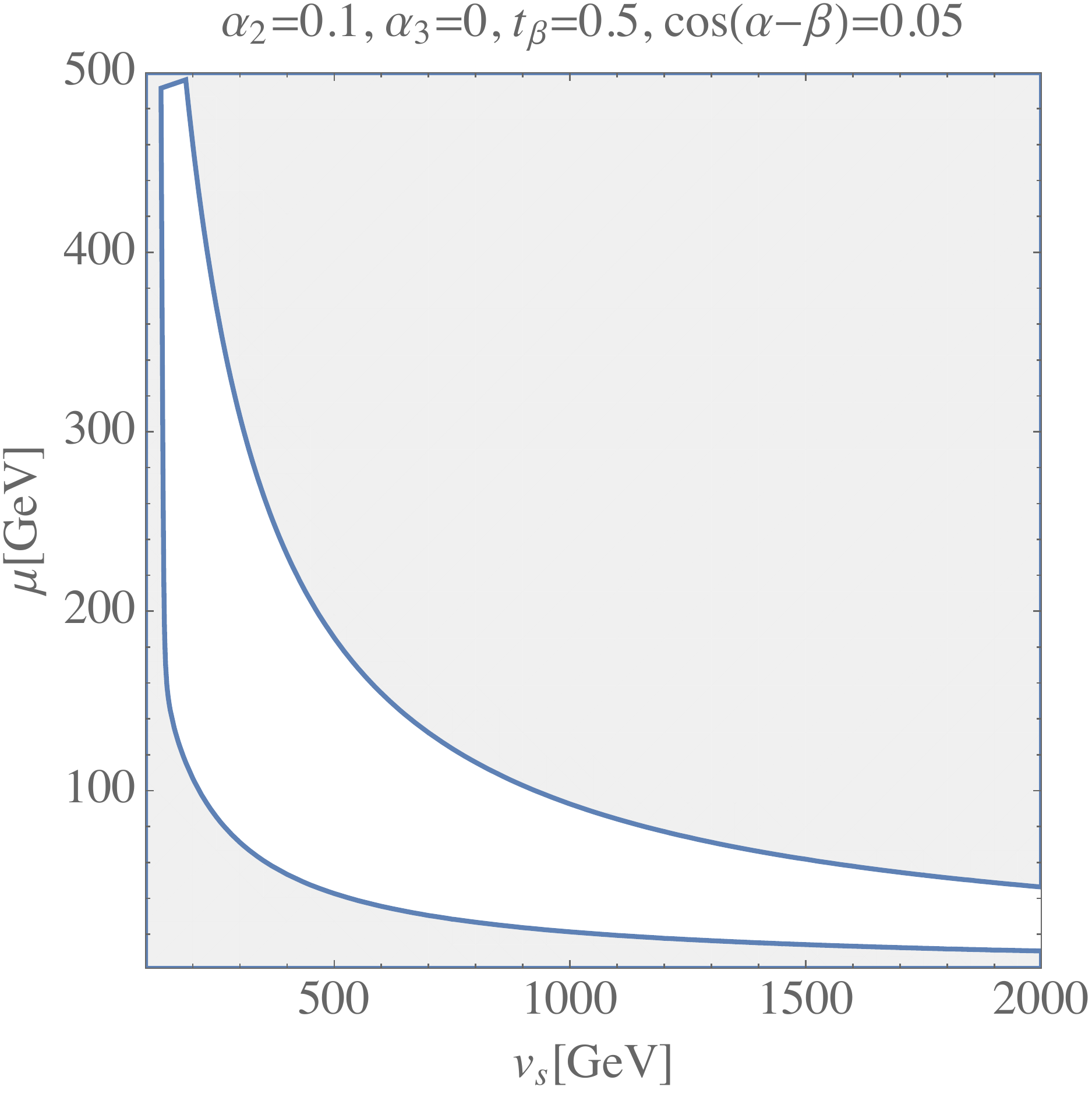}
  \end{center}
  \caption{Parameter space in terms of $v_s$ and $\mu$ for
    $m_{h_3}=m_{H^\pm}=m_{h_2}=0.5$~TeV and $\cos(\alpha-\beta)=0.05$.
    The gray regions are excluded by unitarity and stability bounds.
    $\tan\beta=1$ (0.5) in the left (right) panel.
    The mixing between heavy $CP$-even scalars is taken to be
    zero.\label{fig:unitarity2}}
\end{figure}

The unitarity and stability bounds are depicted in
Figs.~\ref{fig:unitarity1} and~\ref{fig:unitarity2} for the parameter
space in terms of $m_{h_2}$ and $\tan\beta$, or $v_s$ and $\mu$, with assuming
the alignment limit, $\cos(\alpha - \beta) = 0.05$, and zero
mixing between heavy $CP$-even scalars.
In each figure, the gray region corresponds to the parameter space
excluded by the unitarity and stability conditions.
In Fig.~\ref{fig:unitarity1}, we have taken the different choices of
Higgs masses: $m_{h_2} = m_A$ and $m_{H^\pm} = 500$~GeV in the left,
while $m_{h_2} = m_{H^\pm}$ and $m_A = 140$~GeV in the right panel.
On the other hand, the parameter space in terms of $v_s$ and $\mu$ has
been shown in Fig.~\ref{fig:unitarity2}, with setting
$m_{h_3}=m_{H^\pm}=m_{h_2}=0.5$~TeV, but taking different values of
$\tan\beta$.
We note that the unitarity and stability bounds are sensitive to the
choice of $\tan\beta$, while insensitive to the mixing angle of heavy
$CP$-even scalars, in constraining the mass parameters.
The allowed parameter space for mass parameters becomes narrower
as $\tan\beta$ is smaller.


\subsection{Higgs and electroweak precision data\label{sec:higgs_ewpd}}

\begin{figure}[t!]
  \begin{center}
    \includegraphics[height=0.45\textwidth]{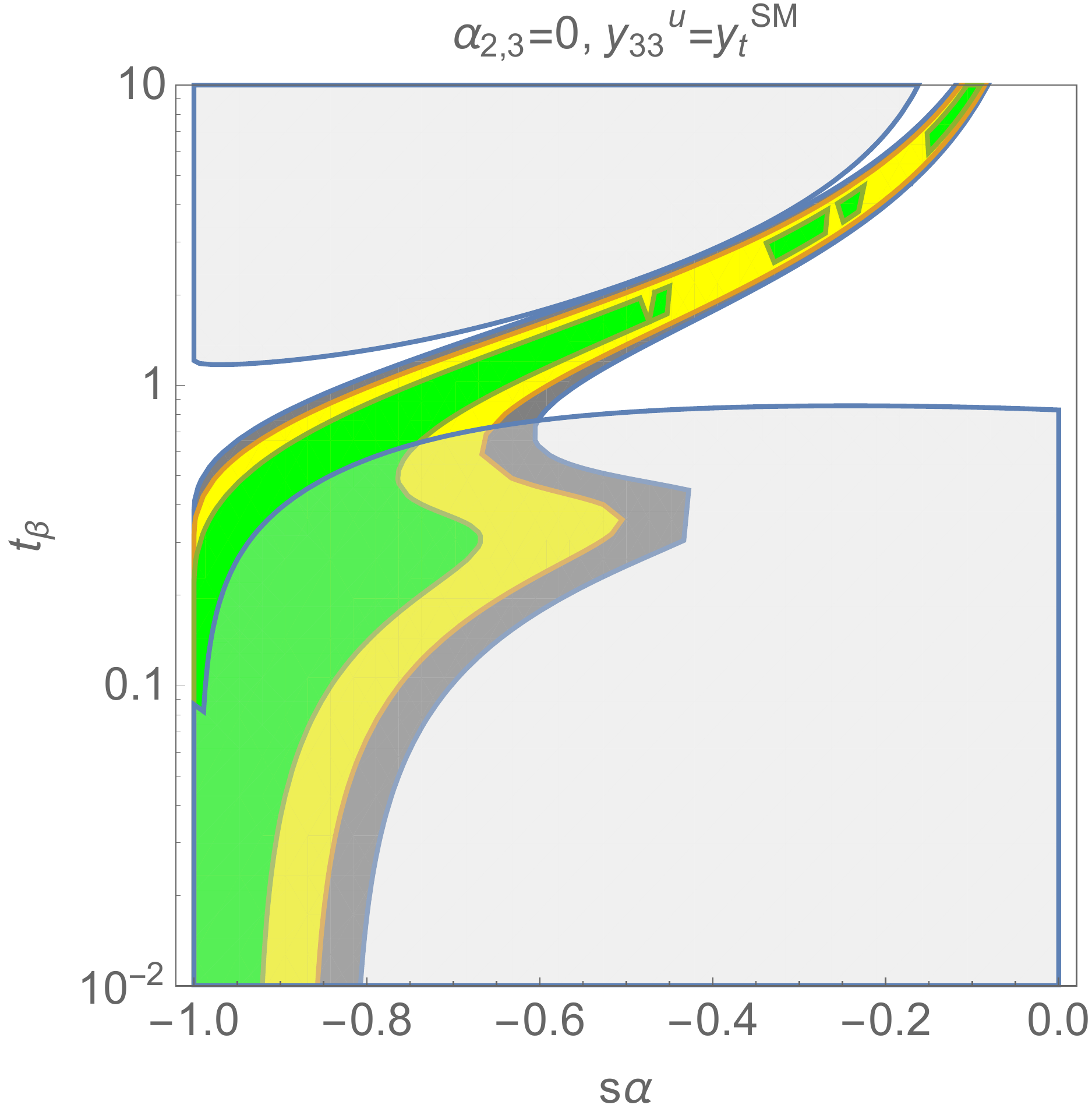}
    \includegraphics[height=0.45\textwidth]{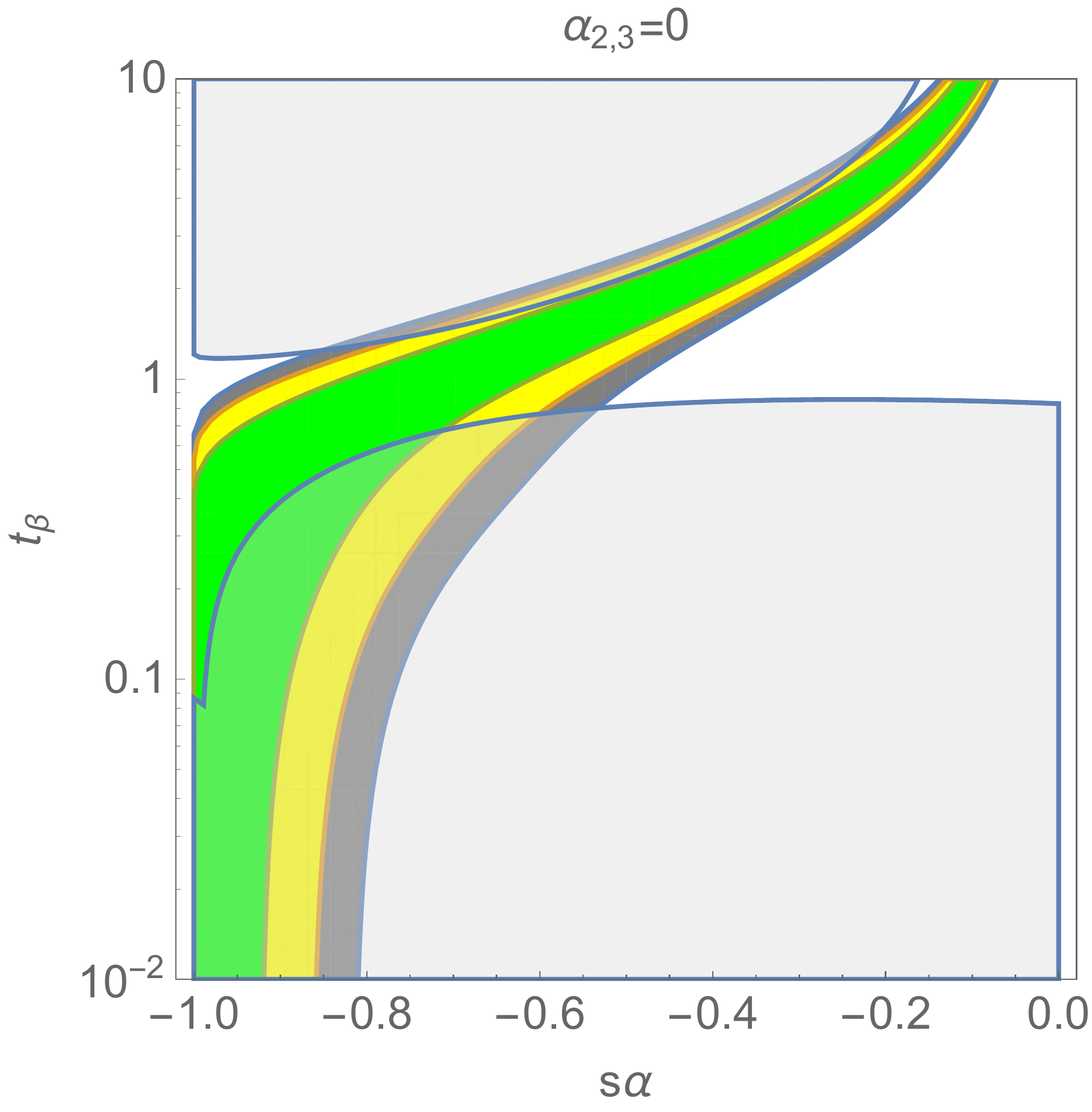}
  \end{center}
  \caption{Parameter space for $\sin\alpha$ and $\tan\beta$ allowed
    by Higgs data within $1\sigma$ (green), $2\sigma$ (yellow), and
    $3\sigma$ (dark gray). The gray regions corresponds to
    the unitarity and stability bounds.
    $y^u_{33}=y^{\rm SM}_t$ in the left and $y^u_{33}=y^{\rm
      SM}_t/\cos\beta$ in the right panel.
    $m_{h_2}=m_{H^\pm}=450$~GeV, $m_A=140$~GeV and $v_s=1$~TeV has
    been taken in all panels.\label{fig:Higgsfit}}
\end{figure}

Provided that the Higgs mixings with the singlet scalar are small,
the mixing angle $\alpha$ between $CP$-even Higgs
scalars are constrained by Higgs precision
data~\cite{WW,ZZ,gammagamma,bb,tautau}.
The parameter space for $\sin\alpha$ and $\tan\beta$ allowed by the
Higgs data is shown in Fig.~\ref{fig:Higgsfit}.
We take the $(33)$ component of the up-type
Higgs Yukawa coupling to be $y^u_{33}=y^{\rm SM}_t$ in the left, and
$y^u_{33}=y^{\rm SM}/\cos\beta$ in the right panel.
For illustration, we have also imposed unitarity and stability bounds
discussed in the previous subsection for $m_{h_2}=m_{H^\pm}=450$ GeV,
$m_A=140$~GeV and $v_s=1$~TeV.
As a result, we find a wide parameter space close to the
  line of the alignment, $\alpha=\beta- \pi/2$, that is consistent with both the Higgs
data and unitarity/stability bounds for $\tan\beta\gtrsim 0.1$. Thus, henceforth, for the phenomenology of the extra Higgs scalars, we focus on the parameter space near the alignment limit, $\cos(\alpha-\beta)\sim 0$.

To see bounds from electroweak precision data,
we obtain effective Lagrangian after integrating out $W$ and $Z$
bosons as follows~\cite{dumitru,kennedy}:
\begin{equation}
  {\cal L}_{\rm eff} = -\frac{4G_F}{\sqrt{2} g^2 \sec^2\theta_W}\,
  \Big(\sec^2\theta_W J^\mu_{W^+} J_{W^-,\mu}+\rho J^\mu_Z J_{Z,\mu}
  +2 a J^\mu_Z J_{Z',\mu}+ b J^\mu_{Z'} J_{Z',\mu}\Big)+\cdots ,
\end{equation}
where $J^\mu_Z=J^\mu_{3}-\sin^2\theta_* J^\mu_{\rm EM}$ with
$\theta_*$ being the modified Weinberg angle. Here the non-oblique
terms, $a$ and $b$, are determined at tree level as
\begin{equation}
  a= \frac{\rho \sin\zeta \sec\xi}{\cos\zeta+\sin\theta_W
    \tan\xi\sin\zeta}, \quad
  b= \frac{a^2}{\rho}.
\end{equation}
From the $Z$-boson like mass given in Eq.~(\ref{Zmass}) and the $Z$--$Z'$
mixing angle in Eq.~(\ref{Zmix}), we find the correction to the $\rho$
parameter as
\begin{align}
  \Delta\rho
  &= \frac{m^2_W}{m^2_{Z_1} \cos^2\theta_W}\, (\cos\zeta+\sin\theta_W
    \tan\xi \sin\zeta)^2-1 \nonumber \\
  &\simeq \frac{\sin^2\theta_W}{\cos^2\xi} \frac{m^2_Z}{m^2_{Z'}}
    \left[\Big(2 Q'_{H_2}\frac{g_{Z'}}{g_Y}
    \Big)^2\sin^4\beta-\sin^2\xi \right] ,
\end{align}
where we assumed that $\tan 2\zeta\simeq 2m^2_{12} / m^2_{Z_2} \ll 1$.
Taking the limit of zero gauge kinetic mixing, {\em i.e.} $\sin\xi=0$,
we have
\begin{align}
  \Delta\rho
  &=\frac{m^2_W}{m^2_{Z}\cos^2\theta_W}-1 \nonumber \\
  &\simeq 10^{-4} \left(\frac{x}{0.05}\right)^2 g^2_{Z'} \sin^4\beta
    \left(\frac{400\,{\rm GeV}}{m_{Z'}}\right)^2,
\end{align}
which is consistent with the result in Ref.~\cite{Bian:2017rpg}.
Therefore, for $\tan\beta\simeq 1$, $g_{Z'}\simeq 1$, and $x\simeq
0.05$, $Z'$ with the mass $m_{Z'}\gtrsim 400\,{\rm GeV}$ is consistent
with electroweak precision data.
The mass splittings between extra Higgs scalars can also be
constrained by the electroweak precision data, but
it can be easily satisfied if we take $m_{h_2}=m_{H^\pm}$ or
$m_{h_2}=m_A$, and a small mixing between $CP$-even scalars.

\subsection{{\boldmath$B$}-meson anomalies from {\boldmath$Z'$}}

Before considering constraints from $B$-meson mixings and decays,
we show how to explain the $B$-meson anomalies in our model and
identify the relevant parameter space for that. This section is based
on the detailed results on $U(1)'$ interactions presented in
Appendix~\ref{app:u1_ints} and phenomenological findings in
Ref.~\cite{Bian:2017rpg}.

From the relevant $Z'$ interactions for $B$-meson anomalies and the
$Z'$ mass term,
\begin{equation}
{\cal L}'_{Z'}= g_{Z'} Z'_\mu \Big(\frac{1}{3}x\,V^*_{ts}
V_{tb}\,{\bar s}\gamma^\mu P_L b+{\rm h.c.}+y {\bar \mu}\gamma^\mu \mu
\Big)+ \frac{1}{2} m^2_{Z'} Z^{\prime 2}_\mu, \label{zpint}
\end{equation}
we get the classical equation of motion for $Z'$ as
\begin{equation}
Z'_\mu= -\frac{g_{Z'}}{m^2_{Z'}} \Big(\frac{1}{3}x\,V^*_{ts} V_{tb}\,{\bar s}\gamma_\mu P_L b+{\rm h.c.}+y {\bar \mu}\gamma_\mu \mu   \Big). \label{zpeq}
\end{equation}
Then, 
by integrating out the $Z'$ gauge boson, we obtain the effective
four-fermion interaction for ${\bar b}\rightarrow {\bar s}\mu^+ \mu^-$
as follows.
\begin{equation}
{\cal L}_{{\rm eff},{\bar b}\rightarrow {\bar s}\mu^+ \mu^-}= -\frac{xy g^2_{Z'}}{3 m^2_{Z'}}\, V^*_{ts} V_{tb}\,  ({\bar s}\gamma^\mu P_L b) ({\bar \mu}\gamma_\mu \mu)+{\rm h.c.}
\end{equation}
Consequently, as compared to the effective Hamiltonian with the SM
normalization,
\begin{equation}
\Delta {\cal H}_{{\rm eff},{\bar b}\rightarrow {\bar s}\mu^+ \mu^-} = -\frac{4G_F}{\sqrt{2}}  \,V^*_{ts} V_{tb}\,\frac{\alpha_\text{em}}{4\pi}\, C^{\mu,{\rm NP}}_9 {\cal O}^\mu_9
\end{equation}
with $ {\cal O}^\mu_9 \equiv ({\bar s}\gamma^\mu P_L b) ({\bar
  \mu}\gamma_\mu \mu)$ and $\alpha_{\rm em}$ being the electromagnetic
coupling, we obtain new physics contribution to the Wilson
coefficient,
\begin{equation}
  C^{\mu, {\rm NP}}_9= -\frac{8 xy \pi^2\alpha_{Z'}}{3\alpha_{\rm em}}\, \left(\frac{v}{m_{Z'}}\right)^2
\end{equation}
with $\alpha_{Z'}\equiv g^2_{Z'}/(4\pi)$,
and vanishing contributions to other operators, $C^{\mu,{\rm
    NP}}_{10}=C^{\prime\mu, {\rm NP}}_9=C^{\prime\mu, {\rm
    NP}}_{10}=0$.
We note that $xy>0$ is chosen for a negative sign of $C^\mu_9$, being
consistent with $B$-meson anomalies.
Requiring the best-fit value, $C^{\mu, \, {\rm
    NP}}_9=-1.10$~\cite{muon}, (while taking $[-1.27,-0.92]$ and
$[-1.43,-0.74]$ within $1\sigma$ and $2\sigma$ errors), to explain the
$B$-meson anomalies yields
\begin{equation}
  m_{Z'}= 1.2~\text{TeV} \times \left(xy\,
    \frac{\alpha_{Z'}}{\alpha_{\rm em}} \right)^{1/2} .
\end{equation}
Therefore, $m_{Z'} \simeq 1\,{\rm TeV}$ for $xy\simeq 1$ and
$\alpha_{Z'}\simeq \alpha_{\rm em}$. For values of
$xy$ less than unity or $\alpha_{Z'}\lesssim \alpha_{\rm em}$, $Z'$
can be even lighter.

Various phenomenological constraints on the $Z'$ interactions coming
from dimuon resonance searches, other meson decays and mixing, tau
lepton decays and neutrino scattering have been studied in
Ref.~\cite{Bian:2017rpg}, leading to the conclusion that the region
of $x g_{Z'}\lesssim 0.05$ for $y g_{Z'}\simeq 1$ and $m_{Z'}\lesssim
1$~TeV is consistent with the parameter space for which the $B$-meson
anomalies can be explained.

\subsection{Bounds from {\boldmath$B$}-meson mixings and decays}

We now consider the bounds from $B$-meson mixings and decays.
After integrating out the heavy Higgs bosons, the effective Lagrangian
for $B_{s(d)}\rightarrow \mu^+\mu^-$ from the flavor-violating
Yukawa interactions in (\ref{FCYukawas}) is
\begin{align}
  \Delta {\cal L}_{{\rm eff},B_{s(d)}\rightarrow \mu^+ \mu^-}=
  &-\frac{\sqrt{2} m_\mu \sin(\alpha-\beta)\cos\alpha}{2m^2_H  v
    \cos\beta} \Big(({\tilde h}^d_{23})^* {\bar b}_R s_L +({\tilde
    h}^d_{13})^* {\bar b}_R d_L+{\rm h.c.} \Big)({\bar \mu} \mu)
    \nonumber \\
  &-\frac{\sqrt{2} m_\mu \tan\beta}{2m^2_A  v \cos\beta} \Big(({\tilde
    h}^d_{23})^* {\bar b}_R s_L +({\tilde h}^d_{13})^* {\bar b}_R
    d_L+{\rm h.c.} \Big)({\bar \mu}\gamma^5 \mu).
\end{align}
The extra contributions to the effective Hamiltonian for
$B_{s}\rightarrow \mu^+\mu^-$ are thus
\begin{equation}
  \Delta {\cal H}_{{\rm eff},B_{s}\rightarrow \mu^+ \mu^-}=-\frac{G^2_F m^2_W}{\pi} \Big[C^{\rm BSM}_S ({\bar b}P_L s)({\bar \mu}\mu) + C^{\rm BSM}_P({\bar b}P_L s)({\bar \mu}\gamma^5\mu)\Big]
\end{equation}
with
\begin{align}
C^{\rm BSM}_S &=  -\frac{\pi}{G^2_F m^2_W}\, \frac{\sqrt{2}m_\mu\sin(\alpha-\beta)\cos\alpha}{2m^2_H v\cos^2\beta}\,\cdot ({\tilde h}^d_{23})^*, \nonumber\\
C^{\rm BSM}_P&=  -\frac{\pi}{G^2_F m^2_W}\, \frac{\sqrt{2}m_\mu\tan\beta}{2m^2_A v\cos\beta}\,\cdot ({\tilde h}^d_{23})^*.
\end{align}
In the alignment limit with $\alpha=\beta-\pi/2$ and $m_A\simeq m_H$,
the Wilson coefficients become identical and suppressed for a small
$\tan\beta$.
The effective Hamiltonian in the above leads to the corrections of the
branching ratio for $B_{s}\rightarrow \mu^+\mu^-$ as
follows~\cite{crivellin2}:
\begin{align}
  {\cal B}(B_{s}\rightarrow \mu^+\mu^-)
  =&~\frac{G^4_F
    m^4_W}{8\pi^5}\left(1-\frac{4m^2_\mu}{m^2_{B_s}}\right)^{1/2}
    m_{B_s} f^2_{B_s} m^2_\mu \, \tau_{B_s} \nonumber \\
   &\times \left[
     \left|\frac{m^2_{B_s}(C_P-C'_P)}{2(m_b+m_s)m_\mu}-(C_A-C'_A)
     \right|^2+ \left|\frac{m^2_{B_s}(C_S-C'_S)}{2(m_b+m_s)m_\mu}
     \right|^2 \left(1-\frac{4m^2_\mu}{m^2_{B_s}}\right)\right],
     \nonumber \\
\end{align}
where $m_{B_s}$, $f_{B_s}$, and $\tau_{B_s}$ are mass, decay constant,
and lifetime of $B_s$-meson, respectively. $C^{(')}_A ,C^{(')}_S,
C^{(')}_P$ are Wilson coefficients of the effective operators, ${\cal
  O}^{(')}_A=[{\bar b}\gamma_\mu P_{L(R)}s][{\bar \mu}\gamma^\mu
\gamma^5 \mu]$, ${\cal O}^{(')}_S=[{\bar b}P_{L(R)}s][{\bar \mu} \mu]$
and ${\cal O}^{(')}_P=[{\bar b}P_{L(R)}s][{\bar \mu}\gamma^\mu\gamma^5
\mu]$, respectively. We note that there is no
contribution from $Z'$ interactions to $B_s\rightarrow \mu^+\mu^-$
since the muon couplings to $Z'$ are vector-like.
On the other hand, in the alignment limit the bounds obtained from
$B_{s,d}\rightarrow \mu^+\mu^-$ in Ref.~\cite{crivellin2} can be
translated to our case as
\begin{align}
  \left\vert{\tilde h}^d_{23}\right\vert
  &< 3.4\times 10^{-2}
    \left(\frac{\cos\beta}{\tan\beta}\right){\left(\frac{m_{H,A}}{500\,{\rm
    GeV}} \right)}^2,  \nonumber \\
  \left\vert{\tilde h}^d_{13}\right\vert
  &< 1.7\times 10^{-2}
    \left(\frac{\cos\beta}{\tan\beta}\right){\left(\frac{m_{H,A}}{500\,{\rm
    GeV}} \right)}^2.  \label{eq:Bmumu}
\end{align}
From Eqs.~(\ref{ht13}) and (\ref{ht23}), we find that flavor
constraints are satisfied as far as
\begin{equation}
  \sin\beta< \sqrt{1-0.033{\left(\frac{500~{\rm GeV}}{m_{H,A}} \right)}^2}.
\end{equation}
This leads to $\tan\beta<5.4$ for $m_{H,A}=500\,{\rm GeV}$.

The flavor-violating Yukawa couplings of heavy Higgs bosons as well as
$Z'$ interactions~\cite{Bian:2017rpg} can modify the $B_s$--${\bar B}_s$ mixing. The additional effective Hamiltonian
relevant for the mixing is given by
\begin{equation}
  \Delta{\cal H}_{\text{eff},B_s-{\bar B}_s}=C'_2 (\bar s_\alpha P_R
  b_\alpha)({\bar s}_\beta P_R b_\beta)+ \frac{G^2_F m^2_W}{16\pi^2}\, (V^*_{ts} V_{tb})^2\,C^{\rm NP}_{VLL}\,  ({\bar s}_\alpha\gamma^\mu P_L b_\alpha)({\bar s}_\beta\gamma_\mu P_L b_\beta),
\end{equation}
with
\begin{align}
  C'_2=&~ \frac{{\tilde h}^d_{23}}{4\cos^2\beta\, m^2_H}
  \left( \frac{m^2_H}{m^2_A}-\sin^2(\alpha-\beta) -\frac{m^2_H \cos^2(\alpha-\beta)}{m^2_h}\right), \label{C2p} \\
  C^{\rm NP}_{VLL}=&~ \frac{16\pi^2}{9} \, \frac{(x g_{Z'})^2 v^4}{m^2_{Z'} m^2_W}  \nonumber \\
=&~ 0.27 \Big(\frac{x g_{Z'}}{0.05}\Big)^2 \left(\frac{300\,{\rm GeV}}{m_{Z'}}\right)^2. \label{CVLL}
\end{align}
The mass difference in the $B_s$ system becomes
\begin{equation}
\Delta M_{B_s}= \frac{2}{3}m_{B_s} f^2_{B_s} B^s_{123}(\mu) \left[ \frac{G^2_F m^2_W}{16\pi^2}\, (V^*_{ts} V_{tb})^2\,\Big(C^{\rm SM}_{VLL}+C^{\rm NP}_{VLL}\Big)+ |C'_2| \right], \label{DMs}
\end{equation}
where $B^s_{123}(\mu)$ is a combination of
bag-parameters~\cite{BsBsbar} and $C^{\rm SM}_{VLL}\simeq 4.95$~\cite{MsSM}.
The SM prediction and the experimental values of $\Delta M_s$ are
given by $(\Delta M_{B_s})^{\rm SM}=(17.4\pm 2.6)\,{\rm
  ps}^{-1}$~\cite{MsSM} and $(\Delta M_{B_s})^{\rm exp}=(17.757\pm
0.021)\,{\rm ps}^{-1}$~\cite{Msexp}, respectively.
Then, taking into account the SM uncertainties, we obtain the bounds
on $\Delta M_{B_s}$ as $16\,(13)\,{\rm ps}^{-1}<\Delta M_{B_s}<21\,(23)\,{\rm
  ps}^{-1}$ or $(\Delta M_{B_s})^{\rm BSM}<3.0\,(5.6)\,{\rm ps}^{-1}$ at
$1\sigma$ ($2\sigma$) level for new physics.
We also note that the most recent lattice calculations show
considerably large values for the bag parameters, leading to $(\Delta
M_{B_s})^{\rm SM}=(20.01\pm 1.25)\,{\rm ps}^{-1}$~\cite{MsSM2}.
It needs an independent confirmation, but if it is true, the new
physics contributions coming from the heavy Higgs bosons and $Z'$
would be constrained more tightly.

Taking the SM prediction as $(\Delta M_{B_s})^{\rm SM}=(17.4\pm 2.6)\,{\rm ps}^{-1}$~\cite{MsSM}, from Eq.~(\ref{DMs}) with
Eqs.~(\ref{C2p}) and (\ref{CVLL}), we get the bound on the
flavor-violating Yukawa coupling  in the alignment limit of heavy Higgs bosons as
\begin{equation}
  \frac{|{\tilde h}^d_{23}|}{\cos\beta} \bigg|\frac{m^2_H}{m^2_A}-1\bigg|^{1/2}\bigg(\frac{500\,{\rm GeV}}{m_H}\bigg)<4.6(6.4)\times 10^{-3}\sqrt{1-0.1(0.06)\Big(\frac{x g_{Z'}}{0.05}\Big)^2 \left(\frac{300\,{\rm GeV}}{m_{Z'}}\right)^2}.  \label{hd23}
\end{equation}
Here, since we need to choose $x g_{Z'}\lesssim 0.05$ for $m_{Z'}\lesssim 1\,{\rm TeV}$ to satisfy the $B$-meson anomalies and the LHC dimuon bounds at the same time as discussed in the previous section, we can safely ignore the contribution of $Z'$ interactions to the $B_s$--${\bar B}_s$ mixing on the right-hand side of Eq.~(\ref{hd23}).
Furthermore, with the $Z'$ contribution ignored, the $B_d$--${\bar B}_d$ mixing leads to a similar
bound~\cite{BsBsbar}:
\begin{equation}
  \left\vert{\tilde h}^d_{13}\right\vert<0.91(1.3)\times 10^{-3} \cos\beta\,
  \bigg|\frac{m^2_H}{m^2_A}-1\bigg|^{-1/2}\bigg(\frac{m_H}{500\,{\rm GeV}}\bigg).
\end{equation}

\begin{figure}[t!]
  \begin{center}
    \includegraphics[height=0.45\textwidth]{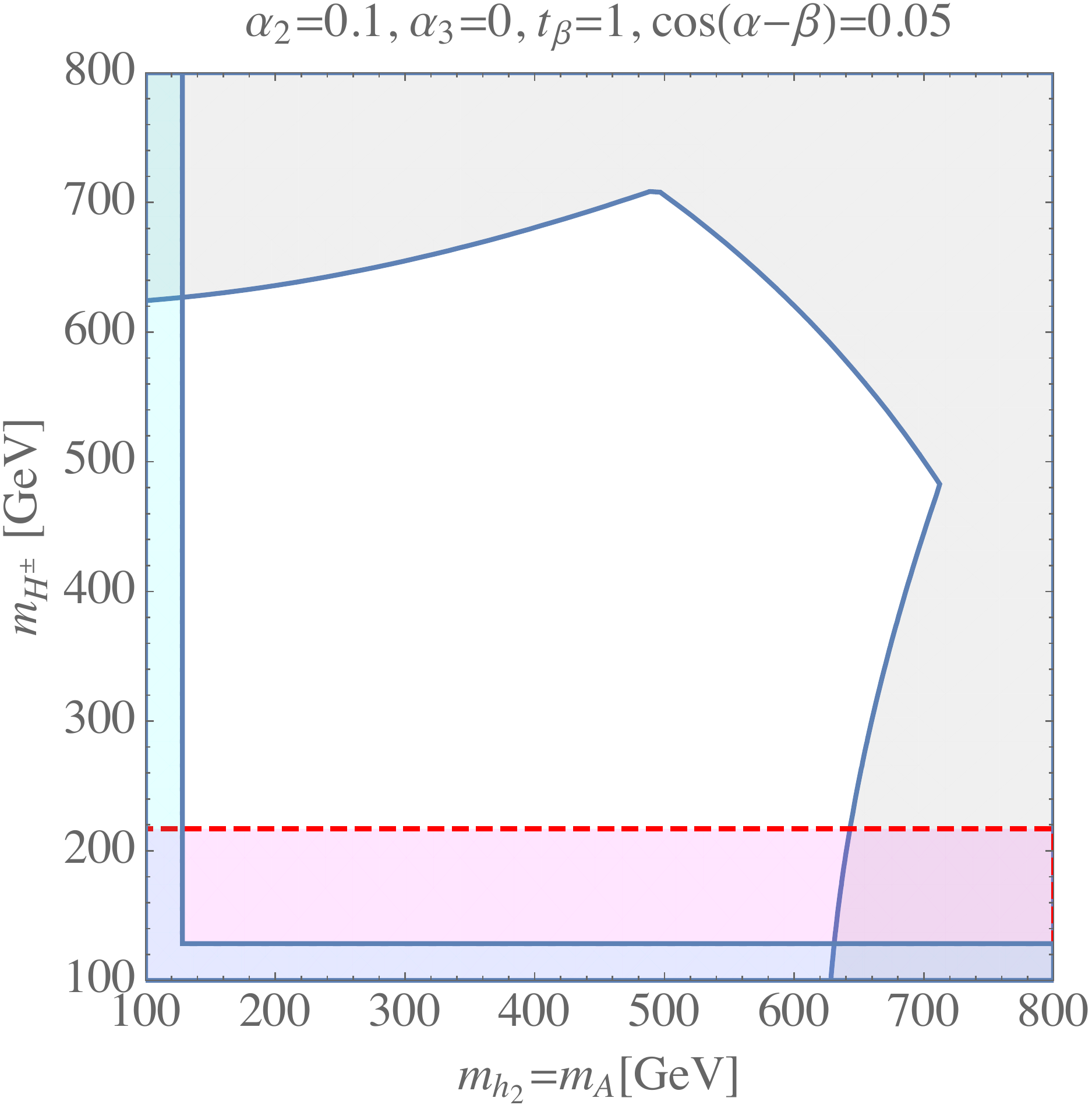}
    \includegraphics[height=0.45\textwidth]{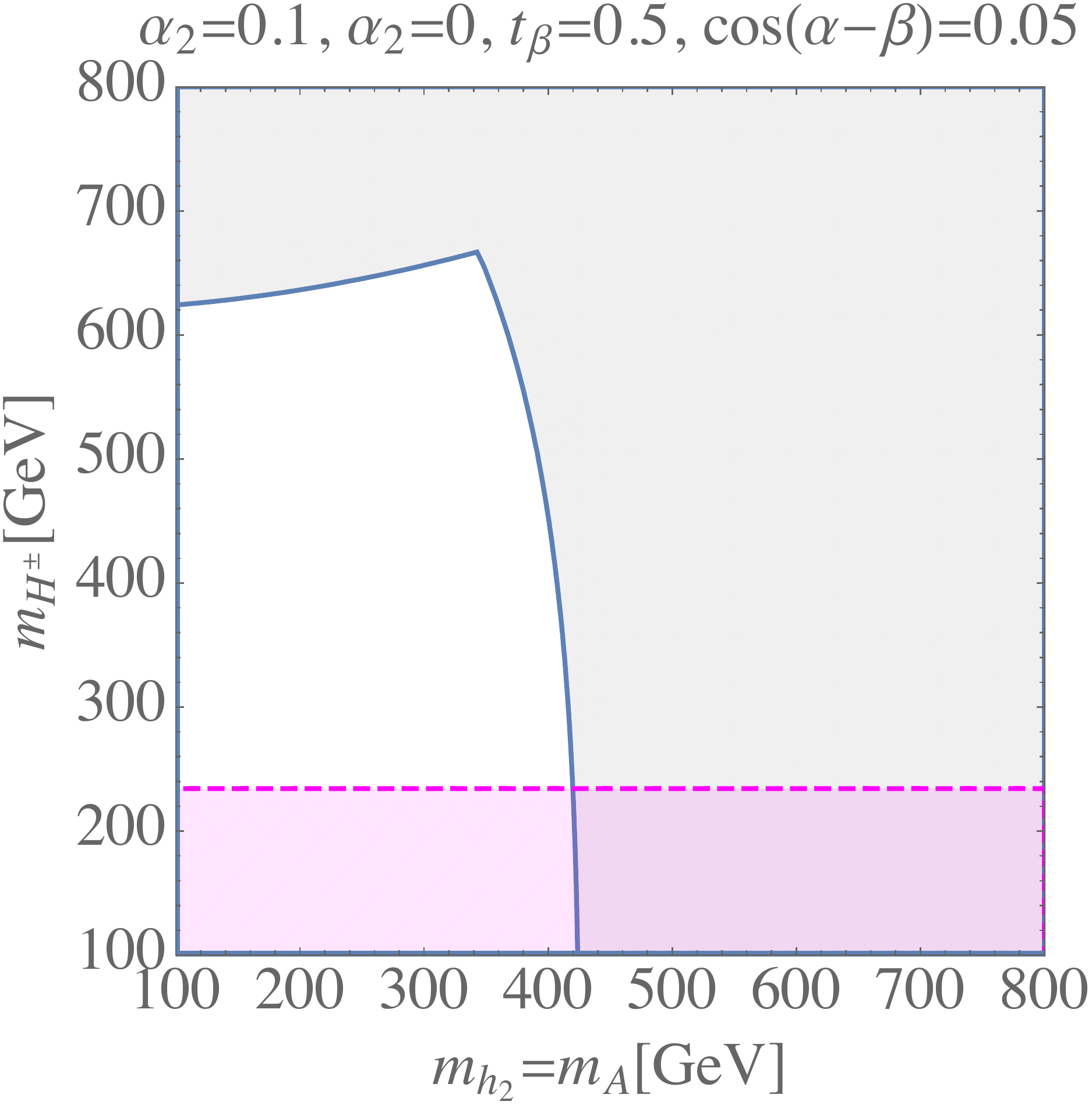}\\[0.2cm]
    \includegraphics[height=0.45\textwidth]{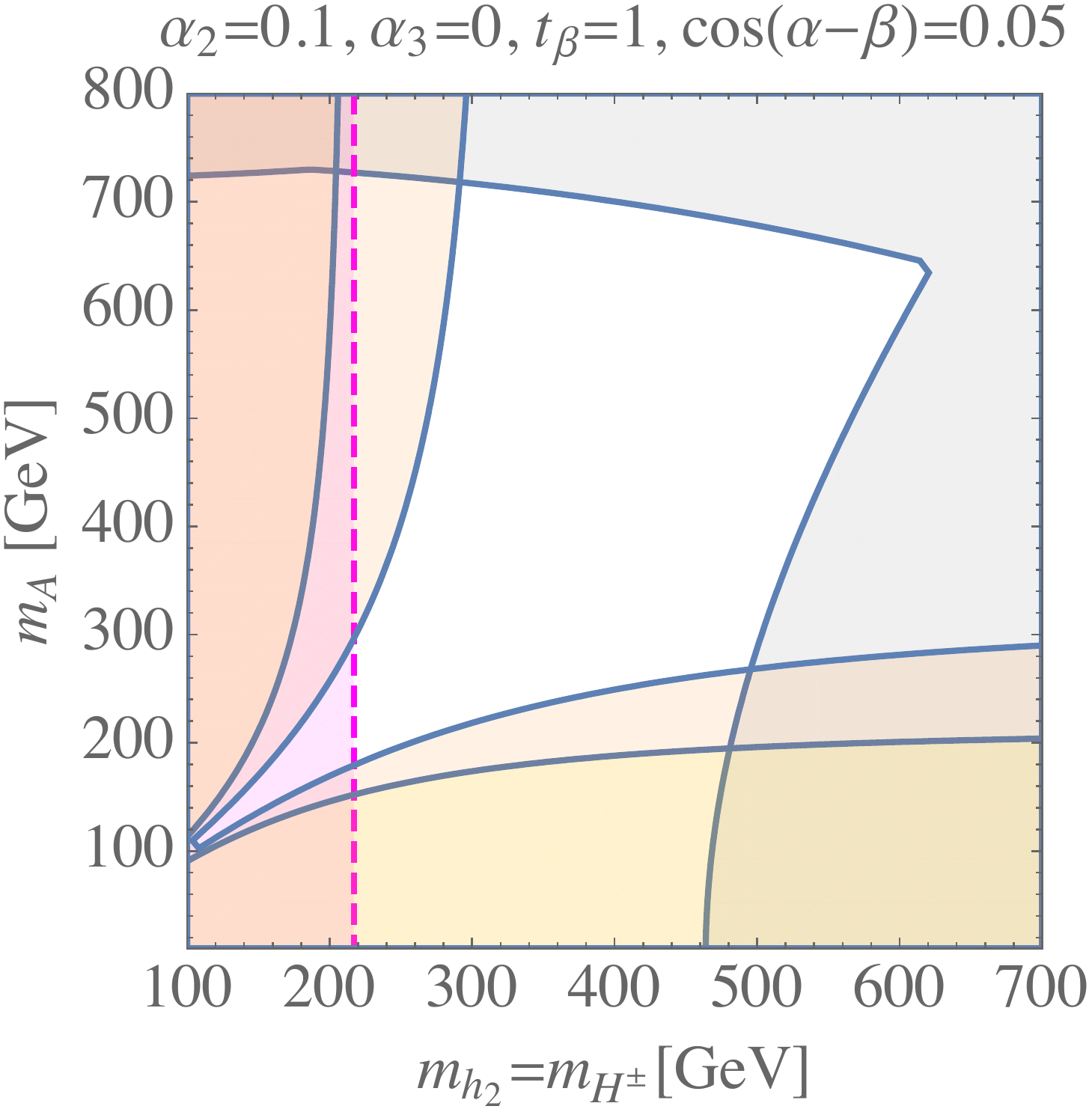}
    \includegraphics[height=0.45\textwidth]{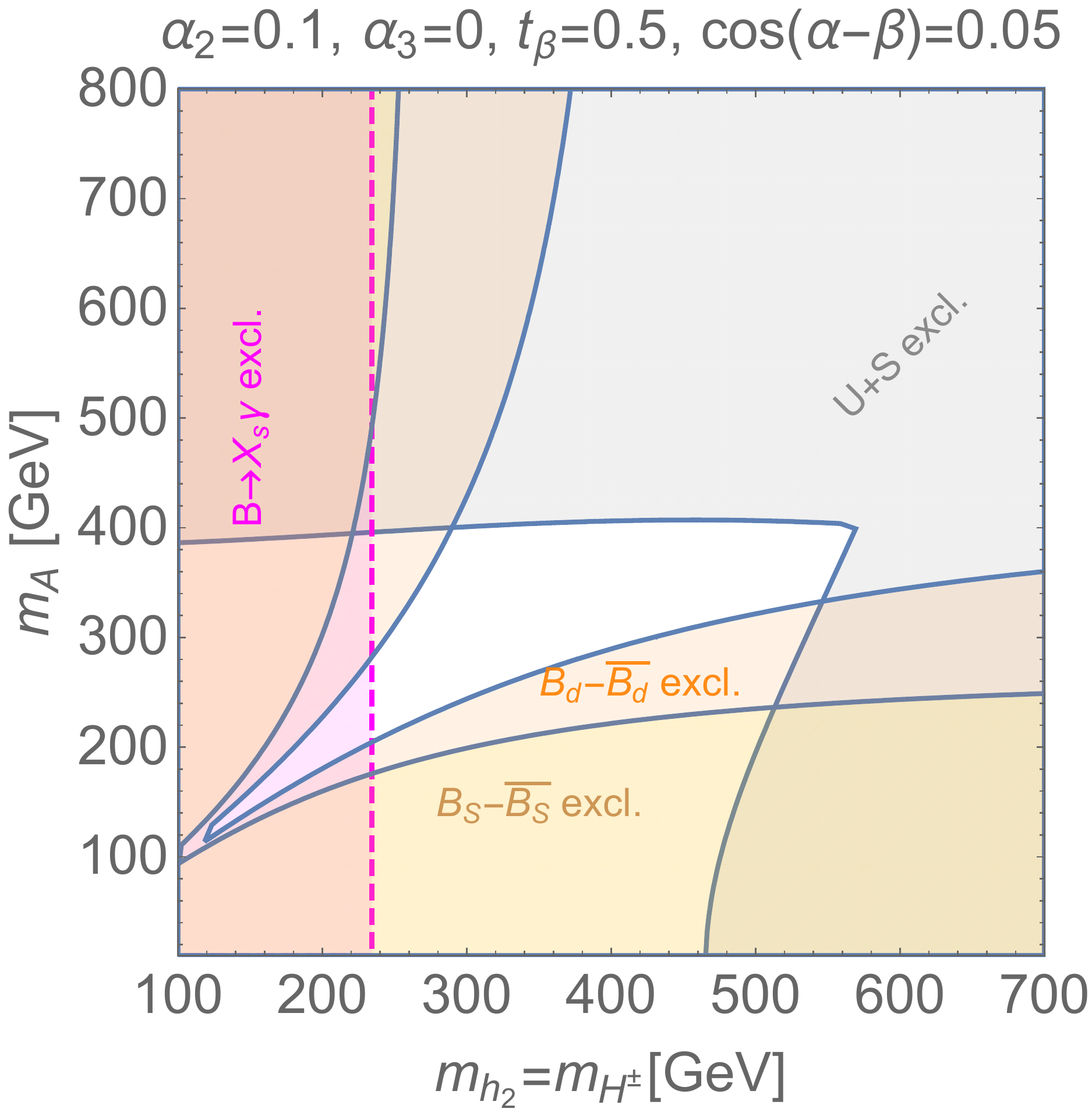}
  \end{center}
  \caption{Parameter space in terms of $m_{h_2}$ and $m_{H^\pm}$ (upper
    frames), and $m_A$ (lower frames). $\tan\beta = 1$ in the
    left and 0.5 in the right panels.
    We have chosen $v_s=2m_{h_3}=1$~TeV, $\cos(\alpha-\beta)=0.05$,
    and $y_{33}^u=y_t^\text{SM}$ in all frames.
    The mixing between heavy $CP$-even scalars is taken to be zero.
    The gray regions are excluded by unitarity and stability bounds.
    The magenta regions are excluded by $B\rightarrow X_s \gamma$, and
    cyan region is excluded by $B_s\rightarrow \mu^+\mu^-$.
    The yellow and orange regions are excluded by $B_s$ and $B_d$
    mixings, respectively.\label{fig:unitarity3}}
\end{figure}

Comparing to the bounds from $B_s\rightarrow \mu^+\mu^-$
in~(\ref{eq:Bmumu}), the $B$--${\bar B}$ mixings could
lead to tighter constraints on the flavor-violating Yukawa couplings
for down-type quarks unless $m_H$ and $m_A$ are almost degenerate.
The upper frames of Fig.~\ref{fig:unitarity3} show that a wide range
of heavy Higgs masses up to 600--700~GeV are allowed for $m_{h_2} =
m_A$ and $\tan\beta = \mathcal{O}(1)$.
On the other hand, for $\tan\beta=0.5$, the neutral Higgs boson
can be as heavy as 400~GeV, but the charged Higgs mass is constrained
as $240~{\rm GeV} \lesssim m_{H^\pm} \lesssim 650$~GeV.
For illustration, the case with $m_{h_2} = m_{H^\pm}$ has also been shown
in the lower frames of Fig.~\ref{fig:unitarity3}, where the narrower
region is allowed as compared with the case with $m_{h_2} = m_A$.

\begin{figure}[t!]
  \begin{center}
    \includegraphics[height=0.45\textwidth]{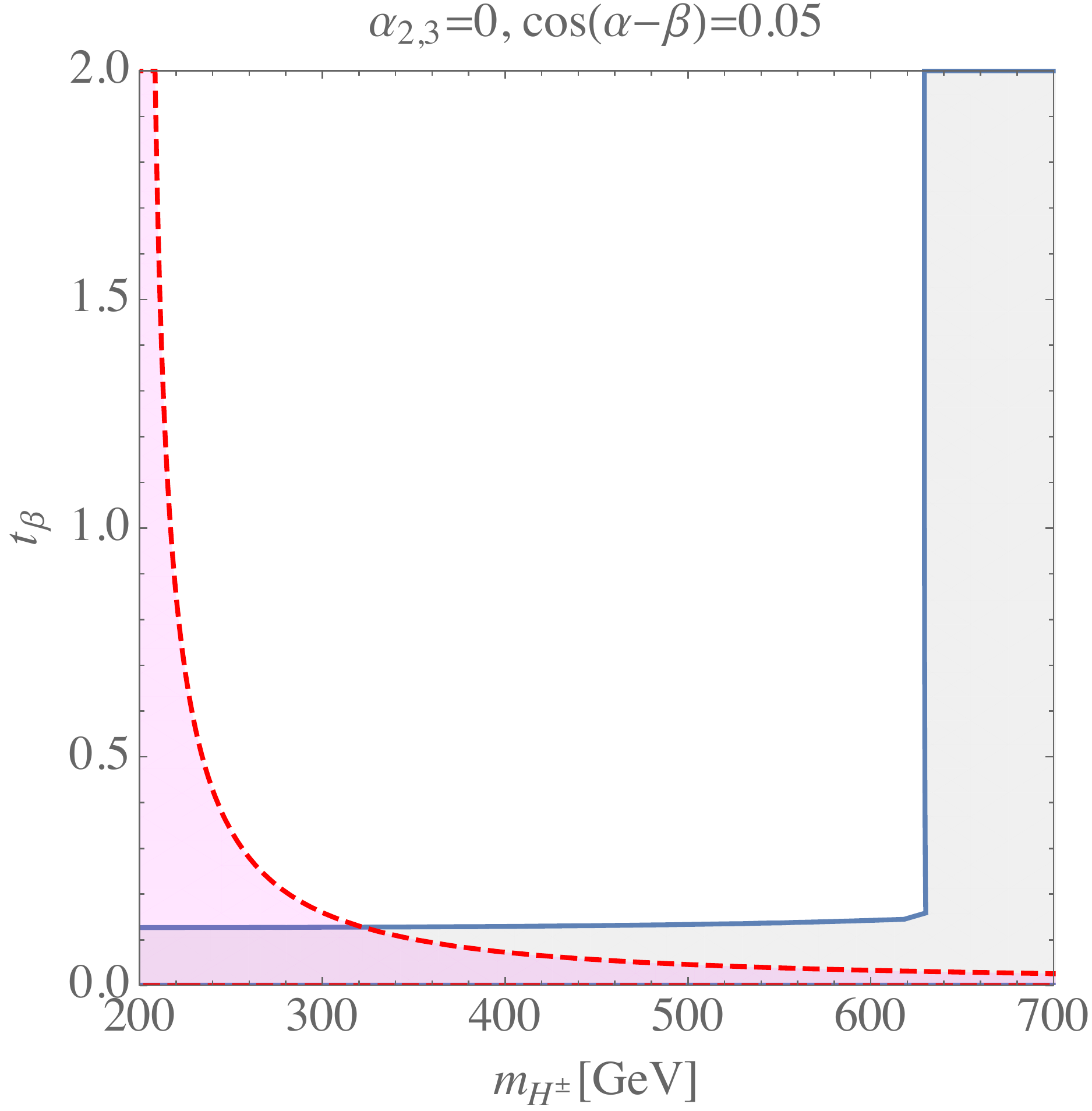}
    \includegraphics[height=0.45\textwidth]{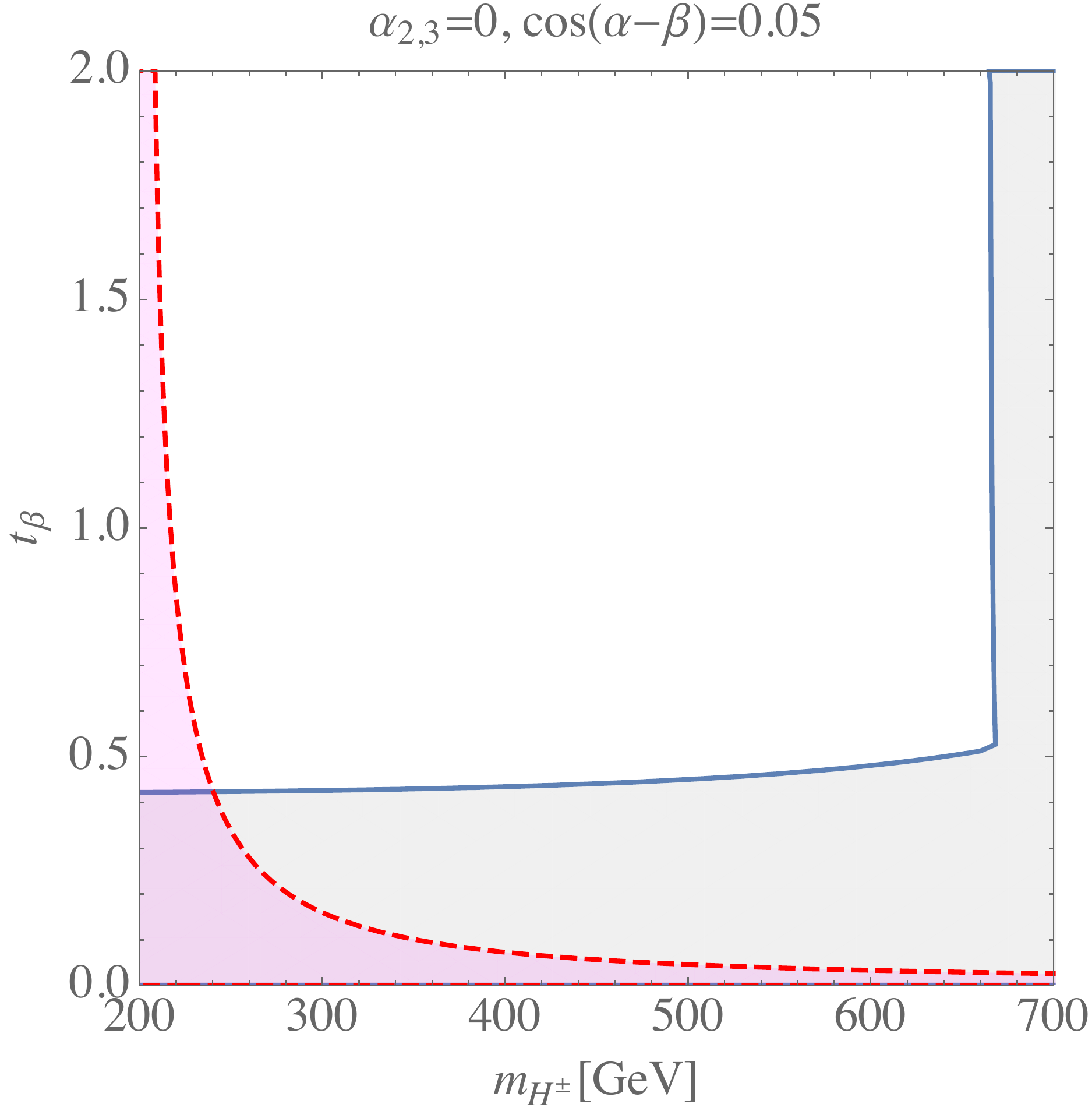}
  \end{center}
  \caption{Parameter space for $m_{H^\pm}$ and $\tan\beta$
    excluded by $B\rightarrow X_s \gamma$ within $2\sigma$ (red) and
    unitarity bounds (gray) with $y_{33}^u=y_t^\text{SM}$ for
    $m_A=m_{h_2}=160$~GeV (left panel) and  $m_A=m_{h_2}=350$~GeV
    (right panel).\label{fig:Bsgamma}}
\end{figure}

Another important bound comes from the inclusive radiative decay, $B
\to X_s\gamma$. The effective Hamiltonian relevant for the
$b\rightarrow s\gamma$ transition is
\begin{equation}
  {\cal H}_{{\rm eff},b\rightarrow s\gamma }= -\frac{4G_F}{\sqrt{2}}
  \, V_{tb} V^*_{ts} \left(C_7 {\cal O}_7+ C_8 {\cal O}_8 \right)
\end{equation}
with
\begin{equation}
  {\cal O}_7
  = \frac{e}{16\pi^2}\, m_b\, {\bar s} \sigma^{\mu\nu} P_R b\, F_{\mu\nu}, \quad
  {\cal O}_8
  = \frac{g_s}{16\pi^2}\, m_b\,  {\bar s} \sigma^{\mu\nu} P_R  T^a b\,  G^a_{\mu\nu}.
\end{equation}
The charged Higgs contributions to the Wilson coefficients are given
by~\cite{crivellin2017,ko2}
\begin{align}
  C^{\rm BSM}_7&= \frac{v^2}{2m^2_t}\frac{(\lambda^{H^-}_{t_R})^*\lambda^{H^-}_{t_R}}{V_{tb }V^*_{ts}}\, C^{(1)}_7(x_t) + \frac{v^2}{2m_t m_b}\frac{(\lambda^{H^-}_{t_L})^* \lambda^{H^-}_{t_R}}{V_{tb}V^*_{ts}}\,   C^{(2)}_7(x_t) , \nonumber\\
  C^{\rm BSM}_8&=\frac{v^2}{2m^2_t}\frac{(\lambda^{H^-}_{t_R})^*\lambda^{H^-}_{t_R}}{V_{tb }V^*_{ts}}\, C^{(1)}_8(x_t) + \frac{v^2}{2m_t m_b}\frac{(\lambda^{H^-}_{t_L})^* \lambda^{H^-}_{t_R}}{V_{tb}V^*_{ts}}\,   C^{(2)}_8(x_t)
\end{align}
with $x_t\equiv (m_t/m_{H^\pm})^2$, and
\begin{align}
C^{(1)}_7(x) &= \frac{x}{72} \bigg\{\frac{-8x^3+3x^2+12x-7+(18x^2-12)\ln x}{(x-1)^4} \bigg\}, \nonumber\\
C^{(2)}_7(x) &= \frac{x}{12}\bigg\{\frac{-5x^2+8x-3+(6x-4)\ln x}{(x-1)^3} \bigg\}, \nonumber\\
C^{(1)}_8(x) &= \frac{x}{24} \bigg\{\frac{-x^3+6x^2-3x-2-6x\ln x}{(x-1)^4} \bigg\}, \nonumber\\
C^{(2)}_8(x) &= \frac{x}{4} \bigg\{\frac{-x^2+4x-3-2\ln x}{(x-1)^3} \bigg\}.
\end{align}
Here $\lambda^{H^-}_{t_{L,R}}$ are given by Eqs.~(\ref{lamtL}) and~(\ref{lamtR}).
The Wilson coefficients in the SM at one loop are given by $C^{\rm
  SM}_7=3 C^{(1)}_7(m^2_t/m^2_W)$ and $C^{\rm SM}_8=3
C^{(1)}_8(m^2_t/m^2_W)$. $C^{\rm BSM}_8$ mixes into the  $C^{\rm BSM}_7$ at the scale of
$\mu_b=m_b$ through the renormalization group equations and contribute to
 ${\cal B}(B\rightarrow X_s \gamma)$~\cite{Borzumati:1998tg}.
The next-to-next-leading order SM prediction for ${\cal
  B}(B\rightarrow X_s \gamma)$ is~\cite{bsg-th}
\begin{equation}
  {\cal B}(B\rightarrow X_s \gamma) = (3.36\pm 0.23)\times 10^{-4},
\end{equation}
whereas the experimentally measured value of ${\cal B}(B\rightarrow
X_s \gamma)$ from HFAG is~\cite{Msexp}
\begin{equation}
{\cal B}(B\rightarrow X_s \gamma) =(3.43\pm 0.21\pm 0.07)\times 10^{-4}.
\end{equation}
As a result, the SM prediction for $B\rightarrow X_s \gamma$ is
consistent with experiments, so we obtain the bounds on the modified
Wilson coefficients as $-0.032<C^{\rm BSM}_7(\mu_b)<0.027$ at
$2\sigma$ level~\cite{Paul:2016urs}.
This constrains $\tan\beta$ in terms of charged Higgs mass as shown
in Fig.~\ref{fig:Bsgamma}, where unitarity and stability bounds are
displayed as well.
We also find that the case with $y_{33}^u = y_t^\text{SM} / \cos\beta$
has been excluded by $B \to X_s\gamma$, hence the case with $y_{33}^u =
y_t^\text{SM}$ is considered in Figs.~\ref{fig:unitarity3}
and~\ref{fig:Bsgamma} and collider studies in the next section.

\subsection{Predictions for {\boldmath$R_D$} and {\boldmath$R_{D^*}$}}

We briefly discuss the implications of flavor-violating couplings with
charged Higgs on $R_D$ and $R_{D^*}$.
The effective Hamiltonian relevant for $B\rightarrow D^{(*)}\tau \nu$
in our model is given as follows:
\begin{equation}
  {\cal H}_{\rm eff}= C^{cb}_{\rm SM} ({\bar c}_L \gamma_\mu b_L)
  ({\bar \tau}_L \gamma^\mu \nu_L) + C^{cb}_R ({\bar c}_L b_R)
  ({\bar\tau}_R \nu_L) +C^{cb}_L ({\bar c}_R b_L) ({\bar\tau}_R \nu_L),
\end{equation}
where the Wilson coefficient in the SM is $C^{cb}_{\rm
  SM}=2V_{cb}/v^2$, and the new Wilson coefficients generated by
charged Higgs exchanges are
\begin{align}
  C^{cb}_R = -\frac{\sqrt{2}m_\tau \tan\beta}{v \,m^2_{H^\pm}} \,
  {(\lambda^{H^-}_{c_L})}^*, \quad
  C^{cb}_L =-\frac{\sqrt{2}m_\tau \tan\beta}{v\, m^2_{H^\pm}} \,
  {(\lambda^{H^-}_{c_R})}^*.
\end{align}
See Eqs.~(\ref{lamcL}) and~(\ref{lamcR}) for $\lambda^{H^-}_{c_{L,R}}$.

\begin{figure}[t!]
  \begin{center}
    \includegraphics[height=0.45\textwidth]{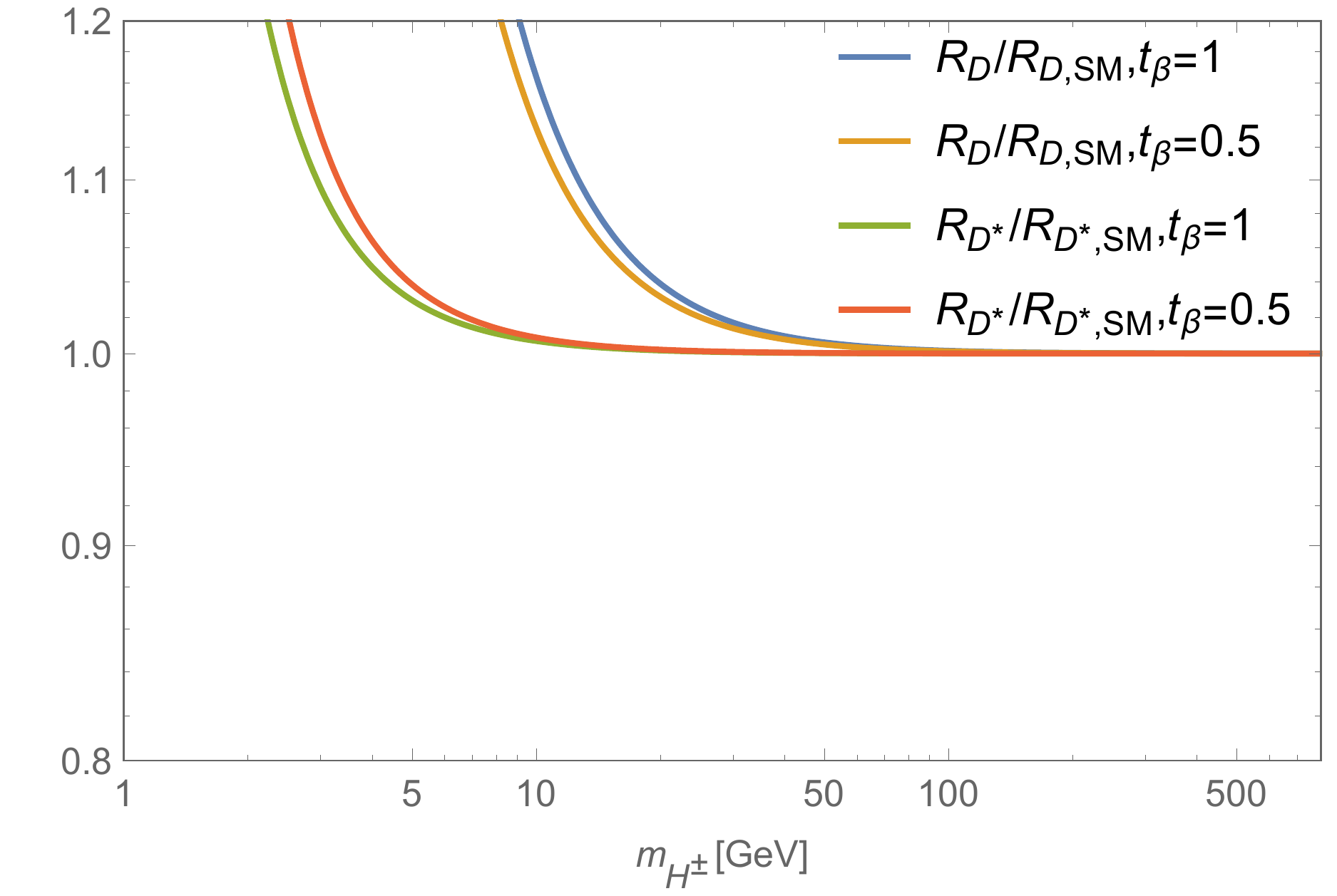}
  \end{center}
  \caption{The ratios of $R_D/R_{D,\text{SM}}$ and $R_{D^\ast}/R_{D^\ast,\text{SM}}$
    as the functions of charged Higgs mass for given
    $\tan\beta$.\label{fig:RD}}
\end{figure}

The ratios of the branching ratios for $B\rightarrow D^{(*)}\tau\nu$
to $B\rightarrow D^{(*)}\ell \nu$ with $\ell=e$, $\mu$ are defined by
\begin{equation}
  R_{D^{(*)}}= \frac{ \mathcal{B} (B\to D^{(*)} \tau \nu)}{ \mathcal{B}
    (B\to D^{(*)} \ell \nu)}.
\end{equation}
The SM expectations are $R_D=0.300\pm 0.008$ and
$R_{D^*}=0.252\pm 0.003$~\cite{RDs}, but the experimental results for
$R_{D^{(*)}}$ are deviated from the SM values by more than
$2\sigma$~\cite{RDexp,RDexp2,RDsexp,RDsexp2}.
Including the additional contributions from charged Higgs exchanges,
we find the simplified forms for $R_D$ and $R_{D^*}$ as
follows~\cite{ko2,crivellin2012}:
\begin{align}
  R_D
  &= R_{D,{\rm SM}} \left[1 +1.5\, {\rm Re}
    \left(\frac{C^{cb}_R+C^{cb}_L}{C^{cb}_{\rm SM}}\right)+
    \left\vert\frac{C^{cb}_R+C^{cb}_L}{C^{cb}_{\rm SM}}
    \right\vert^2\right], \nonumber\\
  R_{D^*}
  &= R_{D^*,{\rm SM}}  \left[1 +0.12 \, {\rm Re} \left(
    \frac{C^{cb}_R-C^{cb}_L}{C^{cb}_{\rm SM}}\right)+0.05
    \Big|\frac{C^{cb}_R-C^{cb}_L}{C^{cb}_{\rm SM}} \Big|^2\right].
\end{align}
As can be seen in Fig.~\ref{fig:RD}, a light charged Higgs is
necessary to have large deviations of $R_D$ and $R_{D^*}$. However, it
is excluded by $B\rightarrow X_s\gamma$. [See Fig.~\ref{fig:Bsgamma}.]
Therefore, our model cannot explain the experimental results for
$R_{D^{(*)}}$ simultaneously with the other bounds.

\section{Productions and decays of heavy Higgs bosons at the LHC\label{sec:higgs_lhc}}

We investigate the main production channels for heavy
Higgs bosons at the LHC, including the contributions from
flavor-violating interactions of quarks.
The decay modes of the heavy Higgs bosons for some benchmark
points are also studied, and we discuss smoking gun signals for heavy
Higgs searches at the LHC\@.
In this section, mixings with singlet scalar have been
neglected and the heavy neutral Higgs boson $H$ denotes
$h_2$. $h \equiv h_1$ is the SM-like Higgs with $m_h = 125$~GeV\@.

\subsection{Heavy neutral Higgs boson}

The main channels for neutral Higgs productions are the gluon fusion $gg
\to H$, bottom-quark fusion $b \bar b \to H$, and additional productions through the
flavor-violating interactions for the bottom quark, $b \bar d_i \to H$
and $d_i \bar b \to H$, where $d_i$ denotes light down-type quarks,
$d_i = d,\,s$. There are bottom quark associated productions, $b g \to
b H$ and $d_i g \to b H$, as well.

The leading-order cross section for the gluon fusion process at
parton level is
\begin{equation}
  \hat\sigma (gg \to H) = \frac{\alpha_s^2 m_H^2 }{576
    \pi v^2} \left\vert \frac{3}{4} \sum_q \left(
      \frac{\cos\alpha}{\cos\beta} + \frac{v \sin(\alpha -
        \beta)}{\sqrt{2} m_q \cos\beta} \tilde h_{33}^q \right)
    A_{1/2}^H (\tau_q) \right\vert^2 \delta (\hat s - m_H^2) ,
\end{equation}
where $\tau_q = m_H^2 / (4m_q^2)$. The loop function $A_{1/2}^H
(\tau)$ is given in Ref.~\cite{Djouadi:2005gi}.
$\hat s$ is the partonic center-of-mass energy. Here the contributions
of only top and bottom quarks have been taken into account. Note that
the top quark contribution is vanishing if one takes $y_{33}^u =
y_t^\text{SM}$ and the alignment limit as can be seen in
Eq.~(\ref{eq:lambda_t_SM}).
The parton-level cross section for bottom-quark fusion
$b \bar b \to H$ is
\begin{equation}
  \hat\sigma (b\bar b \to H)
  = \frac{\pi m_b^2}{18 v^2} \left( \frac{\cos\alpha}{\cos\beta} +
    \frac{v \sin(\alpha - \beta)}{\sqrt{2} m_b \cos\beta} \tilde
    h_{33}^d \right)^2 \left( 1 - \frac{4m_b^2}{m_H^2} \right)^{1/2}
    \delta (\hat s - m_H^2) .
\end{equation}
There are other single Higgs production channels through the
flavor-violating interactions, $b \bar d_i \to H$ and $d_i \bar b \to
H$.
The corresponding cross section is given by
\begin{equation}
  \hat\sigma (d_i \bar b \to H)
  = \frac{\pi |\tilde h_{i3}^d|^2 \sin^2 (\alpha - \beta)}{72
    \cos^2 \beta} \delta (\hat s - m_H^2) ,
\end{equation}
and $\hat\sigma (b \bar d_i \to H) = \hat\sigma (d_i \bar b \to H)$ at
parton level.

\begin{figure}[bt!]
  \begin{center}
    \includegraphics[width=0.48\textwidth]{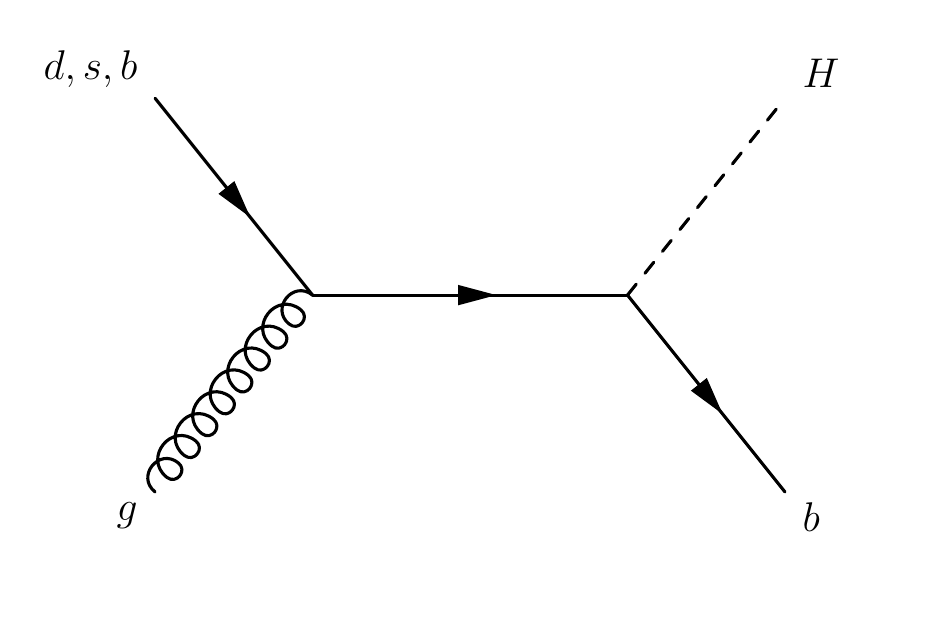}
    \includegraphics[width=0.48\textwidth]{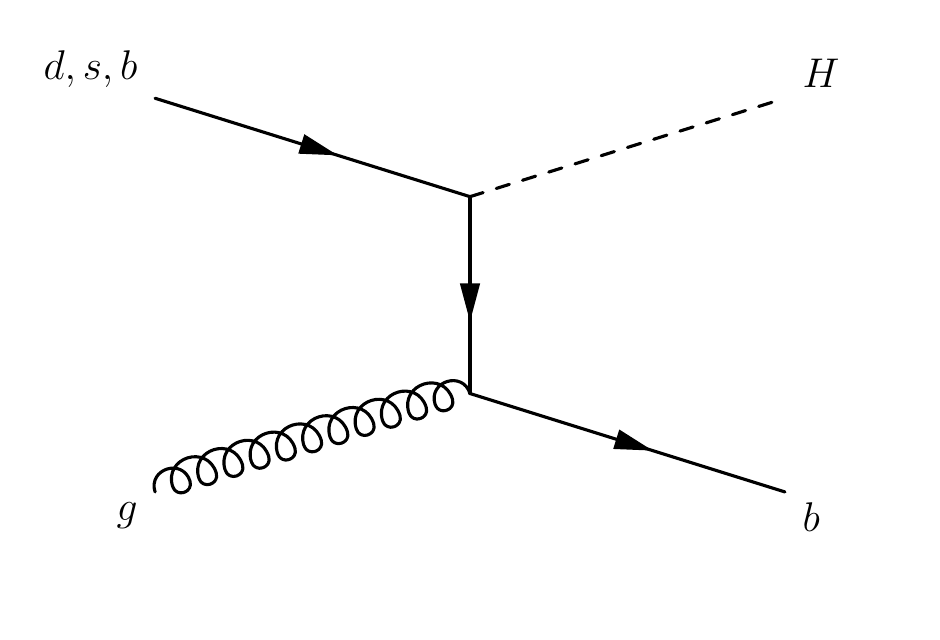}
  \end{center}
  \caption{Diagrams of the bottom quark associated productions of
    neutral Higgs bosons.\label{fig:qgqH_diag}}
\end{figure}

The bottom quark associated production of the Higgs boson can
occur by initial states with a bottom quark, that is, $b g \to b H$,
through the flavor-conserving interactions or initial states
with a light down-type quark, $d_i g \to b H$, via the
flavor-violating interactions. The former is nonvanishing even if all
the components of $\tilde h^d$ are zero.
The diagrams of the bottom quark associated production are
shown in Fig.~\ref{fig:qgqH_diag}.
The differential cross section for $b g \to b H$ at parton level is
\begin{equation}
  \frac{d \hat\sigma}{d\hat t} (b g \to b H)
  = \frac{\alpha_s (\lambda_b^H)^2}{96 (\hat s - m_b^2)^2} \left[
    \frac{2 F_1 - F_2^2 - 2 G_1 G_2}{(\hat s - m_b^2) (\hat t -
      m_b^2)} + 2m_b^2 \left( \frac{G_1}{(\hat s - m_b^2)^2} +
      \frac{G_2}{(\hat t - m_b^2)^2} \right) \right] ,
\end{equation}
where
\begin{equation*}
  F_1 = \hat s \hat t - m_b^4, \quad F_2 = \hat s + \hat t - 2m_b^2 ,
  \quad
  G_1 = m_H^2 - m_b^2 - \hat s, \quad G_2 = m_H^2 - m_b^2 - \hat t,
\end{equation*}
and $\lambda_b^H$ is given in (\ref{eq:lambda_b}). For the $d_i g \to
b H$ process, it is
\begin{align}
  \frac{d\hat\sigma}{d\hat t} (d_i g \to b H)
  = \frac{\alpha_s |\tilde h_{i3}^d|^2}{96
  \hat s^2 (\hat t - m_b^2)} \frac{\sin^2 (\alpha -
  \beta)}{\cos^2\beta} \left[ \frac{2F_1 - F_2^2 - 2G_1 G_2}{\hat s} +
  \frac{2 m_b^2 G_2}{\hat t - m_b^2} \right]
\end{align}
with
\begin{equation*}
  F_1 = \hat s \hat t, \quad F_2 = \hat s + \hat t - m_b^2, \quad
  G_1 = m_H^2 - m_b^2 - \hat s, \quad G_2 = m_H^2 - \hat t .
\end{equation*}
And again, $\hat\sigma (\bar d_i g \to \bar b H) = \hat\sigma (d_i g
\to b H)$ at parton level.

\begin{figure}[bt!]
  \begin{center}
    \includegraphics[width=0.45\textwidth]{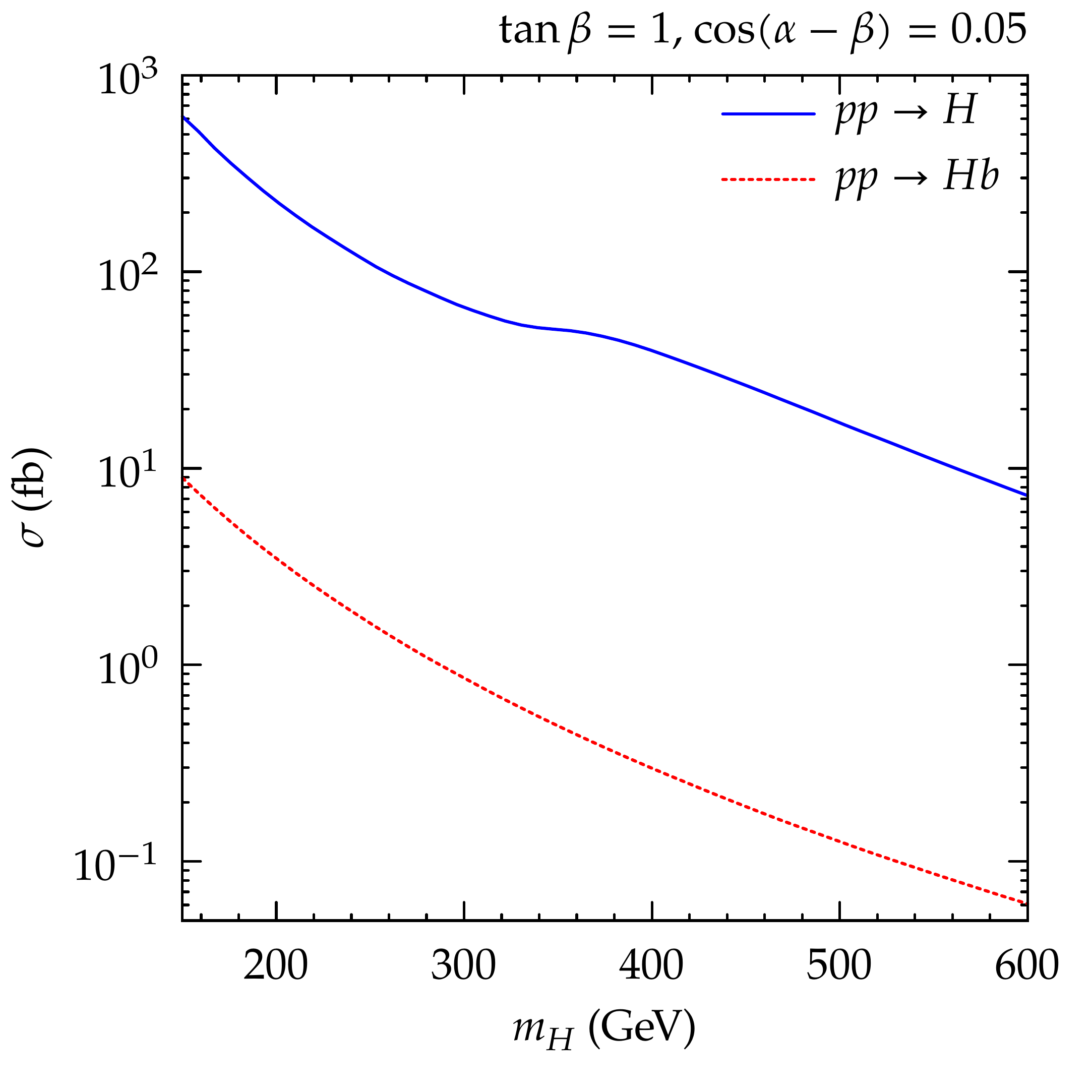}
    \includegraphics[width=0.45\textwidth]{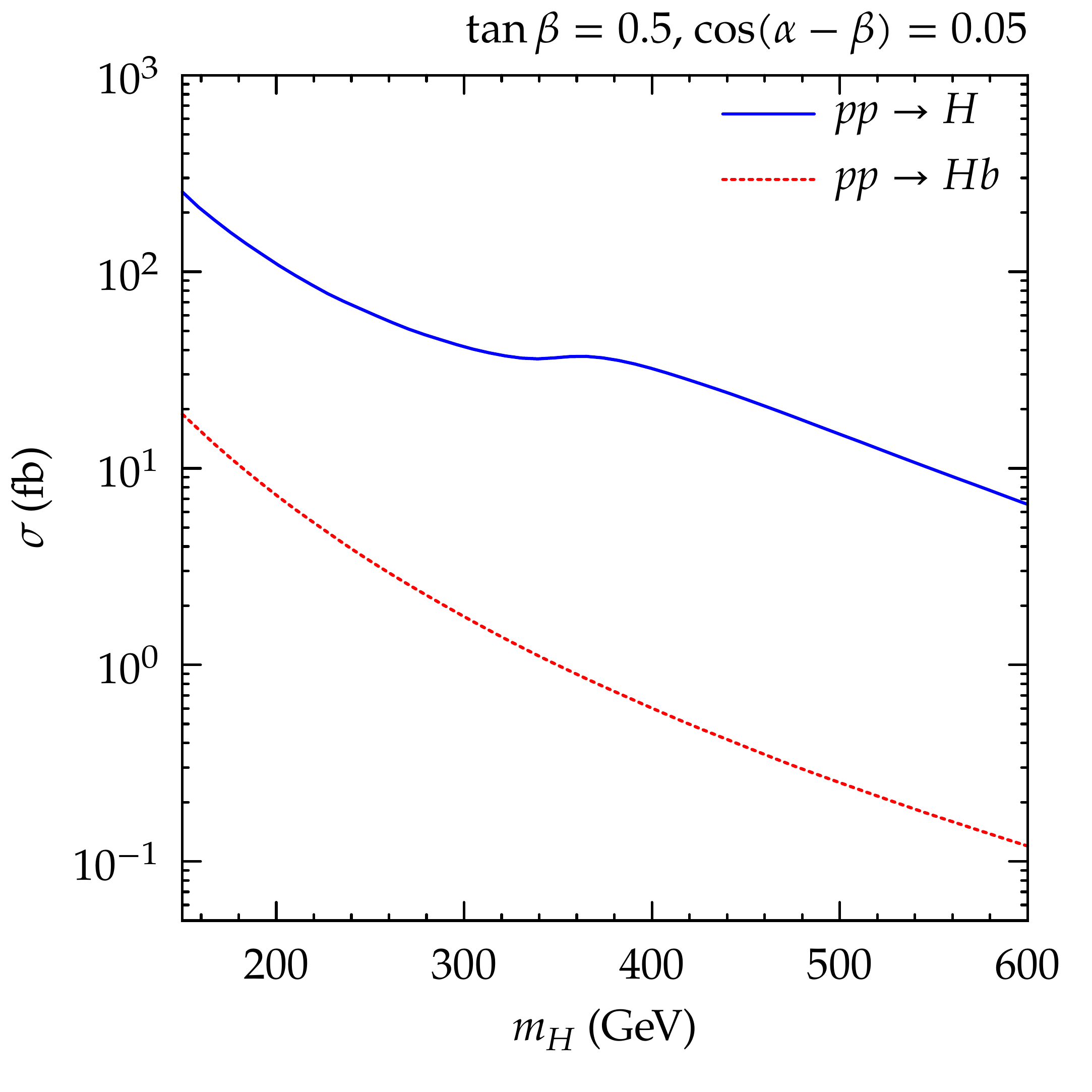}
  \end{center}
  \caption{Production cross sections of the heavy neutral Higgs $H$
    at 14~TeV proton-proton collisions. We have chosen $\tan\beta = 1$
    (left panel) and $\tan\beta = 0.5$ (right panel) with
    $\cos(\alpha - \beta) = 0.05$ and $y_{33}^u =
    y_t^\text{SM}$.\label{fig:xsec_neutral}}
\end{figure}

We perform the integration by using the Monte Carlo method to
obtain the production cross sections at proton-proton collisions of
14~TeV and employ the NNPDF2.3 parton distribution function (PDF)
set~\cite{Ball:2012cx} via the LHAPDF 6
library~\cite{Buckley:2014ana}. The renormalization and factorization
scales are set to $m_H$, and $m_b = 4.7$~GeV. The resulting production
cross sections as a function of $m_H$ are shown in
Fig.~\ref{fig:xsec_neutral}.
In all frames we set $\cos (\alpha - \beta) = 0.05$, close to the
alignment limit, and $y_{33}^u = y_t^\text{SM}$.
A constant $K$-factor of 2.5 has been multiplied to the gluon fusion
production cross section, while the leading-order expressions have
been used for the other production channels.

In the alignment limit, the neutral Higgs coupling to the top quarks
$\lambda_t^H$ is vanishing as can be seen in
Eq.~(\ref{eq:lambda_t_SM}).
In this case, the single Higgs production through the gluon fusion
process is suppressed compared to the SM case, though nonvanishing due
to the bottom quarks in the loop.
Still, the gluon fusion production convoluted with PDF is the most
dominant channel for the single Higgs production and $b \bar b \to H$
is the subdominant one for $\tan\beta \gtrsim \mathcal{O}(0.1)$.
On the other hand, for smaller $\tan\beta$, the
flavor-violating Higgs couplings to light quarks become larger and
contributions from the initial states with the light down-type quarks
$d_i \bar b \to H$ is subdominant, and become even the most dominant
channel in the case of very small $\tan\beta = \mathcal{O}(0.01)$.
However, since we find that such scenarios with very small $\tan\beta$
have been excluded by bounds from the experimental results on $B$-meson
mixings and decays, particularly by $B \to X_s \gamma$ as seen in the
previous section, we have chosen $\tan\beta = 1$ and 0.5 as benchmarks
for this study.
%
%
%
For $m_H = 200$~GeV and $\tan\beta = 1$ (0.5), $\sigma_{pp \to H}
\simeq 225.2$ (110.5)~fb, and
$(\sigma_{b \bar d_i \to H} + \sigma_{d_i \bar b \to H}) / \sigma_{g g
  \to H} = 0.62$\% (1.6\%), while $(\sigma_{b \bar d_i \to H} +
\sigma_{d_i \bar b \to H})/ \sigma_{b \bar b \to H} \simeq 1.6$\%
(10.9\%) at the LHC\@.
As the neutral Higgs gets heavier, the production cross sections rapidly
decreases.
For $m_H = 400$~GeV and $\tan\beta = 1$ (0.5), $\sigma_{pp \to H}
\simeq 38.4$ (31.7)~fb.

As can be seen in Fig.~\ref{fig:xsec_neutral}, the production cross
section of the bottom quark associated process increases as
$\tan\beta$ is smaller since the effect of the flavor-violating
couplings become larger. In particular, if $m_H \lesssim 200$~GeV the
production cross section is $\mathcal{O}(10)$~fb, so it can be served
as a good search channel at the LHC\@. Meanwhile, if $m_H \gtrsim
2m_t$, the cross section decreases down to $\lesssim\mathcal{O}(1)$~fb.

We now turn to the decay widths of the neutral Higgs bosons and obtain
their branching ratios. Ignoring the mixing among the SM-like Higgs
and singlet scalar, the partial decay widths to quarks are
\begin{align}
  \Gamma (H \to b \bar d_i) = \Gamma (H \to d_i \bar b)
  &= \frac{3 |\tilde h_{i3}^d|^2 \sin^2 (\alpha - \beta)}{32 \pi
    \cos^2\beta} m_H {\left( 1 - \frac{m_b^2}{m_H^2} \right)}^2,
    \nonumber\\
  \Gamma (H \to q \bar q)
  &= \frac{3 {(\lambda_q^H)}^2}{16\pi}  m_H {\left( 1 - \frac{4
    m_q^2}{m_H^2} \right)^{3/2}},
\end{align}
where $q = t$, $b$, $c$. $\lambda_b^H$ and
$\lambda_t^H$ are given in (\ref{eq:lambda_b}) and
(\ref{eq:lambda_t}), and
\begin{equation}
  \lambda_c^H = \frac{\sqrt{2} m_c \cos\alpha}{v \cos\beta} .
\end{equation}
On the other hand, the Higgs interactions to the charged leptons are
flavor-conserving and the corresponding decay width is given as
\begin{equation}
  \Gamma (H \to \tau^+ \tau^-) =
  \frac{m_\tau^2 \cos^2 \alpha}{8\pi v^2 \cos^2\beta} m_H \left( 1 -
    \frac{4 m_\tau^2}{m_H^2} \right)^{3/2} .
\end{equation}

The partial widths to electroweak gauge bosons $V = W$, $Z$ are
given as
\begin{equation}
  \Gamma (H \to VV) = \frac{\delta_V m_H^3 \cos^2 (\alpha
  - \beta)}{32\pi v^2} \left( 1 - \frac{4 m_V^2}{m_H^2} \right)^{1/2}
  \left( 1 - \frac{4 m_V^2}{m_H^2} + \frac{12 m_V^4}{m_H^4} \right) ,
\end{equation}
where $\delta_W = 2$ and $\delta_Z = 1$. These partial widths are
vanishing in the alignment limit.
If $m_H > 2 m_{Z^\prime}$, the decay mode of $H \to Z^\prime Z^\prime$
opens. Ignoring the small mixing with the $Z$ boson, the decay width
is
\begin{equation}
  \Gamma (H \to Z^\prime Z^\prime) = \frac{g_{Z^\prime}^4 x^4 m_H^3
    v^2 \sin^2 \beta \sin^2 \alpha}{2592 \pi m_Z^4} \left( 1 - \frac{4
      m_{Z^\prime}^2}{m_H^2} \right)^{1/2} \left( 1 - \frac{4
      m_{Z^\prime}^2}{m_H^2} + \frac{12 m_{Z^\prime}^4}{m_H^4} \right) .
\end{equation}
However, we find that this decay mode is almost negligible for
small $g_{Z^\prime} x \simeq \mathcal{O}(0.05)$ and $m_{Z'} \gtrsim
400$~GeV, which would be necessary to evade constraints from the
$Z^\prime$ searches at the LHC\@.

The neutral Higgs boson can also decay into $\gamma \gamma$ and
$gg$ through fermion or gauge boson loops. At leading order, the
decay widths are given as
\begin{align}
  \Gamma (H \to \gamma\gamma)
  &= \frac{\alpha^2 m_H^3}{256 \pi^3 v^2} \Bigg\vert
    \sum_{q = t, \, b} 3 Q_q^2 \frac{\lambda_q^H v}{\sqrt{2} m_q}
    A_{1/2}^H (\tau_q) + \frac{\cos\alpha}{\cos\beta} A_{1/2}^H
    (\tau_\tau) + \cos (\alpha - \beta) A_1^H (\tau_W) \Bigg\vert^2,
    \nonumber\\
  \Gamma (H \to gg)
  &= \frac{\alpha_s^2 m_H^3}{72 \pi^3 v^2} \left\vert
     \frac{3}{4} \sum_{q = t, \, b} \frac{\lambda_q^H v}{\sqrt{2} m_q}
     A_{1/2}^H (\tau_q) \right\vert^2,
\end{align}
where $Q_q$ is the electric charge of the quark and
$\tau_i \equiv m_H^2 / (4m_i^2)$.
The loop functions $A_{1/2}^H$ and $A_1^H$ can be found in
Ref.~\cite{Djouadi:2005gi}.

If $m_H > 2m_h$, the heavy neutral Higgs can decay into a pair of SM-like Higgs
bosons.\footnote{If the singlet scalar $h_3 = S$ is light enough,
  additional decay modes such as $H \to S h$ can occur and become
  important channels~\cite{vonBuddenbrock:2016rmr}.
  Here we assume that $S$ is heavy, $m_S \gtrsim 0.5$--1~TeV, and the
  mixings with doublet Higgs bosons are negligible.}
The triple interaction comes from the scalar potential in
(\ref{eq:scalar_potential_1}),
\begin{equation}
  V_1 \supset \frac{g_{Hhh} v}{2} H h h,
\end{equation}
where
\begin{align}
  g_{H h h} =
  &~3 ( \lambda_1 \sin\alpha \cos\beta + \lambda_2 \cos\alpha \sin\beta
    ) \sin(2\alpha) \nonumber\\
  & + (\lambda_3 + \lambda_4) \left[ 3 \cos(\alpha + \beta)
    \cos(2\alpha)  - \cos(\alpha - \beta) \right].
    \label{eq:ghhh}
\end{align}
The decay width for the $H \to hh$ process is given as
\begin{equation}
  \Gamma(H \to hh) = \frac{g_{Hhh}^2 v^2}{32 \pi m_H} {\left( 1 -
      \frac{4m_h^2}{m_H^2} \right)}^{1/2}.
\end{equation}
The quartic couplings in the Higgs potential can be evaluated by
choosing values of $\mu v_s$, $\tan\beta$, $\sin\alpha$, and $m_H$ if
mixing with the singlet scalar is negligible, $\alpha_2 \simeq
\alpha_3 \simeq 0$. See Appendix~\ref{app:higgs_sector}.

\begin{figure}[bt!]
  \begin{center}
    \includegraphics[width=0.48\textwidth]{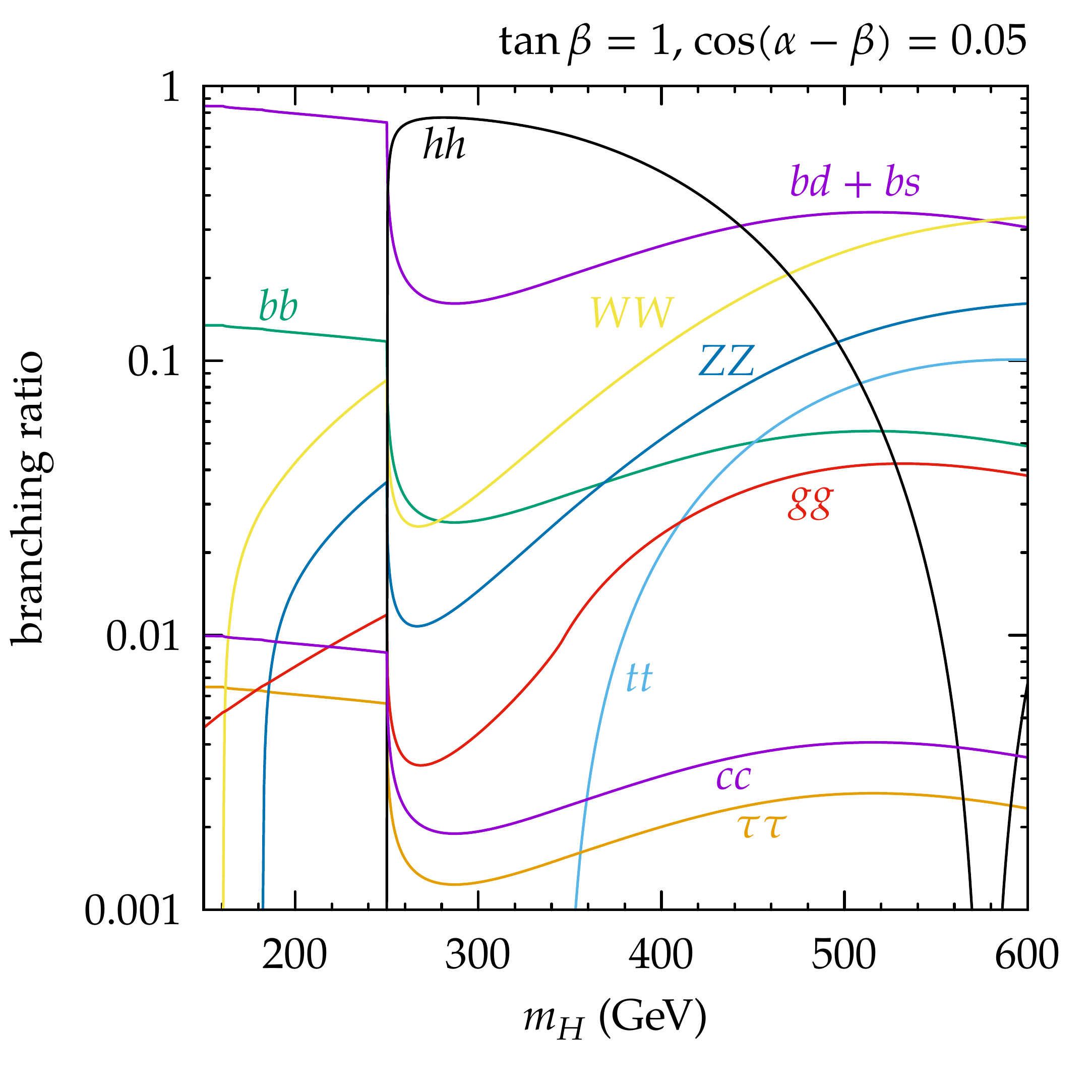}
    \includegraphics[width=0.48\textwidth]{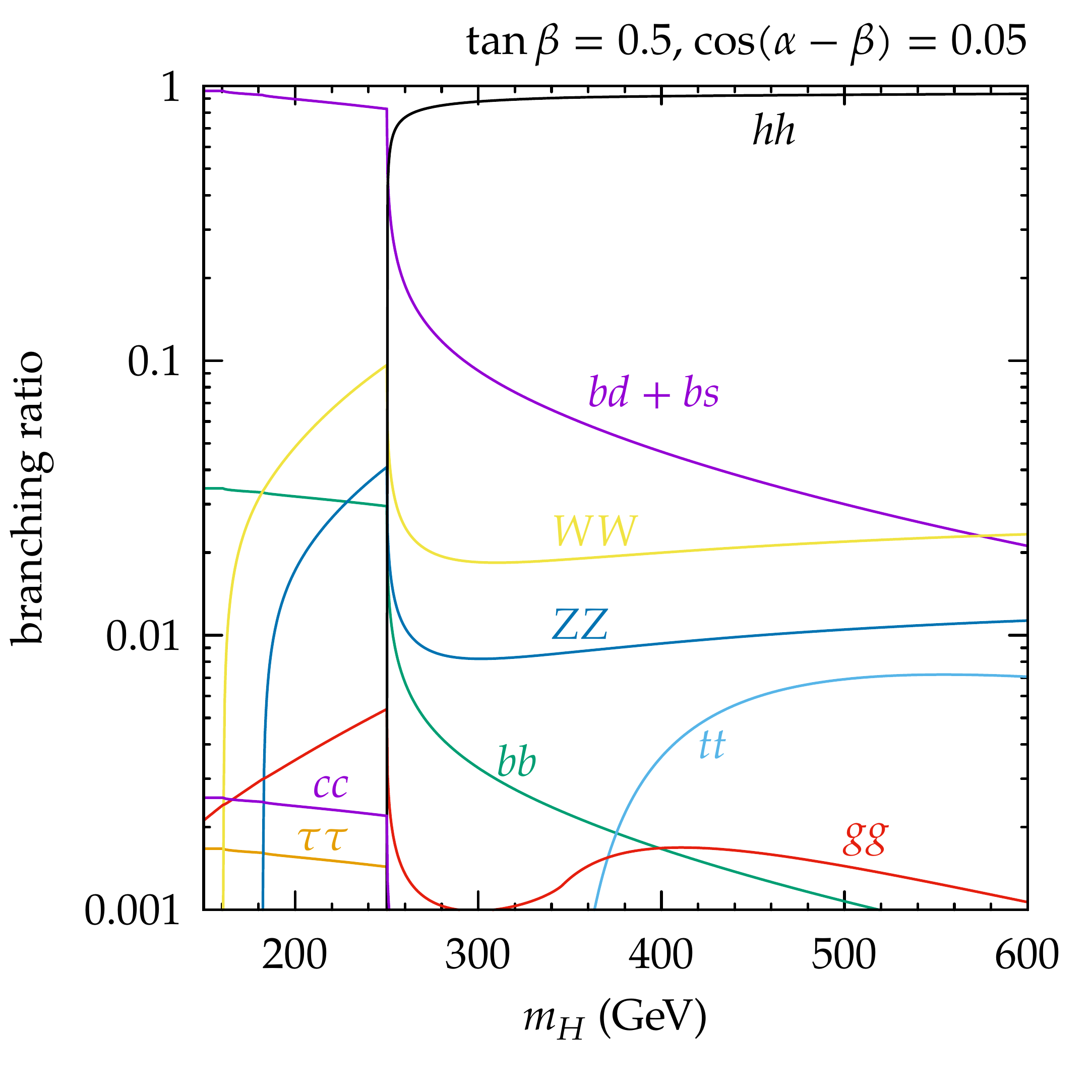}
  \end{center}
  \caption{Branching ratios of the heavy neutral Higgs $H$.
    $\tan\beta =1$ and $\mu = 200$~GeV (left panel), and $\tan\beta =
    0.5$ and $\mu = 50$~GeV (right panel) have been taken.
    $v_s = 1$~TeV and $\cos (\alpha - \beta) = 0.05$ for both
    panels.\label{fig:br_neutral}}
\end{figure}

By combining all the decay widths, we obtain the branching ratio
of each decay mode. Fig.~\ref{fig:br_neutral} shows the branching
ratios of the neutral Higgs boson $H$ for $\cos (\alpha - \beta) =
0.05$ and $v_s = 1$~TeV, but with different values of $\mu$ to satisfy
the unitary and stability bounds studied in
Subsec.~\ref{sec:unitarity_bounds}.
We observe that $H \to b \bar d_i$/$d_i \bar b$ is the predominant
decay mode if $m_H < 2m_h$, whereas the di-Higgs mode $H \to hh$
becomes the most important if the mode is kinematically allowed,
irrespective of $\tan\beta$.
In practice, the branching ratio of di-Higgs mode $\mathcal{B}(H \to
hh)$ depends on the choice of $\mu v_s$ value.
If we take a smaller $\mu v_s$ value, for instance, $\mu = 200$~GeV
and $v_s = 500$~GeV with $\tan\beta = 1$, we find that
$H \to b \bar d_i$/$d_i \bar b$ is always the most dominant decay mode.
The dip near $m_H = 580$~GeV in the left panel of
Fig.~\ref{fig:br_neutral} is due to the accidental cancellation
in the Higgs triple coupling (\ref{eq:ghhh}).
The position of dip also depends on the value
of $\mu v_s$ for given $\tan\beta$ and $\cos(\alpha - \beta)$.
On the other hand, the $b{\bar b}$ mode and diboson modes such as
$WW/ZZ$ are subdominant.

From these observations, we expect that the search strategies would be
different depending on the mass of the heavy Higgs boson. For $m_H <
2m_h$, $p p \to H \to b \bar d_i$/$d_i \bar b$, {\em i.e.}, dijet
final states containing one $b$ jet is the most important, but for
$m_H > 2m_h$, the di-Higgs channel, and possibly in conjunction with
the dijet channel with one $b$ jet, is important to search the heavy
neutral Higgs boson at the LHC\@.
Thus, the neutral Higgs boson with $m_H < 250$~GeV can receive
constraints from dijet searches~\cite{LHC-dijet}.
Although the dijet channel has typically been used to seek for heavy
resonances in a few TeV scales, it can probe lower scales if it is
associated with a hard photon or jet from initial state
radiations. The ATLAS collaboration has searched light resonance
with dijet invariant mass down to 200~GeV in the final states of dijet
in association with a photon~\cite{ATLAS:2016jcu}.
In our case, gluon fusion production is the most dominant channel and
it is not associated with a hard photon.
It can have a hard jet from the gluons in the initial states, but the
mass region below 250~GeV has not been searched yet in the final
states of dijet in association with a hard jet.
For $m_H > 250$~GeV, bounds from di-Higgs searches can be imposed, but
we find that they do not have enough sensitivities for heavy neutral
Higgs bosons in our model yet~\cite{LHC-hh}.

\subsection{Heavy charged Higgs boson}

One of the conventional search channels for the heavy charged Higgs
with $m_{H^\pm} > m_t$ at hadron colliders is the top quark associated
production, $b g \to t H^-$, by the similar diagrams as
$b g \to b H$.
Since the charged Higgs boson can have enhanced couplings with the
light up-type quarks due to nonzero components of $\tilde h^d$,
we can also have a sizable production cross section of
the bottom quark associated process from the initial
states with light up-type quarks, $u_i g \to b H^+$ where $u_i = u$,
$c$.\footnote{
  We note that there have been collider studies on the
  production of heavy Higgs bosons due to flavor-violating interactions
  for up-type quarks. See, for instance, Ref.~\cite{FChiggs}.
}

The differential cross section for $b g \to t H^-$ at parton level is
\begin{align}
  \frac{d \hat\sigma}{d\hat t} =
  &~\frac{\alpha_s}{48 (\hat s - m_b^2)^2} \left[ \left(
    | \lambda_{t_L}^{H^-} |^2 + | \lambda_{t_R}^{H^-} |^2 \right)
    \left( \frac{2 F_1 - F_2^2 - 2G_1 G_2}{(\hat s - m_b^2)
    (\hat t - m_t^2)} + \frac{2 m_b^2 G_1}{(\hat s - m_b^2)^2}
    + \frac{2 m_t^2 G_2}{(\hat t - m_t^2)^2} \right)
    \right. \nonumber\\
  &\left .
    + \left( \lambda_{t_L}^{H^-} (\lambda_{t_R}^{H^-})^\ast +
    \lambda_{t_R}^{H^-} (\lambda_{t_L}^{H^-})^\ast \right)
    \frac{4 m_b m_t m_{H^\pm}^2}{(\hat s - m_b^2) (\hat t - m_t^2)}
    \left( 1 - \frac{F_1 F_2}{m_{H^\pm}^2 (\hat s - m_b^2) (\hat t -
    m_t^2)} \right) \right],
\end{align}
where
\begin{align}
  F_1
  &= \hat s \hat t - m_b^2 m_t^2, \quad
    F_2 = \hat s + \hat t - m_b^2 - m_t^2, \nonumber\\
  G_1
  &= m_{H^\pm}^2 - m_t^2 - \hat s , \quad
    G_2 = m_{H^\pm}^2 - m_b^2 - \hat t.
\end{align}
Since the diagrams contributing to bottom quark associated processes
has the same Lorentz structure as those for $b g \to t H^-$, we can obtain
their parton-level cross sections by replacing
$\lambda_{t_{L,R}}^{H^-}$ with $\lambda_{u_{iL,R}}^{H^-}$, $m_b$ with
$m_{u_i} \simeq 0$, and $m_t$ with $m_b$. They are given as
\begin{equation}
  \frac{d \hat\sigma}{d\hat t} (u_i g \to b H^+) =
  \frac{\alpha_s (| \lambda_{u_{iL}}^{H^-} |^2 + |
    \lambda_{u_{iR}}^{H^-} |^2)}{48 \hat s^2 (\hat t - m_b^2)}
  \left[ \frac{2 F_1 - F_2^2 - 2G_1 G_2}{\hat s} + \frac{2 m_b^2
      G_2}{\hat t - m_b^2} \right]
\end{equation}
with
\begin{equation}
  F_1 = \hat s \hat t, \quad
    F_2 = \hat s + \hat t - m_b^2, \quad
  G_1 = m_{H^\pm}^2 - m_b^2 - \hat s , \quad
    G_2 = m_{H^\pm}^2 - \hat t.
\end{equation}

\begin{figure}[tb!]
  \begin{center}
    \includegraphics[width=0.45\textwidth]{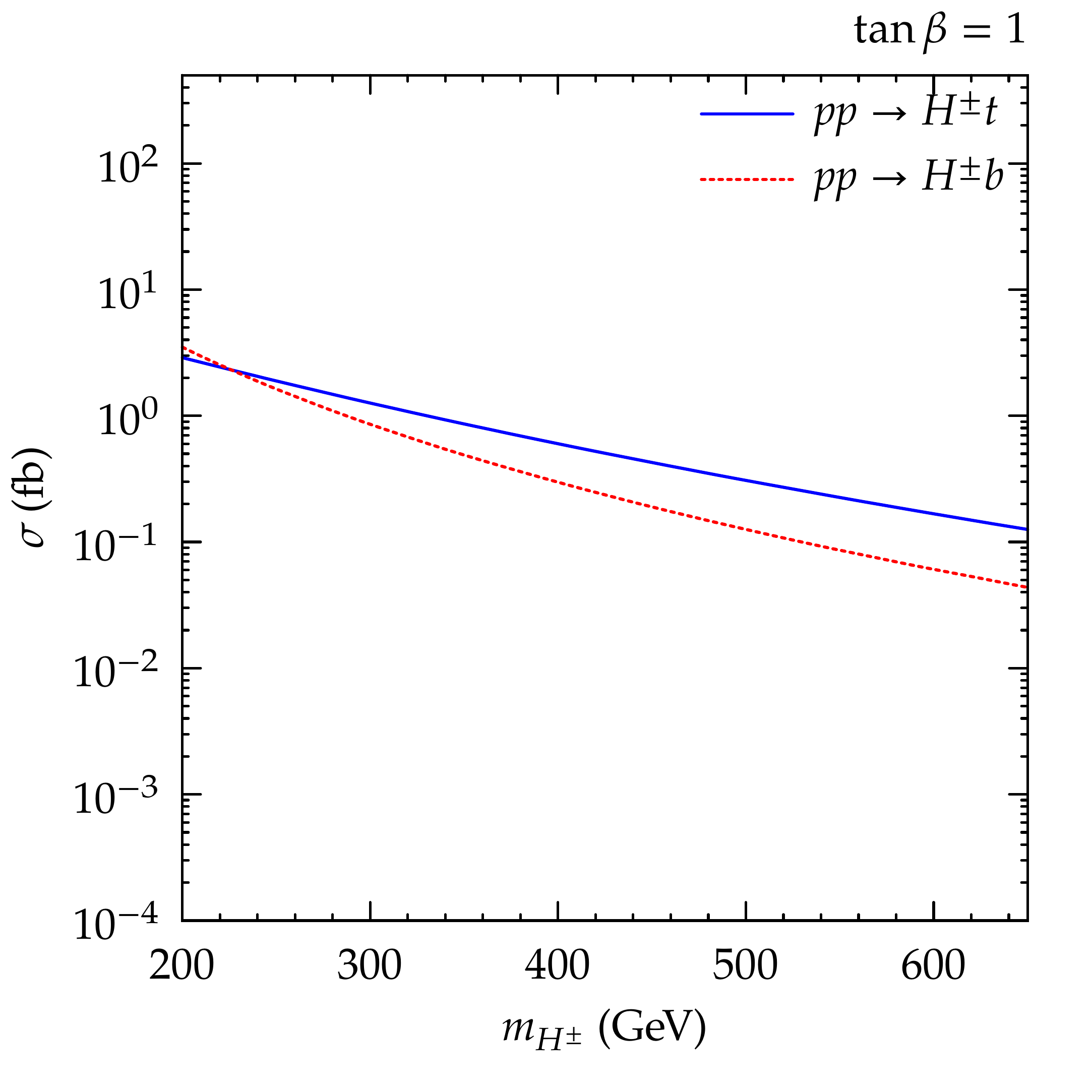}
    \includegraphics[width=0.45\textwidth]{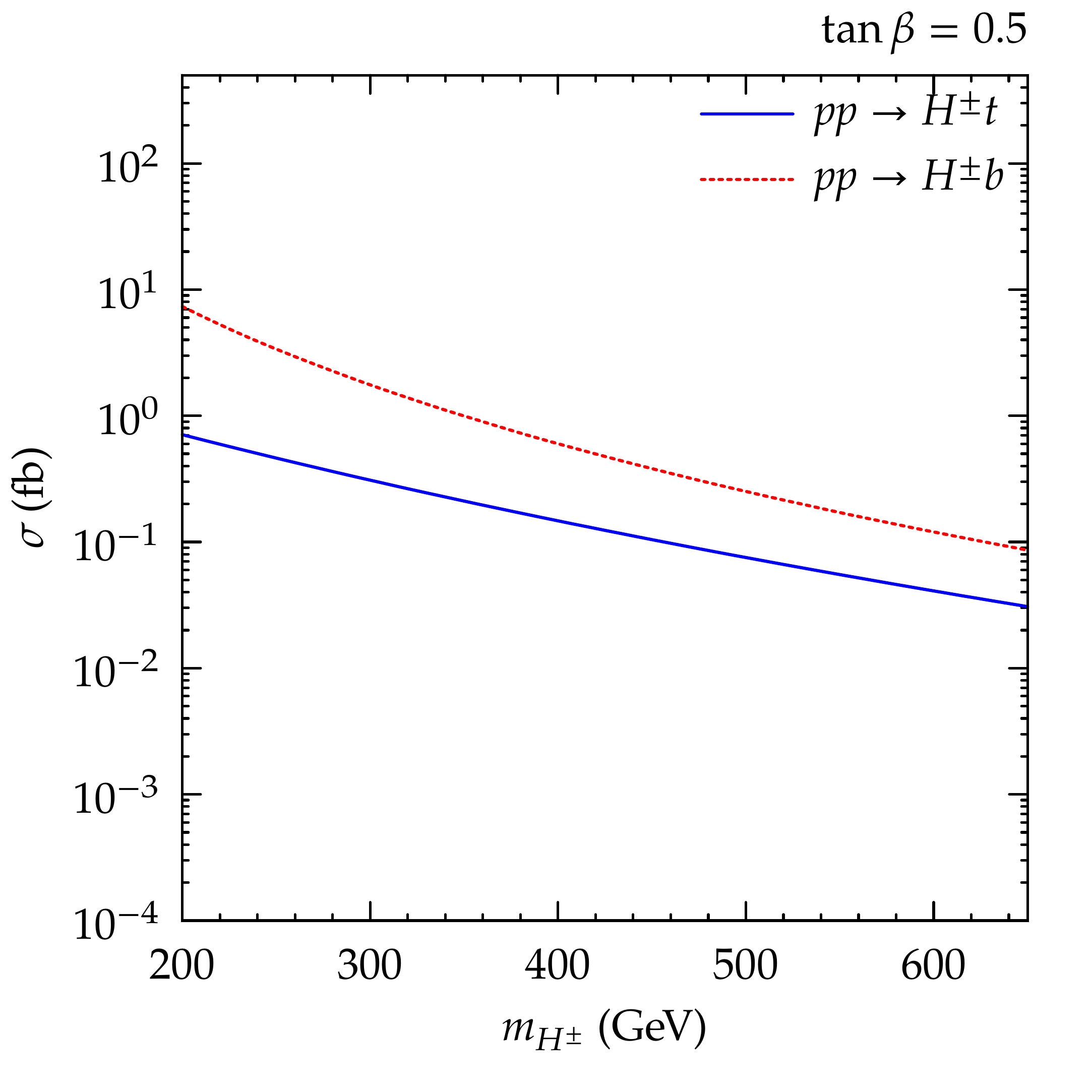}
   \end{center}
  \caption{Production cross sections of the heavy charged Higgs $H^\pm$
    at 14~TeV proton-proton collisions. We have chosen $\tan\beta = 1$
    (left panel) and $\tan\beta = 0.5$ (right panel) with $y_{33}^u =
    y_t^\text{SM}$.\label{fig:xsec_charged}}
\end{figure}

The leading-order cross sections evaluated by convoluting the partonic
cross section with the PDFs at proton-proton collisions of 14~TeV are
shown in Fig.~\ref{fig:xsec_charged}. In each figure, $\sigma(pp \to
H^\pm q) = \sigma(pp \to H^+ q) + \sigma(pp \to H^- q)$.
The production cross sections are quite sensitive to $\tan\beta$. For
$\tan\beta = 1$, the top quark associated production, $p p \to H^\pm t$,
is the dominant channel, while the bottom quark associated
production, $p p \to H^\pm b$, which is the characteristic channel of
our model, can also be served as a good channel to search the charged
Higgs boson at the LHC\@.
On the other hand, for smaller $\tan\beta$, the bottom quark associated
production becomes the dominant channel due to the enhanced charged-Higgs
couplings with light up-type quarks.
The suppression of top quark associated production is also due to the
partial cancellation of two terms in $\lambda_{t_L}^{H^-}$.

Concerning the decays of charged Higgs, the most important fermionic
decay mode is $H^+ \to t \bar b$. The decay width is
\begin{align}
  \Gamma(H^+ \to t \bar b)
  =&~\Gamma(H^- \to b \bar t) \nonumber\\
  =&~\frac{3}{16\pi} m_{H^\pm}
     \left[ \left( 1 - \frac{(m_t + m_b)^2}{m_{H^\pm}^2} \right)
     \left( 1 - \frac{(m_t - m_b)^2}{m_{H^\pm}^2} \right)
     \right]^{1/2} \nonumber\\
   & \times
     \left[ \left( | \lambda_{t_L}^{H^-} |^2 + | \lambda_{t_R}^{H^-}
     |^2 \right) \left( 1 - \frac{m_t^2 + m_b^2}{m_{H^\pm}^2} \right)
     \right . \nonumber\\
   & \qquad \left .
     - 2 \left( \lambda_{t_L}^{H^-} (\lambda_{t_R}^{H^-})^\ast +
    \lambda_{t_R}^{H^-} (\lambda_{t_L}^{H^-})^\ast \right) \frac{m_t
    m_b}{m_{H^\pm}^2} \right] .
\end{align}
By replacing $m_t$ with $m_c$ or $m_u$ and $\lambda_{t_{L,R}}^{H^-}$
with $\lambda_{c_{L,R}}^{H^-}$ or $\lambda_{u_{L,R}}^{H^-}$, one can
obtain the decay widths of $H^+ \to c \bar b$ and $H^+ \to u \bar b$.
The other fermionic decay modes are $H^+ \to c \bar s$ and $c \bar d$,
whose decay widths are proportional to $\tan^2\beta |V_{cs}|^2$ and
$\tan^2\beta |V_{cd}|^2$, respectively.
The decay widths of leptonic decay modes are given as
\begin{equation}
  \Gamma (H^+ \to \ell^+ \nu)
  = \Gamma (H^- \to \ell^- \bar\nu)
  =\frac{m_\ell^2 \tan^2 \beta}{8\pi v^2} m_{H^\pm} \left( 1 -
     \frac{m_\ell^2}{m_{H^\pm}^2} \right)^2 .
\end{equation}
Meanwhile, if $H^+ \to W^+ A$ and $W^+ H$ are kinematically forbidden,
the only non-fermionic decay mode is $H^+ \to W^+ h$. The decay width
is
\begin{align}
  \Gamma(H^+ \to W^+ h)
  &= \Gamma(H^- \to W^- h) \nonumber\\
  &= \frac{g^2 \cos^2 (\alpha-\beta) m_{H^\pm}^3}{64\pi m_W^2}
    \left[ \left( 1 - \frac{m_W^2}{m_{H^\pm}^2} -
    \frac{m_h^2}{m_{H^\pm}^2}\right)^2 - \frac{4 m_W^2
    m_h^2}{m_{H^\pm}^4}\right]^{3/2} .
\end{align}

\begin{figure}[tb!]
  \begin{center}
    \includegraphics[width=0.45\textwidth]{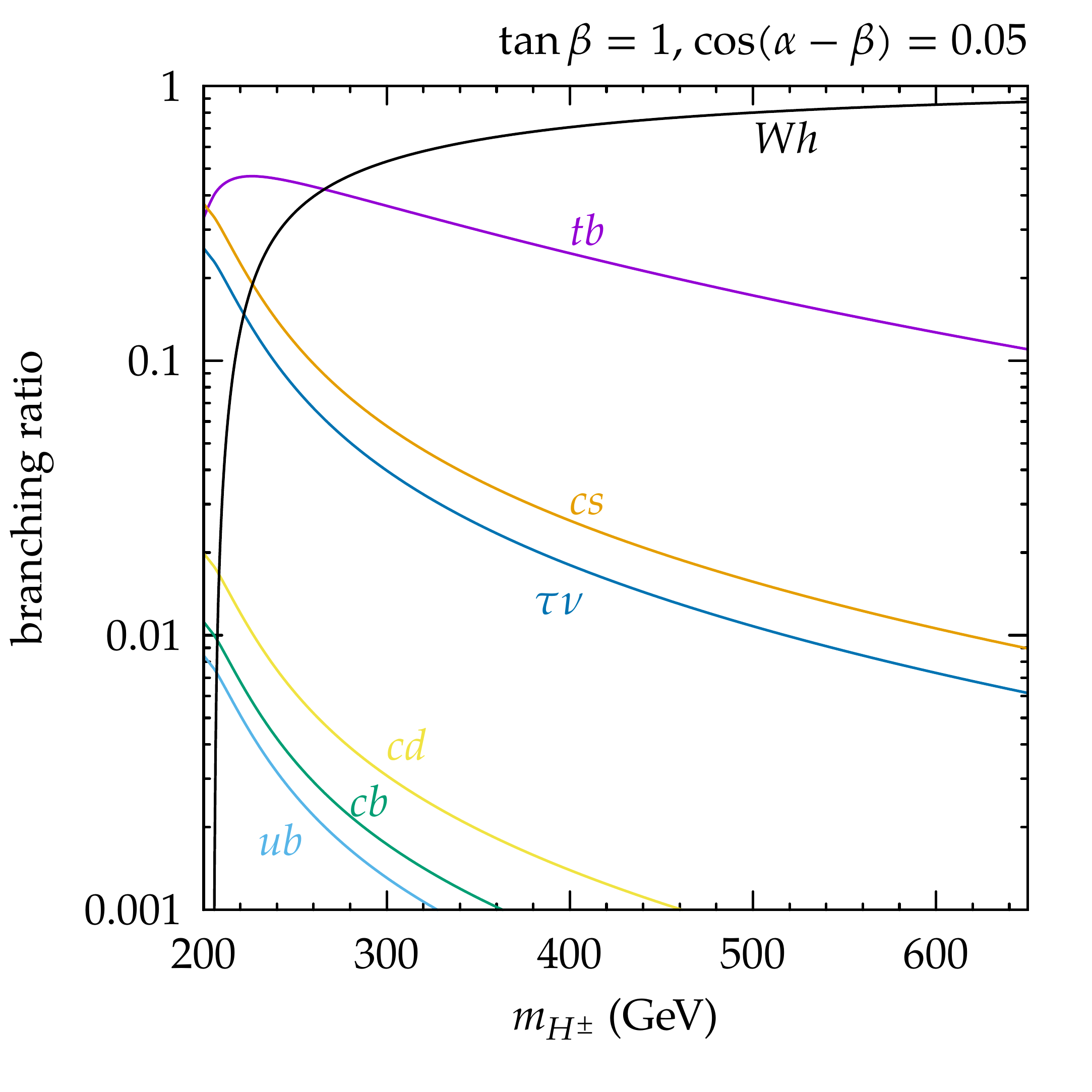}
    \includegraphics[width=0.45\textwidth]{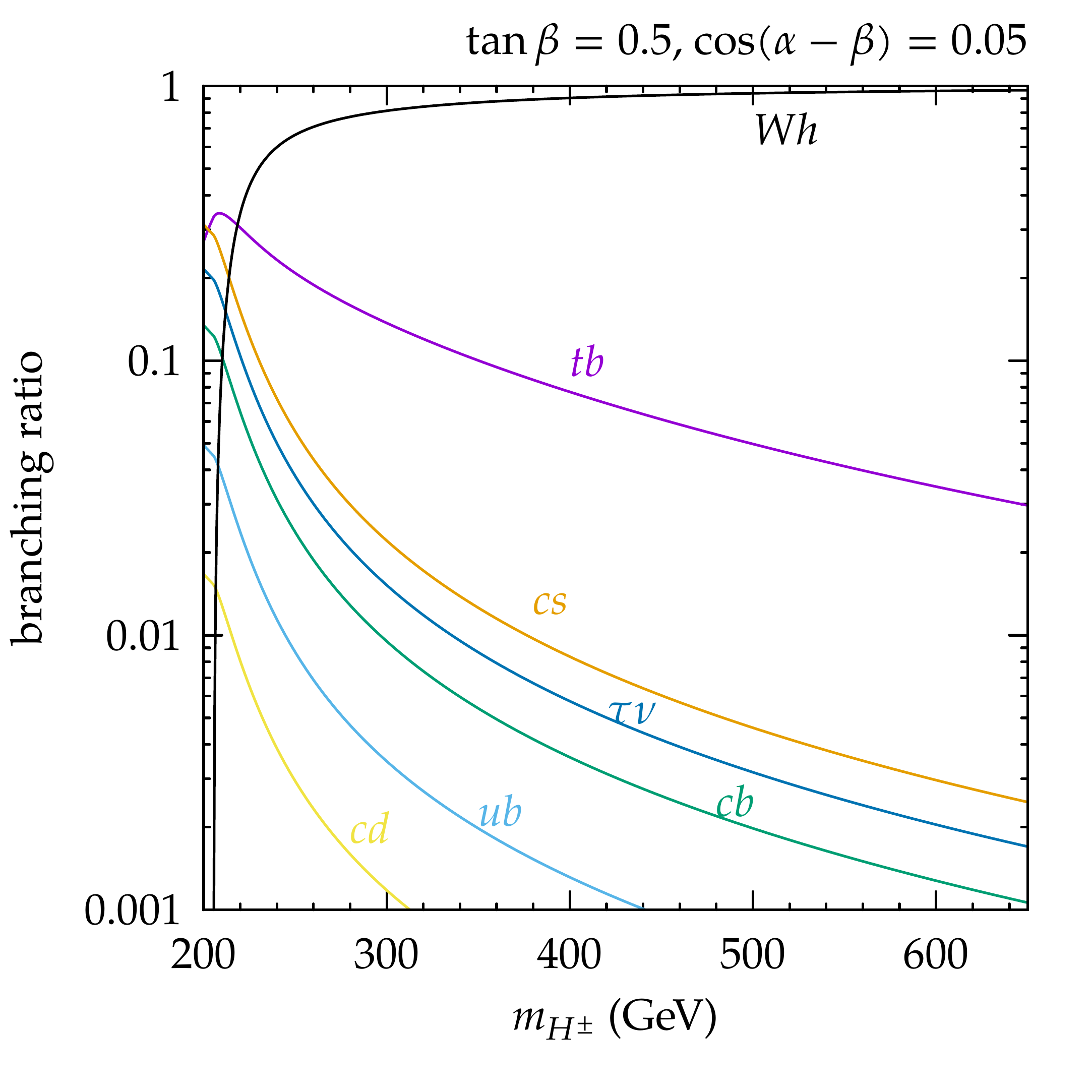}
  \end{center}
  \caption{Branching ratios of the heavy charged Higgs $H^\pm$.
    $\tan\beta = 1$ (left panel) and $\tan\beta = 0.5$ (right panel)
    have been taken. $\cos (\alpha - \beta) = 0.05$ and $y_{33}^u =
    y_t^\text{SM}$ for both panels.\label{fig:br_charged}}
\end{figure}

By combining all the decay modes in the above we obtain the branching
ratios of the heavy charged Higgs, which are shown in
Fig.~\ref{fig:br_charged}.
Interestingly, the dominant decay mode of the charged Higgs boson is
$H^+ \to W^+ h$ if it is kinematically allowed, although we have taken
the alignment limit. $H^+ \to t \bar b$ is subdominant. Together with
the production, we expect that $p p \to H^\pm b \to W^\pm h + b$ can
be served as the important process to probe the charged Higgs boson at
the LHC and future hadron colliders.
Most LHC searches for $W^+ h$ have been dedicated to heavy
resonances~\cite{LHC-cH1} that decay directly into $W^+ h$, so our
model is not constrained by $W^+ h$ at the moment.
On the other hand, the $t{\bar b}$ mode is next-to-dominant and this
is not constrained by the current LHC data~\cite{LHC-cH2}, because the
production cross section for the heavy charged Higgs in our model is
less than 10~fb in most of the parameter space.

\section{Conclusions}
We have considered an extra local $U(1)$ with flavor-dependent
couplings as a linear combination of $B_3-L_3$ and $L_\mu-L_\tau$,
that has been recently proposed to explain the $B$-meson anomalies. In
our model, we have reproduced the correct flavor structure of the
quark sector due to the VEV of the second Higgs
doublet, at the expense of new flavor violating couplings for quarks and the violation of lepton universality.

The extra gauge boson leads to flavor violating interactions
for down-type quarks appropriate for explaining $B$-meson anomalies in $R_{K^{(*)}}$
whereas heavy Higgs bosons render up-type quarks have modified
flavor-conserving Yukawa couplings and down-type quarks receive
flavor-violating Yukawa couplings.
We also found that the $B$-meson anomalies in $R_{D^{(*)}}$ cannot be explained by the charged Higgs boson in our model, due to small flavor-violating couplings.

We showed how the extended Higgs sector can be constrained by
unitarity and stability, Higgs and electroweak precision data,
$B$-meson decays/mixings. Taking the alignment limit of heavy Higgs
bosons from Higgs precision data, we also
investigated the production of heavy Higgs bosons at the LHC\@. We found
that there are reductions in the cross sections of the usual
production channels in 2HDM, such as $pp\rightarrow H$ and
$pp\rightarrow H^\pm t$ at the LHC\@. In addition, new production
channels such as $pp\rightarrow Hb$ and $pp\rightarrow H^\pm b$ become
important for $\tan\beta\lesssim 1$. Decay products of heavy
Higgs bosons lead to interesting collider signatures due to large
branching fractions of $bd+bs$ modes for neutral Higgs bosons and
$W^\pm h$ mode for charged Higgs boson if kinematically allowed, thus
requiring a more dedicated analysis for the LHC\@.

\section*{Acknowledgments}

The work is supported in part by Basic Science Research Program
through the National Research Foundation of Korea (NRF) funded by the
Ministry of Education, Science and Technology (NRF-2016R1A2B4008759).
The work of LGB is partially supported by the National Natural Science
Foundation of China (under Grant No. 11605016), Korea Research
Fellowship Program through the National Research Foundation of Korea
(NRF) funded by the Ministry of Science and ICT (2017H1D3A1A01014046).
The work of CBP is supported by IBS under the project code, IBS-R018-D1.

\appendices%

\section{The extended Higgs sector\label{app:higgs_sector}}

By using the minimization condition of the Higgs potential given by
\begin{align}
  \mu_1^2
  &=\frac{\sqrt{2} \mu v_2 v_s-2 \lambda_1 v_1^3-2 \lambda_3 v_1
    v_2^2-2 \lambda_4 v_1 v_2^2-2 \kappa_1 v_1 v_s^2}{2 v_1},
    \nonumber\\
  \mu_2^2
  &= \frac{\sqrt{2} \mu  v_1 v_s -2 \lambda_3 v_1^2 v_2-2\lambda_4
    v_1^2 v_2-2\lambda_2 v_2^3-2 \kappa_2 v_2 v_s^2}{2 v_2},
    \nonumber\\
  m_s^2
  &= \frac{\sqrt{2} \mu v_1 v_2 -2 \kappa_1 v_1^2 v_s-2 \kappa_2 v_2^2
    v_s -2 \lambda_S v_s^3}{2 v_s},
\end{align}
the mass matrix for $CP$-even scalars can be written as
\begin{equation}
  M_S =
  \begin{pmatrix}
    2 \lambda_1 v_1^2+\frac{\mu v_2 v_s}{\sqrt{2} v_1}&2 v_1 v_2 (\lambda_3+\lambda_4)-\frac{\mu v_s}{\sqrt{2}} & 2\kappa_1 v_1 v_s-\frac{\mu v_2}{\sqrt{2}} \\
    2 v_1 v_2 (\lambda_3+\lambda_4)-\frac{\mu v_s}{\sqrt{2}} & 2 \lambda_2 v_2^2+\frac{\mu v_1 v_s}{\sqrt{2}v_2} & 2\kappa_2 v_2 v_s-\frac{\mu v_1}{\sqrt{2}} \\
    2 \kappa_1 v_1  v_s-\frac{\mu v_2}{\sqrt{2}} & 2  \kappa_2 v_2  v_s-\frac{\mu v_1}{\sqrt{2}} & 2\lambda_S  v_s^2+\frac{\mu  v_1 v_2}{\sqrt{2} v_s}
  \end{pmatrix}.  \label{h0matrix}
\end{equation}
We introduce a rotation matrix $R$ to change the interaction basis
$(\rho_1, \rho_2, S_R)$ to the physical mass eigenstates, $h_1$, $h_2$
and $h_3$ as
\begin{equation*}
  \begin{pmatrix}
    h_1 \\ h_2 \\ h_3
  \end{pmatrix} = R
  \begin{pmatrix}
    \rho_1 \\ \rho_2 \\ S_R
  \end{pmatrix}.
\end{equation*}
The mass matrix $M_S$ can be then diagonalized as
\begin{equation}
  R M_S R^\mathsf{T} = \mbox{diag}(m_{h_1}^2,m_{h_2}^2,m_{h_3}^2).
\end{equation}
We use a convention such that the mass eigenstates are ordered as
$m_{h_1} < m_{h_2} < m_{h_3}$.
Here, the orthogonal matrix $R$ is parametrized in terms of the mixing
angles $\alpha_1$ to $\alpha_3$ as
\begin{equation}
  R =
  \begin{pmatrix}
    c_{\alpha_1} c_{\alpha_2} & s_{\alpha_1} c_{\alpha_2} & s_{\alpha_2}\\
    -(c_{\alpha_1} s_{\alpha_2} s_{\alpha_3} + s_{\alpha_1} c_{\alpha_3})
    & c_{\alpha_1} c_{\alpha_3} - s_{\alpha_1} s_{\alpha_2} s_{\alpha_3}
    & c_{\alpha_2} s_{\alpha_3} \\
    - c_{\alpha_1} s_{\alpha_2} c_{\alpha_3} + s_{\alpha_1} s_{\alpha_3} &
    -(c_{\alpha_1} s_{\alpha_3} + s_{\alpha_1} s_{\alpha_2} c_{\alpha_3})
    & c_{\alpha_2}  c_{\alpha_3}
  \end{pmatrix},
\end{equation}
where $s_{\alpha_i} \equiv \sin\alpha_i$ and $c_{\alpha_i} \equiv \cos\alpha_i$.
Without loss of generality the angles can be chosen in the range of
\begin{equation*}
- \frac{\pi}{2} \le \alpha_{1,2,3} < \frac{\pi}{2}.
\end{equation*}
In the text we focus mainly on the situation where mixings
between $\rho_{1,2}$ and $S_R$ are small.

The mass eigenvalues of $CP$-even neutral scalars are given by
\begin{align}
    m_{h_1}^2&=\frac{1}{2} (a+b -\sqrt{D})\equiv m_h^2, \nonumber \\
    m_{h_2}^2&=\frac{1}{2} (a+b+\sqrt{D})\equiv m_H^2,  \nonumber  \\
    m_{h_3}^2&=2\lambda_S  v_s^2+\frac{\mu  v_1 v_2}{\sqrt{2} v_s}\equiv
             m_s^2 ,  \label{h0s}
\end{align}
where
\begin{equation}
  a\equiv 2 \lambda_1 v_1^2+\frac{\mu v_2 v_s}{\sqrt{2} v_1} ,\quad
  b\equiv  2 \lambda_2 v_2^2+\frac{\mu v_1 v_s}{\sqrt{2}v_2}, \quad
  D\equiv (a-b)^2+4 d^2
\end{equation}
with $d\equiv 2 v_1 v_2 (\lambda_3+\lambda_4)-\mu v_s / \sqrt{2}$.
We can trade off quartic couplings, $\lambda_{1,2,3,4}$ and
$\kappa_{1,2}$, for mixing angles and Higgs masses.
\begin{align}
    \lambda_1&=\frac{2 \sum_i m_{h_i}^2 R_{i1}^2-\sqrt{2} \mu v_s
      \tan\beta}{4 v^2 \cos^2 \beta} ,\nonumber\\
    \lambda_2&=\frac{2 \sum_i m_{h_i}^2 R_{i2}^2-\sqrt{2} \mu v_s
      \cot\beta}{4 v^2 \sin^2 \beta} ,\nonumber\\
    \lambda_3+\lambda_4&=\frac{\sqrt{2} \mu v_s+2 \sum_i m_{h_i}^2
      R_{i1} R_{i2}}{4 v^2 \sin 2\beta} ,\nonumber\\
    \lambda_S&=\frac{2 v_s  \sum_i m_{h_i}^2 R_{i3}^2-\sqrt{2} \mu v^2
      \sin\beta \cos\beta}{4 v_s^3} ,\nonumber\\
    \kappa_1&=\frac{\sqrt{2}\mu v \sin\beta+2 \sum_i m_{h_i}^2
      R_{i1}R_{i3}}{4 v v_s \cos\beta}, \nonumber\\
    \kappa_2&=\frac{\sqrt{2}\mu v \cos\beta+2 \sum_i m_{h_i}^2
      R_{i2}R_{i3}}{4 v v_s \sin\beta}. \label{lams0}
\end{align}
In the case when the Higgs mixings with the singlet scalar are
negligible, $\alpha_2 \simeq \alpha_3 \simeq 0$, the rotation matrix
can be simplified as
\begin{equation}
  R \approx
  \begin{pmatrix}
    \cos\alpha & \sin\alpha & 0\\
    -\sin\alpha & \cos\alpha & 0\\
    0 & 0 & 1
  \end{pmatrix},
\end{equation}
where $\alpha = \alpha_1$. Then the Higgs quartic couplings are given
by
\bea
  \lambda_1
  & \approx &
  \frac{2 (m_h^2\cos^2\alpha  + m_H^2 \sin^2\alpha ) -
    \sqrt{2} \mu v_s \tan\beta}{4 v^2 \cos^2 \beta} ,  \nonumber \\
  \lambda_2
  & \approx &
   \frac{2 (m_h^2\sin^2\alpha  + m_H^2 \cos^2\alpha ) -
    \sqrt{2} \mu v_s \cot\beta}{4 v^2 \sin^2 \beta} ,  \nonumber  \\
  \lambda_3 + \lambda_4
  &\approx &
  \frac{(m_h^2 - m_H^2) \sin 2\alpha + \sqrt{2} \mu
    v_s}{4 v^2 \sin 2\beta}.  \label{lams}
\eea
Here $h = h_1$ with $m_h = 125$~GeV and $H = h_2$.
This relations show that the values of quartic couplings can be
evaluated solely by $m_H$ if one chooses a benchmark point in terms of
$\mu v_s$, $\tan\beta$, and $\sin\alpha$.

\section{Unitarity bounds}
\label{app:unitarity_bounds}

The initial scattering states can be classified by hypercharges and
isospins~\cite{Akeroyd:2000wc,Ginzburg:2005dt,Kanemura:2015ska}.
In the basis of $(\phi_1^+ \phi_1^-$, $\phi_2^+
\phi_2^-$, $\eta_1\eta_1/\sqrt{2}$, $\rho_1\rho_1/\sqrt{2}$,
$\eta_2\eta_2/\sqrt{2}$, $\rho_2\rho_2/\sqrt{2}$, $S_R S_R/\sqrt{2}$,
$S_I S_I/\sqrt{2})$, the scattering amplitude is
\begin{equation}
\mathcal{M}_1=
\begin{pmatrix}
 4 \lambda_1 & 2 (\lambda_3+\lambda_4) & \sqrt{2}\lambda_1 & \sqrt{2}\lambda_1 & \sqrt{2} \lambda_3 & \sqrt{2} \lambda_3 & \sqrt{2} \kappa_1 & \sqrt{2} \kappa_1 \\
 2 (\lambda_3+\lambda_4) & 4 \lambda_2 & \sqrt{2} \lambda_3 & \sqrt{2} \lambda_3 & \sqrt{2} \lambda_2 & \sqrt{2}\lambda_2 & \sqrt{2}\kappa_2 & \sqrt{2} \kappa_2 \\
 \sqrt{2}\lambda_1 & \sqrt{2}\lambda_3 & 3 \lambda_1 & \lambda_1 & \lambda_3+\lambda_4 & \lambda_3+\lambda_4 & \kappa_1 &\kappa_1 \\
 \sqrt{2} \lambda_1 & \sqrt{2} \lambda_3 & \lambda_1 & 3 \lambda_1 & \lambda_3+\lambda_4 & \lambda_3+\lambda_4 & \kappa_1 & \kappa_1 \\
 \sqrt{2} \lambda_3& \sqrt{2} \lambda_2&  \lambda_3+\lambda_4 &\lambda_3+\lambda_4& 3\lambda_2 & \lambda_2 &\kappa_2 & \kappa_2 \\
 \sqrt{2}\lambda_3 & \sqrt{2}\lambda_2 & \lambda_3+\lambda_4 & \lambda_3+\lambda_4 &\lambda_2 & 3 \lambda_2 & \kappa_2 &\kappa_2 \\
 \sqrt{2}  \kappa_1 & \sqrt{2} \kappa_2 &  \kappa_1 & \kappa_1 & \kappa_2 & \kappa_2 & 3 \lambda_S &  \lambda_S \\
 \sqrt{2} \kappa_1 & \sqrt{2} \kappa_2 & \kappa_1 &\kappa_1 & \kappa_2
 &\kappa_2 &\lambda_S & 3 \lambda_S
 \end{pmatrix},
\end{equation}
whose eigenvalues are $2\lambda_1$, $2\lambda_2$,
$\lambda_1+\lambda_2\pm\sqrt{(\lambda_1-\lambda_2)^2+4\lambda_4^2}$,
and $2\lambda_S$.

In the basis of $(\phi_1^+ S_R$, $\phi_2^+ S_R$, $\phi_1^+ S_I$,
$\phi_2^+ S_I)$, the submatrix is given by
\begin{equation}
\mathcal{M}_2=
\begin{pmatrix}
 2  \kappa_1 & 0 & 0 & 0 \\
 0 & 2\kappa_2& 0 & 0 \\
 0 & 0 & 2 \kappa_1 & 0 \\
 0 & 0 & 0 & 2 \kappa_2
\end{pmatrix}
\end{equation}
with eigenvalues being $2\kappa_{1,2}$.

In the basis of $(\rho_1 \eta_1$, $\rho_2\eta_2$, $S_R S_I)$, the
matrix is
\begin{equation}
\mathcal{M}_3=
\begin{pmatrix}
 2  \lambda_1 & 0 & 0 \\
 0 & 2\lambda_2 & 0 \\
 0 & 0 & 2  \lambda_S
\end{pmatrix}
\end{equation}
with eigenvalues being $2\lambda_{1,2,s}$.

In the basis of $(\phi_1^+\phi_2^-$, $\phi_2^+\phi_1^-$,
$\rho_1\eta_2$, $\rho_2\eta_1$, $\eta_1\eta_2$, $\rho_1\rho_2)$, we
have
\begin{equation}
\mathcal{M}_4=
\begin{pmatrix}
 0 & 2  \lambda_3+2 \lambda_4 & i \lambda_4 & -i \lambda_4 &\lambda_4& \lambda_4 \\
 2\lambda_3+2 \lambda_4 & 0 & -i \lambda_4 & i \lambda_4 &  \lambda_4& \lambda_4 \\
 i \lambda_4& -i \lambda_4 &  2  \lambda_3+2 \lambda_4& 0 & 0 & 0 \\
 -i \lambda_4 & i \lambda_4 & 0 &2  \lambda_3+2 \lambda_4 & 0 & 0 \\
 \lambda_4  &\lambda_4  & 0 & 0 & 2  \lambda_3+2 \lambda_4 & 0 \\
\lambda_4  & \lambda_4  & 0 & 0 & 0 & 2  \lambda_3+2 \lambda_4
\end{pmatrix}
\end{equation}
with eigenvalues being
$2 \lambda _3$, $2 (\lambda_3+\lambda_4)$, $2 (\lambda_3+2
\lambda_4)$, and $\pm 2 \sqrt{\lambda_3 (\lambda_3+2 \lambda_4)}$.

Finally, in the basis of $(\rho_1 \phi_1^+$, $\rho_2\phi_1^+$,
$\eta_1\phi_1^+$, $\eta_2\phi_1^+$, $\rho_1\phi_2^+$,
$\rho_2\phi_2^+$, $\eta_1\phi_2^+$, $\eta_2\phi_2^+)$,
we obtain the matrix as
\begin{equation}
\mathcal{M}_5=
\begin{pmatrix}
 2 \lambda_1 & 0 & 0 & 0 & 0 & \lambda_4 & 0 & i \lambda_4 \\
 0 & 2 \lambda_3 & 0 & 0 &\lambda_4 & 0 & -i \lambda_4 & 0 \\
 0 & 0 & 2 \lambda_1 & 0 & 0 & -i \lambda_4 & 0 & \lambda_4 \\
 0 & 0 & 0 & 2 \lambda_3 & i \lambda_4 & 0 & \lambda_4 & 0 \\
 0 & \lambda_4 & 0 & -i \lambda_4 & 2\lambda_3 & 0 & 0 & 0 \\
 \lambda_4 & 0 & i \lambda_4 & 0 & 0 & 2 \lambda_2 & 0 & 0 \\
 0 & i \lambda_4 & 0 & \lambda_4 & 0 & 0 & 2\lambda_3 & 0 \\
 -i \lambda_4 & 0 & \lambda_4 & 0 & 0 & 0 & 0 & 2 \lambda_2
\end{pmatrix},
\end{equation}
with eigenvalues being
$2 \lambda_1$, $2 \lambda_2$, $2 \lambda_3$, $2
(\lambda_3\pm\lambda_4)$, and
$\lambda_1+\lambda_2\pm\sqrt{(\lambda_1-\lambda_2)^2+4\lambda_4^2}$.

The eigenvalues obtained in the above are constrained by unitarity as
\begin{align}
  |2\lambda_{1,2,3,S}|\leq8\pi ,\quad
  |2\kappa_{1,2}| \leq 8\pi ,& \nonumber\\
  |2(\lambda_3\pm\lambda_4)| \leq 8\pi ,\quad
  |2(\lambda_3+2\lambda_4)| \leq 8\pi ,\quad
  |2\sqrt{\lambda_3(\lambda_3+2\lambda_4)}| \leq 8 \pi ,& \nonumber\\
  |\lambda_1+\lambda_2\pm\sqrt{(\lambda_1-\lambda_2)^2+4\lambda_4^2}|
  \leq 8 \pi, &\nonumber\\
  a_{1,2,3} \leq 8 \pi. &
\end{align}
Here $a_{1,2,3}$ are three other solutions of the following equation:
\begin{align}
& x^3-2 x^2 (3 \lambda_1+3 \lambda_2+2 \lambda_S)-4 x \left(2 \kappa_1^2+2 \kappa_2^2-9  \lambda_1 \lambda_2-6 \lambda_1 \lambda_S-6 \lambda_2 \lambda_S+4 \lambda_3^2+4 \lambda_3 \lambda_4+ \lambda_4^2\right)\nonumber\\
& +16 \left(3 \kappa_1^2\lambda_2-2 \kappa_1\kappa_2 (2\lambda_3+ \lambda_4)+3 \kappa_2^2 \lambda_1+\lambda_S \left((2 \lambda_3+\lambda_4)^2-9  \lambda_1 \lambda_2\right)\right)=0. \label{cubic}
\end{align}

\section{The quark Yukawa couplings}
\label{app:quark_yukawa}

The quark Yukawa couplings in the interaction basis are given by
\begin{align}
  -{\cal L}_{Y}^q =
  &~ \frac{1}{\sqrt{2}} {\bar u}_L \bigg( (\rho_1-i\eta_1)y^u+ (\rho_2-i\eta_2)h^u \bigg) u_R \nonumber \\
  &+ \frac{1}{\sqrt{2}} {\bar d}_L \bigg( (\rho_1+i\eta_1)y^u+ (\rho_2+i\eta_2)h^u \bigg) d_R   \nonumber\\
  &- {\bar d}_L \Big( y^u (\phi^+_1)^*+h^u (\phi^+_2)^*\Big) u_R + {\bar u}_L \Big(y^d\phi^+_1+h^d \phi^+_2 \Big) d_R +{\rm h.c.}
\end{align}
In the basis of mass eigenstates the quark Yukawa interactions of the $CP$-even neutral scalars are
\begin{equation}
  -{\cal L}_{Y}^q = ({\bar u}'_L Y^u_{H_i} u'_R + {\bar d}'_L Y^d_{H_i}
  d'_R) H_i +{\rm h.c.},
\end{equation}
where primed fields are mass eigenstates, and
\begin{align}
Y^u_{H_1} &=-\frac{R_{11}}{v\cos\beta}\, M^D_u +\frac{R_{11}\tan\beta-R_{12}}{\sqrt{2}}\, {\tilde h}^u,   \nonumber\\
Y^d_{H_1} &=-\frac{R_{11}}{v\cos\beta}\, M^D_u+\frac{R_{11}\tan\beta-R_{12}}{\sqrt{2}}\, {\tilde h}^d,   \nonumber\\
Y^u_{H_2} &=- \frac{R_{21}}{v\cos\beta}\, M^D_u +\frac{R_{21}\tan\beta-R_{22}}{\sqrt{2}}\, {\tilde h}^u,    \nonumber\\
Y^d_{H_2} &=-\frac{R_{21}}{v\cos\beta}\,  M^D_d+\frac{R_{21}\tan\beta-R_{22}}{\sqrt{2}}\, {\tilde h}^d,    \nonumber\\
Y^u_{H_3} &=- \frac{R_{31}}{v\cos\beta}\, M^D_u +\frac{R_{31}\tan\beta-R_{22}}{\sqrt{2}}\, {\tilde h}^u,    \nonumber\\
Y^d_{H_3} &=-\frac{R_{31}}{v\cos\beta}\,  M^D_d+\frac{R_{31}\tan\beta-R_{22}}{\sqrt{2}}\, {\tilde h}^d,
\end{align}
Assuming the singlet scalars are decoupled and using Eqs.~(\ref{base-h})
to~(\ref{base-chargedH}), the above quark Yukawa interactions become
\begin{align}
  -{\cal L}_{q,Y}
  =&~({\bar u}'_L Y^{u}_h u'_R  + {\bar d}'_L Y^{d}_h d'_R) h +({\bar
     u}'_L Y^u_H u'_R + {\bar d}'_L Y^d_H d'_R) H\nonumber \\
   &+i( {\bar u}'_L Y^u_A u'_R+ {\bar d}'_L Y^d_A d'_R  )A^0
     \nonumber \\
   & +{\bar u}'(Y_{2,H^+}P_R+Y_{1,H^+}P_L) d' H^+ +{\rm h.c.} ,
\end{align}
where
\begin{align}
Y^u_h &=-\frac{\sin\alpha}{v\cos\beta}\, M^D_u +\frac{\cos(\alpha-\beta)}{\sqrt{2}\cos\beta}\, {\tilde h}^u,   \nonumber\\
Y^d_h &=-\frac{\sin\alpha}{v\cos\beta}\, M^D_d+\frac{\cos(\alpha-\beta)}{\sqrt{2}\cos\beta}\, {\tilde h}^d,   \nonumber\\
Y^u_H &= \frac{\cos\alpha}{v\cos\beta}\, M^D_u +\frac{\sin(\alpha-\beta)}{\sqrt{2}\cos\beta}\, {\tilde h}^u,    \nonumber\\
Y^d_H &=\frac{\cos\alpha}{v\cos\beta}\, M^D_d+\frac{\sin(\alpha-\beta)}{\sqrt{2}\cos\beta}\, {\tilde h}^d,    \nonumber\\
Y^u_A &=-\frac{\tan\beta}{v} M^D_u+\frac{1}{\sqrt{2}\cos\beta}\, {\tilde h}^u,  \nonumber\\
Y^d_A &=\frac{\tan\beta}{v} M^D_d-\frac{1}{\sqrt{2}\cos\beta}\, {\tilde h}^d,  \nonumber\\
Y_{1,H^+} &= -\bigg(\frac{\sqrt{2} \tan\beta}{v} M^D_u-\frac{1}{\cos\beta}\, ({\tilde h}^u)^\dagger \bigg)V_{\rm CKM},   \nonumber\\
Y_{2,H^+} &=V_{\rm CKM}  \bigg(\frac{\sqrt{2} \tan\beta}{v} M^D_d-\frac{1}{\cos\beta}\, {\tilde h}^d \bigg)
\end{align}
with
\begin{equation}
  {\tilde h}^u \equiv U^\dagger_L h^u U_R , \quad
  {\tilde h}^d \equiv D^\dagger_L h^d D_R.
\end{equation}

For $U_L=1$, we have ${\tilde h}^u=h^u U_R$. As a result,
\begin{align}
    {\tilde h}^u_{31}&= \frac{1}{\sqrt{2}} \frac{v\cos\beta}{m_u}
    \Big(h^u_{31} (y^u_{11})^*+ h^u_{32} (y^u_{12})^* \Big)=0, \nonumber\\
    {\tilde h}^u_{32}&= \frac{1}{\sqrt{2}} \frac{v\cos\beta}{m_c}
    \Big(h^u_{31} (y^u_{21})^*+h^u_{32} (y^u_{22})^* \Big)=0, \nonumber\\
    {\tilde h}^u_{33}&=  \frac{1}{\sqrt{2}} \frac{v\sin\beta}{m_t}
    \Big( |h^u_{31}|^2+|h^u_{32}|^2 \Big) \nonumber\\
    &=\frac{\sqrt{2} m_t}{v\sin\beta}\Big(1-\frac{v^2\cos^2\beta}{2m^2_t}\,|y^u_{33}|^2 \Big),
\end{align}
where use is made of Eqs.~(\ref{UR2}), (\ref{UR4}), and (\ref{UR5}). Other components of ${\tilde h}^u$ are vanishing.
Moreover, with ${\tilde h}^d=V^\dagger_{\rm CKM} h^d$ and using
Eq.~(\ref{hd}) for $h^d_{13}$ and $h^d_{23}$, we obtain nonzero
components of ${\tilde h}^u$ as
\begin{align}
    {\tilde h}^d_{13} &= V^*_{ud} h^d_{13}+ V^*_{cd} h^d_{23}=\frac{\sqrt{2}m_b}{v\sin\beta}\Big(V^*_{ud}V_{ub} +V^*_{cd}V_{cb}\Big)=1.80\times 10^{-2}\Big(\frac{m_b}{v\sin\beta}\Big),  \nonumber\\
    {\tilde h}^d_{23}&= V^*_{us} h^d_{13}+ V^*_{cs} h^d_{23}= \frac{\sqrt{2}m_b}{v\sin\beta}\Big(V^*_{us} V_{ub}+V^*_{cs} V_{cb}\Big)=5.77\times 10^{-2}\Big(\frac{m_b}{v\sin\beta}\Big), \nonumber\\
    {\tilde h}^d_{33}&= V^*_{ub}h^d_{13}+ V^*_{cb} h^d_{23}=\frac{\sqrt{2}m_b}{v\sin\beta}\Big( V^*_{ub} V_{ub}+ V^*_{cb}V_{cb}\Big)=2.41\times 10^{-3}\Big(\frac{m_b}{v\sin\beta}\Big).
\end{align}

\section{{\boldmath$U(1)'$} interactions}
\label{app:u1_ints}

The gauge kinetic terms and mass terms for $U(1)'$ and $U{(1)}_Y$ are
\begin{align}
  {\cal L}_{\rm g. kin}=
  & -\frac{1}{4} B_{\mu\nu} B^{\mu\nu}-\frac{1}{4} Z'_{\mu\nu} Z^{\prime\mu\nu}-\frac{1}{2} \sin\xi Z'_{\mu\nu} B^{\mu\nu} \nonumber \\
  &-\frac{1}{2}  V^T_\mu M^2_V V^\mu  ,
\end{align}
where $V_\mu= (B_\mu, W^3_\mu, Z'_\mu)^\mathsf{T}$, and
\begin{equation}
M^2_V=
\begin{pmatrix}
  m^2_Z s^2_W & -m^2_Z c_W s_W & \frac{1}{2}c^{-1}_W e g_{Z'} Q'_{H_2} v^2_2 \\ -m^2_Z c_W s_W & m^2_Z c^2_W & -\frac{1}{2}s^{-1}_W e g_{Z'} Q'_{H_2} v^2_2  \\  \frac{1}{2}c^{-1}_W e g_{Z'} Q'_{H_2} v^2_2 & -\frac{1}{2}s^{-1}_W e g_{Z'} Q'_{H_2} v^2_2 &    m^2_{Z'}
\end{pmatrix}.
\end{equation}
After diagonalizing the terms simultaneously with
\begin{align}
  \begin{pmatrix}
    B_\mu \\ W_\mu^3 \\ Z_\mu^\prime
  \end{pmatrix}
  &= \begin{pmatrix}
      c_W & -s_W & -t_\xi \\ s_W & c_W & 0  \\  0 & 0 & 1/c_\xi
    \end{pmatrix} \begin{pmatrix}
      1 & 0 & 0 \\ 0 & c_\zeta & s_\zeta  \\  0 & -s_\zeta &    c_\zeta
    \end{pmatrix} \begin{pmatrix}
      A_\mu \\ Z_{1\mu} \\ Z_{2\mu}
    \end{pmatrix} \nonumber\\
  &= \begin{pmatrix}
      c_W & -s_W c_\zeta+t_\xi s_\zeta & -s_W s_\zeta-t_\xi c_\zeta \\ s_W & c_W c_\zeta & c_W s_\zeta  \\  0 & -s_\zeta/c_\xi &    c_\zeta /c_\xi
    \end{pmatrix} \begin{pmatrix}
      A_\mu \\ Z_{1\mu} \\ Z_{2\mu}
    \end{pmatrix} ,
\end{align}
where $\zeta$ is the mass mixing angle and $s_W\equiv \sin\theta_W,
c_W\equiv \cos\theta_W$, etc, we obtain the mass eigenvalues for
massive gauge bosons:
\begin{equation}
  m^2_{Z_{1,2}}= \frac{1}{2} \Big(m^2_Z+m^2_{22}\mp
  \sqrt{(m^2_Z-m^2_{22})^2+4 m^4_{12}} \Big).
\end{equation}
Here $m^2_Z\equiv (g^2+g^2_Y) v^2/4$ and
\begin{align}
    m^2_{22}&\equiv m^2_Z s^2_W t^2_\xi + m^2_{Z'}/c^2_\xi - c^{-1}_W e g_{Z'} Q'_{H_2} v^2_w t_\xi/c_\xi, \nonumber\\
    m^2_{12}&\equiv m^2_Z s_W t_\xi - \frac{1}{2} c^{-1}_W s^{-1}_W e g_{Z'} Q'_{H_2} v^2_2/c_\xi.
\end{align}
We can rewrite the $Z$-boson like mass in terms of the heavy $Z'$ mass
and the mixing angle $\zeta$ as
\begin{equation}
m^2_{Z_1}=\frac{2m^2_Z \sec2\zeta+ m^2_{Z_2}(1-\sec 2\zeta)}{1+\sec2\zeta}, \label{Zmass}
\end{equation}
and the mixing angle as
\begin{equation}
\tan2\zeta = \frac{2m^2_{12} (m^2_{Z_2}-m^2_Z)}{(m^2_{Z_2}-m^2_Z)^2-m^4_{12}}.  \label{Zmix}
\end{equation}
We note that the modified $Z$-boson mass is constrained by electroweak
precision data, in particular, $\Delta\rho$ or $T$ parameter.

The current interactions including $Z'$ are given by
\begin{align}
  \mathcal{L}_g =
  &~B_\mu J_B^\mu + W_\mu^3 J_3^\mu + Z_\mu^\prime J_{Z^\prime}^\mu
    \nonumber\\
  =&~A_\mu J_\text{EM}^\mu + Z_{1\mu} \Big( t_\xi s_\zeta c_W
     J_\text{EM}^\mu + ( c_\zeta - t_\xi s_\zeta s_W) J_Z^\mu -
     s_\zeta J_{Z^\prime}^\mu / c_\xi \Big) \nonumber\\
  & + Z_{2\mu} \Big( -t_\xi c_\zeta c_W J_\text{EM}^\mu + (s_\zeta -
    t_\xi c_\zeta s_W) J_Z^\mu + c_\zeta J_{Z^\prime}^\mu / c_\xi \Big)
\end{align}
with
\begin{align}
    J^\mu_{\rm EM} &=  e {\bar f}\gamma^\mu Q_f f, \nonumber\\
    J^\mu_Z &=  \frac{e}{2c_W s_W} {\bar f} \gamma^\mu (\sigma^3-2s^2_W Q_f) f, \nonumber\\
    J^\mu_{Z'} &=  g_{Z'} {\bar f} \gamma^\mu Q_f' f.
\end{align}
Here $Q_f$ is the electric charge and $Q_f'$ is the $U(1)'$ charge of
fermion $f$.
For a small gauge kinetic mixing and/or the mass mixing $\zeta$, the
$Z'$-like gauge boson $Z_{2\mu}$ couples to the electromagnetic
current with the overall coefficient of $\varepsilon = t_\xi c_\zeta c_W$.

Ignoring the $Z$--$Z'$ mixing, the interaction terms for $Z'$
interactions is
\begin{align}
  {\cal L}_{Z'} =
  &~g_{Z'} Z'_\mu \Big( \frac{1}{3}x\, {\bar t}\gamma^\mu t+\frac{1}{3}x\, {\bar b}\gamma^\mu b+y {\bar\mu}\gamma^\mu \mu+y\, {\bar \nu}_\mu \gamma^\mu P_L \nu_\mu-(x+y)\,{\bar \tau}\gamma^\mu \tau \nonumber \\
  &-(x+y)\,{\bar \nu}_\tau \gamma^\mu P_L \nu_\tau+y\, {\bar \nu}_{2R}\gamma^\mu P_R \nu_{2R}-(x+y)\,{\bar \nu}_{3R} \gamma^\mu P_R \nu_{3R}\Big).
\end{align}
Now we change the basis into the one with mass eigenstates by $d_R=D_R d'_R$, $u_R=U_R u'_R$, $d_L=D_L d'_L$ and $u_L=U_L u'_L$ such that $V_{\rm CKM}=U^\dagger_L D_L$.  Taking $D_R=U_L=1$ and $D_L=V_{\rm CKM}$, the above $Z'$ interactions become
\begin{align}
  {\cal L}_{Z'}=
  &~g_{Z'} Z'_\mu \Big( \frac{1}{3}x\, {\bar t}'\gamma^\mu P_L t'+ \frac{1}{3}x\,\frac{v^2\cos^2\beta |y^u_{33}|^2}{2m^2_t} {\bar t}'\gamma^\mu P_R t'+\frac{1}{3}x\, {\bar d}'_i \gamma^\mu \Gamma^{dL}_{ij}P_L d'_j+\frac{1}{3}x\, {\bar b}'\gamma^\mu P_R b'  \nonumber \\
  &\qquad\quad
    +y {\bar\mu}\gamma^\mu \mu-(x+y)\,{\bar \tau}\gamma^\mu \tau+ y\, {\bar \nu}_\mu \gamma^\mu P_L \nu_\mu-(x+y)\,{\bar \nu}_\tau \gamma^\mu P_L \nu_\tau \nonumber \\
  &\qquad\quad
    +y\, {\bar \nu}_{2R}\gamma^\mu P_R \nu_{2R}-(x+y)\,{\bar \nu}_{3R} \gamma^\mu P_R \nu_{3R}\Big),\end{align}
where
\begin{align}
  \Gamma^{dL}
  &\equiv V^\dagger_{\rm CKM}
    \begin{pmatrix}
      0 & 0 & 0 \\ 0 & 0 & 0 \\ 0 & 0 & 1
    \end{pmatrix} V_{\rm CKM} \nonumber \\
  &= \begin{pmatrix}
    |V_{td}|^2 & V^*_{td} V_{ts} & V^*_{td} V_{tb} \\
    V^*_{ts} V_{td} & |V_{ts}|^2 & V^*_{ts} V_{tb} \\
    V^*_{tb} V_{td} & V^*_{tb} V_{ts} & |V_{tb}|^2
  \end{pmatrix}.
\end{align}

Considering the general mixing of $CP$-even scalars while ignoring the
$Z$--$Z'$ mixing, we obtain the interaction between neutral massive
electroweak gauge bosons ($V=W$, $Z$) and $Z'$ as
\begin{align}
  {\cal L}_{V}^{h_i} =
  &~\frac{2m^2_W}{v}\left[ (\cos\beta R_{i1}+\sin\beta R_{i2})h_i+ \frac{1}{2v}\,h^2_i \right]W_\mu W^\mu \nonumber \\
  &+ \frac{m^2_Z}{v} \left[(\cos\beta R_{i1}+\sin\beta R_{i2})h_i+
   \frac{1}{2v}\, h^2_i\right] Z_\mu Z^\mu.
\end{align}
For a negligible mixing with the singlet scalar, the above couplings become
\begin{align}
  {\cal L}_{V}^{h/H/A^0} =
  &~\frac{2m^2_W}{v} \left[-\sin(\alpha-\beta)h+\cos(\alpha-\beta)H+
    \frac{1}{2v}( h^2+H^2+(A^0)^2) \right] W_\mu W^\mu \nonumber \\
  &+ \frac{m^2_Z}{v} \left[-\sin(\alpha-\beta)h+\cos(\alpha-\beta)H
    +\frac{1}{2v}(h^2+H^2+(A^0)^2) \right] Z_\mu Z^\mu.
\end{align}
One can see that in the alignment limit with $\alpha=\beta-\pi/2$ the
gauge interactions of $h$ are the same as for the SM Higgs while the
triple couplings of heavy Higgs boson to gauge bosons vanish.

\end{document}